\documentclass[aps,twocolumn,superscriptaddress,amsmath,longbibliography]{revtex4-1}
\usepackage{graphicx,color,hyperref,xcolor}
\usepackage{soul}
\usepackage{amsfonts}
\usepackage{amssymb}
\usepackage{amsmath}
\usepackage{amsthm}
\usepackage{bm}
\usepackage{braket}
\usepackage{enumerate}

\begin{document}

\title{Non-bosonic moir\'e excitons}

\author{Tsung-Sheng Huang}
\affiliation{Joint Quantum Institute, University of Maryland, College Park, MD 20742, USA}

\author{Peter Lunts}
\affiliation{Department of Physics, Harvard University, Cambridge MA 02138, USA}
\affiliation{Joint Quantum Institute, University of Maryland, College Park, MD 20742, USA}

\author{Mohammad Hafezi}
\affiliation{Joint Quantum Institute, University of Maryland, College Park, MD 20742, USA}

\date{\today}

\begin{abstract}
Optical excitations in moir\'e transition metal dichalcogenide bilayers lead to the creation of excitons, as electron-hole bound states, that are generically considered within a Bose-Hubbard framework. 
Here, we demonstrate that these composite particles obey an angular momentum commutation relation that is generally non-bosonic.
This emergent spin description of excitons indicates a limitation to their occupancy on each site, which is substantial in the weak electron-hole binding regime.
The effective exciton theory is accordingly a spin Hamiltonian, which further becomes a Hubbard model of emergent bosons subject to an occupancy constraint after a Holstein-Primakoff transformation.
We apply our theory to three commonly studied bilayers (MoSe\textsubscript{2}/WSe\textsubscript{2}, WSe\textsubscript{2}/WS\textsubscript{2}, and WSe\textsubscript{2}/MoS\textsubscript{2}) and show that in the relevant parameter regimes their allowed occupancies never exceed three excitons.
Our systematic theory provides guidelines for future research on the many-body physics of moir\'e excitons.
\end{abstract}

\maketitle

\textit{Introduction.} --- 
Quantum simulation of the paradigmatic Bose-Hubbard (BH) model has recently become a powerful approach to investigate the many-body physics of interacting bosons, including incompressible states, superfluidity, and spatial coherence~\cite{Carusotto2013,Hartmann2016,Gross2017,Carusotto2020}.
These phenomena are believed to exist in different parameter regimes of the model Hamiltonian, and their study requires the ability of the platform to scan over large ranges of energy and filling fractions.
One of the candidate simulators is exciton (electron-hole bound state) degrees of freedom in moir\'e transition metal dichalcogenide (TMD) bilayers~\cite{Rivera2018Interlayer,Wu2018Theory,Seyler2019Signatures,Tran2019Evidence,Yuan2020Twist,Karni2022Structure,Naik2022Intralayer,Yu2022Moire,Zeng2022Strong}, owing to their high tunability in tunneling and interaction strengths via twisting angle~\cite{Wu2018Hubbard,Gotting2022}, and in filling by pump power~\cite{Miao2021Strong,Park2023Dipole,Gao2023,Xiong2023}.
These composite particles have recently been theoretically studied in the BH framework to investigate various many-body phenomena~\cite{Gotting2022,CamachoGuardian2022Moire,CamachoGuardian2022Optical,Remez2022Leaky}.

\begin{figure}[t]
\centering
\includegraphics[width=\columnwidth]{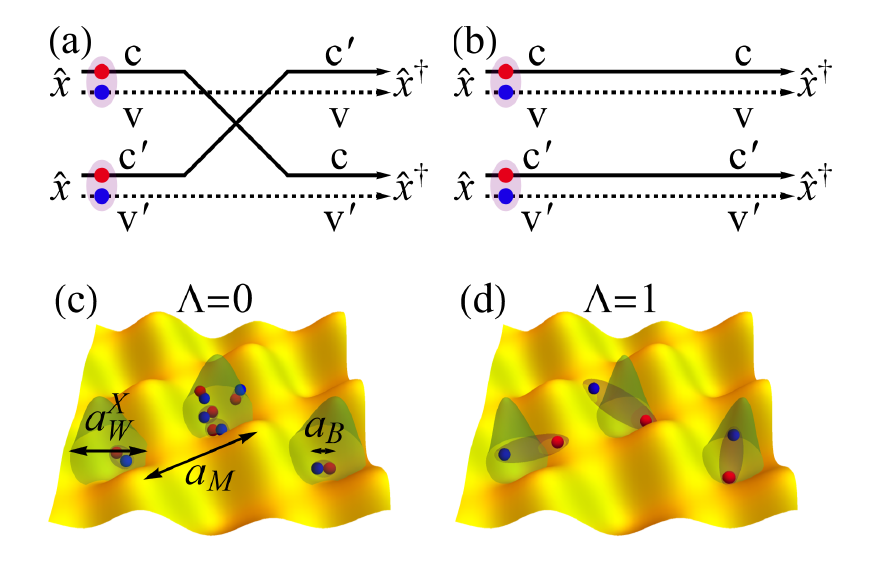}
\caption{
(a) Diagram for the charge exchange scattering between two excitons $\hat{x}$. The strength is captured by the exchange integral $\Lambda$.
In such a process, two incoming excitons (purple-shaded) swap their electrons ($c$ and $c'$, blue dots) while keeping the hole ones ($v$ and $v'$, red dots).
Note that the hole-exchanged diagram is topologically equivalent. 
(b) Disconnected diagram for a two-exciton process reminiscent of two free bosons.
(c) Illustration of moir\'e excitons on top of a superlattice potential (yellow) when charge exchange is suppressed ($\Lambda=0$), allowing for arbitrary exciton occupation.
This situation occurs at $a_W^x\gg a_B$, where $a_W^x$ is the center-or-mass width of the exciton Wannier orbital (Green) and $a_B$ is the Bohr radius.
$a_M$ is the superlattice constant.
(d) Moir\'e excitons with the strongest charge exchange ($\Lambda=1$), occuring at $a_W^x\ll a_B$.
Double occupancies (per supersite and valley) are prohibited in this case.
}
\label{Fig_illustration}
\end{figure}

However, a fundamental assumption of this agenda is that moir\'e excitons are \textit{bosonic} degrees of freedom. 
This is not always true since (generically) two-fermion states can differ from elementary bosons via their non-trivial commutation relations ~\cite{Haug1984Electron,Combescot2007,Combescot2008,Combescot2009}.
This difference results from the process illustrated in Fig.~\ref{Fig_illustration}(a), where electrons from two excitons exchange without swapping their holes~\footnote{In general, holes from two excitons could also exchange without swapping their electrons. Nevertheless, it is equivalent to the electron-exchange process if all incoming and outgoing excitons are the same.}.
Excitonic bound states thus inherit Pauli blockade from such exchange processes, limiting their occupation~\cite{Imamoglu1998,Thilagam2013,Thilagam2015}.
This effect is referred to as \textit{phase space filling} (PSF) and becomes more important as the filling of the (composite) excitons increases.
Two limiting scenarios of the PSF effect for excitons are (a) nearly-bosonic Wannier-Mott excitons in large systems where excitons are dilute ~\cite{Haug2004,Laussy2006} and (b) quantum dots~\cite{Laussy2006} and Frenkel excitons~\cite{Agranovich1968} in organic semiconductors~\cite{Betzold2020,Yagafarov2020} where the exciton wavefunctions overlap significantly and therefore deviate largely from simple bosons.

In this work, we highlight the importance of PSF for moir\'e excitons and show that the currently main moir\'e TMDs platforms are in the regime of intermediate exciton statistics (between the two limiting scenarios noted above).
Specifically, we find that PSF forbids the lowest exciton state at a supersite $\bm{R}$ and valley pseudospin $\tau$ from having an occupancy of more than $\nu_{\mathrm{max}}$.
This occupancy bound $\nu_{\mathrm{max}}$ is smaller for composite particles with stronger charge exchange, which competes with the four-fermion processes that keep the composite objects fully intact (thus as if they are elementary particles), as shown in Figs.~\ref{Fig_illustration}(a) and (b), respectively. 
The strength of the charge-swapping process is captured by the exchange integral $\Lambda$, which depends primarily on the ratio of the Bohr radius $a_B$ and the size of exciton Wannier orbital $a_W^x$. 
The limit $a_B\ll a_W^x$ yields $\Lambda\to 0$, allowing generic occupation and nearly bosonic statistics [see Fig.~\ref{Fig_illustration}(c)]. The other limit $a_B \gg a_W^x$ gives $\Lambda \to 1$, blocking double occupancy at a given $(\bm{R},\tau)$ [see Fig.~\ref{Fig_illustration}(d)]. We referred to these limits as the strong Coulomb and deep moir\'e regimes, respectively.
All of this is made possible by the moir\'e potential, which generates a large lattice spacing $a_M \gg a_B, a_W^x$ so that both limiting behaviors can occur inside a single supercell.
Meanwhile, competition between the electron-hole correlation and the moir\'e potential tunes the ratio $a_B / a_W^x$.

\begin{figure}[t]
\centering
\includegraphics[width=\columnwidth]{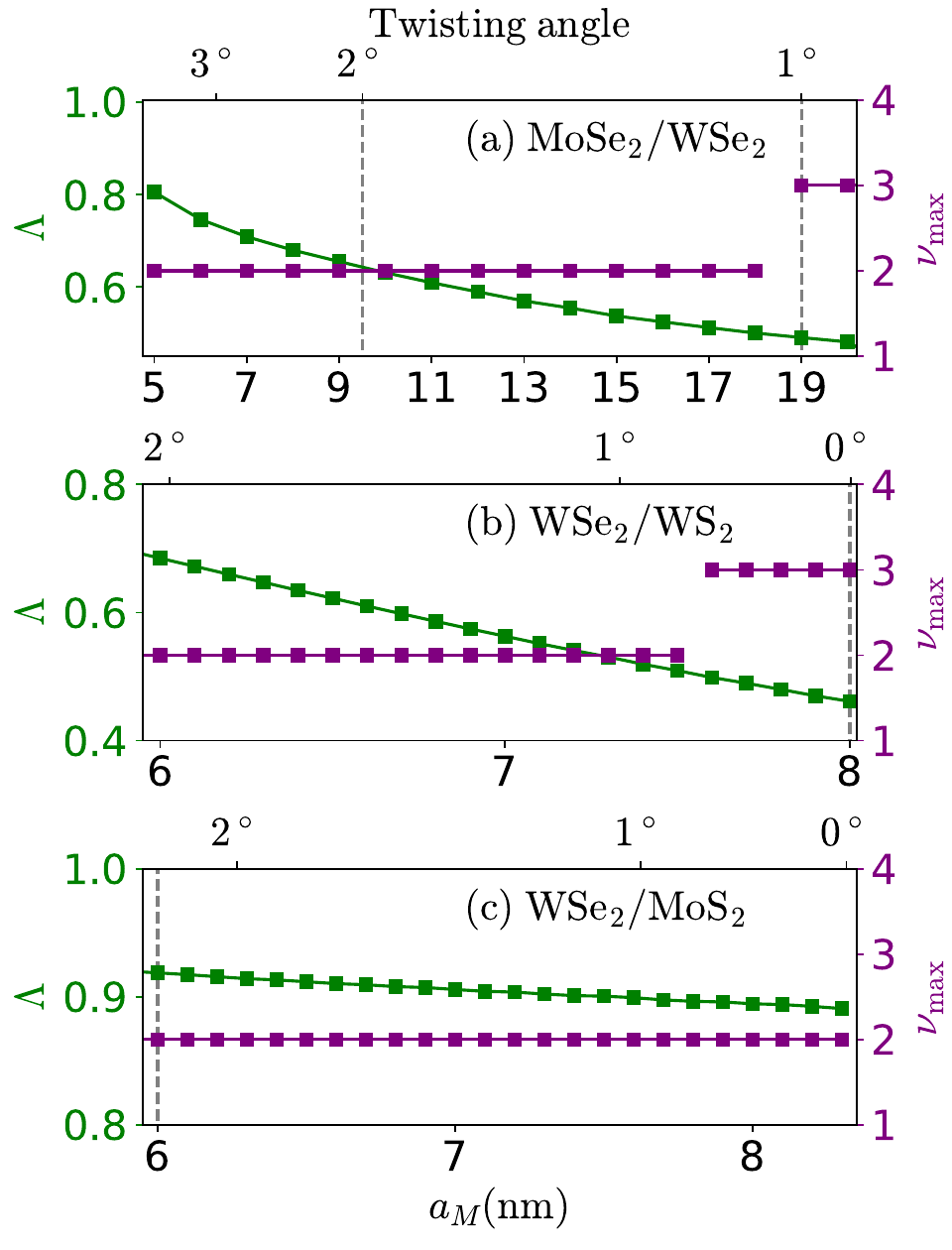}
\caption{
Exchange integral $\Lambda$ and the corresponding occupancy upper bound $\nu_{\mathrm{max}}$ from parameters relevant to different R-stacked TMD bilayers at various twisting angles (see Supplementary material~\cite{Supplement} and also references~\cite{Wu2018Theory,Tran2019Evidence,Conti2020,Park2023Dipole,Zhang2020,Karni2022Structure}).
Dashed vertical lines correspond to twisting angles realized in literature~\cite{Tran2019Evidence,Park2023Dipole,Karni2022Structure}.
}
\label{Fig_exchange_integral_exp}
\end{figure}

Utilizing the experimentally relevant parameters from R-stacked MoSe\textsubscript{2}/WSe\textsubscript{2}, WSe\textsubscript{2}/WS\textsubscript{2}, and WSe\textsubscript{2}/MoS\textsubscript{2} over a range of realistic twisting angles, we find $1/3<\Lambda<1$, corresponding to $3\geq\nu_{\mathrm{max}}\geq2$ (see Fig.~\ref{Fig_exchange_integral_exp}).
Such a restrictive occupation demonstrates the presence of strong PSF effects for moir\'e excitons and is consistent with recent experimental observations in WSe\textsubscript{2}/WS\textsubscript{2}~\cite{Park2023Dipole}.

Moreover, we find an emergent spin description of exciton that captures their non-bosonic features.
In particular, as the Hilbert space of angular momentum operators is truncated, a spin model as the effective exciton theory naturally incorporates the occupancy constraint.
We derive this Hamiltonian explicitly, using the exciton wavefunctions obtained from the solution of the two-body electron-hole Schr\"{o}dinger equation, with parameters being in the experimentally relevant regimes.

These emergent spins are further mappable into $(\nu_{\mathrm{max}}+1)$-order hardcore bosons~\cite{Paredes2007,Daley2009,Diehl2010,Mazza2010,Bonnes2011,Kapit2013,Hafezi2014,Singh2014} utilizing the Holstein-Primakoff (HP) transformation. 
Transforming the effective Hamiltonian accordingly, we find these effective bosons interact through a two-body repulsion and an infinite $(\nu_{\mathrm{max}}+1)$-body interaction, which captures their hardcore nature. 
Such a high-order repulsion between bosons can lead to exotic many-body effects in various systems.
For instance, three-body interaction could yield fractional Chern physics such as Pfaffian states in one-dimensional lattice~\cite{Paredes2007} and non-Abelian anyons in two dimensions~\cite{Kapit2013,Hafezi2014}.
Together with two-particle attraction, it is predicted to give stable droplet-like condensates with scale-invariant density~\cite{Petrov2014,Bulgac2022} and pair (dimer) superfluidity~\cite{Diehl2010,Bonnes2011,Singh2014}.
Even higher-order interactions could emerge from spin models~\cite{Kapit2013}.
These exotic hardcore particles have not been realized in experiments (to the best of our knowledge) and our work points to moir\'e excitons as a more natural platform to explore them.

\textit{Microscopic model.} --- Stacking two monolayers at a distance $d_z$ with a twist angle or a lattice mismatch leads to a bilayer with an enlarged periodicity $a_M$ compared to those of the monolayers.
Accordingly, in addition to the band energies, charges therein feel emergent superlattice potentials, which are invariant under translation with superlattice vectors $\bm{a}_M$.
Incorporating Coulomb interactions in addition to single-charge dynamics, we have the microscopic two-body electron-hole Hamiltonian $\hat{H}_{\mathrm{eh}}=\hat{H}_{\mathrm{eh}}^0+\hat{V}$. 
The non-interacting sector is:
\begin{equation}
\hat{H}_{\mathrm{eh}}^0
=
\sum_{\lambda,\tau}
\int d\bm{r}
\hat{\psi}_{\lambda,\tau}^\dagger(\bm{r})
\hat{h}_\lambda(\bm{r})
\hat{\psi}_{\lambda,\tau}(\bm{r})
,
\end{equation}
where $\lambda\in\{c,v\}$ labels the bands, $\tau\in\{+,-\}$ denotes the valley pseudospin (spin index is absent due to spin-valley locking in TMDs~\cite{Rivera2018Interlayer}) and $\bm{r}$ is the position variable.
$\hat{\psi}_{c,\tau}(\bm{r})$ and $\hat{\psi}_{v,\tau}(\bm{r})$ are the annihilation operators for conduction band electrons and valence band holes, respectively. 
$\hat{h}_\lambda(\bm{r})=-\frac{\hbar^2\nabla_{\bm{r}}^2}{2m_\lambda}+\Delta_\lambda(\bm{r})$ is the energy operator describing a single $\lambda$-band charge with mass $m_\lambda$ moving within moir\'e potential $\Delta_\lambda(\bm{r})$.
These charges interact through interaction $\hat{V}$, which we model by the following density-density interaction:
\begin{equation}
\hat{V}
=
\frac{e^2}{\epsilon_r}
\int d\bm{r} d\bm{r}'
\left[
\frac{\sum_{\lambda}
\hat{\rho}_{\lambda}(\bm{r})
\hat{\rho}_{\lambda}(\bm{r}')
}{2|\bm{r}-\bm{r}'|}
-
\frac{
\hat{\rho}_c(\bm{r})
\hat{\rho}_v(\bm{r}')
}{|\bm{r}-\bm{r}'+d_z\bm{e}_z|}
\right]
,
\end{equation}
with electric charge $e$ and dielectric constant $\epsilon_r$ characterizing the Coulomb potential.
$\hat{\rho}_{\lambda}(\bm{r})=\sum_{\tau}\hat{\psi}_{\lambda,\tau}^\dagger(\bm{r})\hat{\psi}_{\lambda,\tau}(\bm{r})$ captures the charge density at $\lambda$ band.
The displacement between layers $d_z\bm{e}_z$ is present in the electron-hole attraction because opposite charges localize at different layers.
Finally, note that we neglect intervalley scattering~\cite{Huang2023,Rivera2018Interlayer} for simplicity~\cite{Huang2023}. 

\textit{Single exciton states.} ---
A conduction band electron can bind to a valence band hole and form an exciton.
To find the corresponding two-particle energies and eigenfunctions, we perform the ladder-diagram calculation~\cite{Supplement} from $\hat{H}_{\mathrm{eh}}$.
Summation of these diagrams corresponds to the single exciton states described by the following Schr\"{o}dinger equation:
\begin{equation}
\label{eq:Moire_exciton_eigen_eqn}
\left[
\sum_\lambda
\hat{h}_\lambda(\bm{r}_\lambda)
-
\frac{e^2}{\epsilon_r \sqrt{\bm{r}_l^2 + d_z^2}}
-E_{n,\bm{Q}}
\right]
\phi_{n,\bm{Q}}(\bm{r}_c,\bm{r}_v)
=
0
,
\end{equation}
where $\bm{r}_c$ and $\bm{r}_v$ are the positions of the electron and hole, respectively, and $\bm{r}_l=\bm{r}_c-\bm{r}_v$ is the relative coordinate.
$n\in\{0,1,2,...\}$ labels all internal states such as the excitonic moir\'e bands and levels from the relative motion.
$\bm{Q}$ is the total superlattice momentum. 
$\phi_{n,\bm{Q}}(\bm{r}_c,\bm{r}_v)$ is the corresponding exciton Bloch wavefunction (valley index suppressed for simplicity) with energy $E_{n,\bm{Q}}$.
Note that $\hat{h}_\lambda(\bm{r}_\lambda)$ includes the moir\'e potential, making this two-body Schr\"{o}dinger equation distinct from the one for hydrogenic excitons~\cite{Haug1984Electron}.
Fourier transforming the Bloch wavefunctions gives the Wannier orbitals:
\begin{equation}
\label{eq:Exciton_Wannier_state}
W_{n,\bm{R}}(\bm{r}_c,\bm{r}_v)
=
\frac{1}{\sqrt{N}}
\sum_{\bm{Q}}
e^{-i\bm{Q}\cdot\bm{R}}
\phi_{n,\bm{Q}}(\bm{r}_c,\bm{r}_v)
,
\end{equation}
where $N$ denotes the number of supersites in the system.
$\bm{R}$ is any of the (periodically spaced) minima of the overall moir\'e potential~\cite{Supplement} for the center-of-mass position $\bm{r}_x=(m_c\bm{r}_c+m_v\bm{r}_v)/M$, where $M=m_c+m_v$.
We work with these localized orbitals instead of Bloch wavefunctions to capture correlations within a moir\'e unit cell and focus on the lowest state $w_{\bm{R}}(\bm{r}_c,\bm{r}_v)\equiv W_{0,\bm{R}}(\bm{r}_c,\bm{r}_v)$ for simplicity.
The corresponding exciton creation operator is:
\begin{equation}
\label{eq:exciton_operator}
\hat{x}_{\tau;\bm{R}}^\dagger
=
\int
d\bm{r}_c
d\bm{r}_v
w_{\bm{R}}(\bm{r}_c,\bm{r}_v)
\hat{\psi}_{c,\tau}^\dagger(\bm{r}_c)
\hat{\psi}_{v,\tau}^\dagger(\bm{r}_v)
.
\end{equation}

\textit{Exciton statistics.} --- With these composite operators in hand, we proceed to their commutation relations, starting with the states with distinct $\tau$ or $\bm{R}$ labels.
Excitons at opposite $\tau$ commute by definition, whereas rigorously speaking, this is not the case for those at different $\bm{R}$.
Nevertheless, off-site statistics are negligible because of the suppressed orbital overlap, due to the typically large $a_M$ compared to the Wannier orbital size $a_W^x$, defined as root mean square of $\bm{r}_x-\bm{R}$ computed with probability density $|w_{\bm{R}}(\bm{r}_c,\bm{r}_v)|^2$, and Bohr radius $a_B=\frac{\epsilon_r\hbar^2}{\mu e^2}$ (with reduced mass $\mu=\frac{m_cm_v}{M}$)~\cite{Park2023Dipole,Karni2022Structure}.
Combining these arguments, we find $[\hat{x}_{\tau;\bm{R}},\hat{x}_{\tau';\bm{R}'}^\dagger] \propto \delta_{\tau,\tau'}\delta_{\bm{R},\bm{R}'}$.

In contrast, the equal-site-valley commutator is non-trivial.
In particular, we evaluate $[\hat{x}_{\tau;\bm{R}},\hat{x}_{\tau;\bm{R}}^\dagger]-1$ in the charge basis and find it nonzero but an operator, which further yields:
\begin{equation}
\label{eq:x_comm_2nd}
[
[\hat{x}_{\tau;\bm{R}},\hat{x}_{\tau;\bm{R}}^\dagger]
,\hat{x}_{\tau;\bm{R}}^\dagger]
\simeq
-2\Lambda
\hat{x}_{\tau;\bm{R}}^\dagger
,
\end{equation}
when higher orbitals are dropped (which leads to a self-consistent treatment, justified in the Supplementary material ~\cite{Supplement}).
The exchange integral $\Lambda$ has the following expression (denoting $d^8r\equiv d\bm{r}_cd\bm{r}_vd\bm{r}'_cd\bm{r}'_v$):
\begin{equation}
\label{eq:Lambda_0000}
\begin{aligned}
\Lambda
=
\int
d^8r
w_{\bm{R}}^\ast(\bm{r}_c,\bm{r}'_v)
w_{\bm{R}}(\bm{r}_c,\bm{r}_v)
w_{\bm{R}}(\bm{r}'_c,\bm{r}'_v)
w_{\bm{R}}^\ast(\bm{r}'_c,\bm{r}_v)
,
\end{aligned}
\end{equation}
which captures the strength of charge exchange processes between two excitons [see Fig~\ref{Fig_illustration}(a)].
Notably, $|\Lambda|^2\leq1$ from completeness of the orbitals~\cite{Combescot2008} and becomes smaller with wider orbitals until $\Lambda\simeq0$, which yields a bosonic commutation relation for $\hat{x}_{\tau;\bm{R}}$.

\textit{Emergent spins and bosons.} --- Eq.~\eqref{eq:x_comm_2nd} yields the standard relations for angular momentum operators $[\hat{\mathcal{S}}_{\tau;\bm{R}}^+,\hat{\mathcal{S}}_{\tau;\bm{R}}^-]=2\hat{\mathcal{S}}_{\tau;\bm{R}}^z$ and $[\hat{\mathcal{S}}_{\tau;\bm{R}}^z,\hat{\mathcal{S}}_{\tau;\bm{R}}^-]=-\hat{\mathcal{S}}_{\tau;\bm{R}}^-$ upon the following substitution:
\begin{equation}
\label{eq:effective_spin_rep}
\frac{\hat{x}_{\tau;\bm{R}}^\dagger}{\sqrt{\Lambda}}
=
\hat{\mathcal{S}}_{\tau;\bm{R}}^-
,\;
\frac{\hat{x}_{\tau;\bm{R}}}{\sqrt{\Lambda}}
=
\hat{\mathcal{S}}_{\tau;\bm{R}}^+
,\;
\frac{
[\hat{x}_{\tau;\bm{R}},\hat{x}_{\tau;\bm{R}}^\dagger]
}{2\Lambda}
=
\hat{\mathcal{S}}_{\tau;\bm{R}}^z
.
\end{equation}
We note that the largest eigenvalue of $\hat{\mathcal{S}}_{\tau;\bm{R}}^z$, $(2\Lambda)^{-1}$, does not have to be integer multiples of $\frac{1}{2}$ because these emergent angular momentum operators are not generators of rotations.
Besides this spin representation, the HP transformation~\cite{Auerbach1994} indicates the following emergent boson description $\hat{a}_{\tau;\bm{R}}$ for $\hat{x}_{\bm{R},\tau}$:
\begin{equation}
\label{eq:effective_boson_rep}
\hat{x}_{\tau;\bm{R}}
\simeq
\theta(1-\Lambda\hat{a}_{\tau;\bm{R}}^\dagger\hat{a}_{\tau;\bm{R}}) \,
\sqrt{1-\Lambda\hat{a}_{\tau;\bm{R}}^\dagger\hat{a}_{\tau;\bm{R}}}
\; \hat{a}_{\tau;\bm{R}}
,
\end{equation}
where $\theta(x)$ is the step function.
We refer to the Supplementary Material~\cite{Supplement} for the derivation of these representations.

\textit{Phase space filling.} --- Both Eq.~\eqref{eq:effective_spin_rep} and Eq.~\eqref{eq:effective_boson_rep} suggest a limit for the exciton Hilbert space size.
To obtain such bound, we compute $C^{(\nu)}\equiv|(\hat{x}_{\tau;\bm{R}}^\dagger)^{\nu}|\mathrm{vac}\rangle|^2$ for generic positive integer $\nu$, which becomes: 
\begin{equation}
\label{eq:C_nu_approx}
C^{(\nu)}
\simeq
\theta(1-\Lambda(\nu-1))
\nu!\prod_{j=0}^{\nu-1}
(1-\Lambda j)
.
\end{equation}
The physical condition $C^{(\nu)}>0$ suggest exciton occupancy per $(\bm{R},\tau)$ not exceed an upper bound $\nu_{\mathrm{max}}$, where:
\begin{equation}
\label{eq:nu_bound}
\nu_{\mathrm{max}}
=
\mathrm{ceil}(\Lambda^{-1})
,\quad
(\hat{x}_{\tau;\bm{R}})^{\nu_{\mathrm{max}}+1}=0
,
\end{equation}
with $\mathrm{ceil}(x)$ denoting the least integer not smaller than $x$.
Such a restriction exists as long as excitons deviate from bosons ($\Lambda\neq0$), and the extreme case $\nu_{\mathrm{max}}=1$ corresponds to $\Lambda=1$.

\textit{Effective models.} --- We derive an effective exciton Hamiltonian $\hat{\mathcal{H}}_{\mathrm{eff}}$ from $\hat{H}_{\mathrm{eh}}$ in both the emergent spins and boson representations (see Supplementary material~\cite{Supplement} for the spin representation).
For generic $\Lambda$~\footnote{In the derivation of $\hat{\mathcal{H}}_{\mathrm{eff}}$ in the emergent boson representation, we do not implement the large-spin (small $\Lambda$) approximation that is typically applied after the Holstein-Primakoff map}, such a model has the expression below in terms of $\hat{a}_{\bm{R},\tau}$:
\begin{equation}
\label{eq:H_X_plus_tilde_U}
\begin{aligned}
\hat{\mathcal{H}}_{\mathrm{eff}}
&=
E_0
\sum_{\bm{R},\tau}\hat{a}_{\tau;\bm{R}}^\dagger\hat{a}_{\tau;\bm{R}}
-
t
\sum_{\langle\bm{R}',\bm{R}\rangle}
(
\hat{a}_{\tau;\bm{R}'}^\dagger
\hat{a}_{\tau;\bm{R}}
+
\mathrm{H.c.}
)
\\
&
+
\frac{U}{2}
\sum_{\bm{R},\tau,\tau'}
\hat{a}_{\tau;\bm{R}}^\dagger
\hat{a}_{\tau';\bm{R}}^\dagger
\hat{a}_{\tau';\bm{R}}
\hat{a}_{\tau;\bm{R}}
\\
&+
\lim_{\tilde{U}\to\infty}
\tilde{U}
\sum_{\bm{R},\tau}
(\hat{a}_{\tau;\bm{R}}^{\dagger})^{\nu_{\mathrm{max}}+1}
(\hat{a}_{\tau;\bm{R}})^{\nu_{\mathrm{max}}+1}
.
\end{aligned}
\end{equation}
Here, the hopping $t$ and on-site repulsion $U$ are assumed to satisfy $t/U \ll 1$ (otherwise, off-site exciton commutator plays an essential role in the hopping  --- see supplementary material~\cite{Supplement}).
These quantities (and the single exciton occupation energy $E_0$) could vary with different bilayers and twisting angles because they are integrals involving Wannier orbitals (see Supplementary material~\cite{Supplement} for details).
Note the presence of the infinite high-order interaction besides the two-body interaction $U$, which results from the highly nonlinear transformation Eq.~\eqref{eq:effective_boson_rep} and is a manifestation of the bound $\nu_{\mathrm{max}}$.

Such an inaccuracy of a two-body interaction could lead to a qualitative change in optical spectra when the exciton occupancy crosses $\nu_{\mathrm{max}}$. 
Below this critical filling, energy differences between $\nu$- and $(\nu+1)$-boson states (valley-polarized) are roughly $E_0+\nu U$, and the corresponding transitions provide a series of peaks in optical spectra separated by a splitting $U$~\cite{Park2023Dipole}.
In contrast, adding additional excitons onto a $\nu_{\mathrm{max}}$-filled site would populate higher energy states such that this splitting is generally not $U$.

Finally, we note that PSF also affects light-matter interaction in moir\'e TMDs.
More specifically, under dipole approximation~\cite{Haug2004}, photons hybridize linearly with $\hat{x}_{\tau;\bm{R}}$, or equivalently the emergent spins, because the absorption of each photon generates an additional electron-hole pair.
Such an effect from non-bosonic excitons has not been considered in the state-of-the-art models (to the best of our knowledge) for optical properties of moir\'e excitons~\cite{CamachoGuardian2022Moire,CamachoGuardian2022Optical,Remez2022Leaky,Santiago-Garcia2023}, in which photon couples linearly with bosonic degrees of freedom.

\textit{Numerical results.} --- We compute $\Lambda$ and $\nu_{\mathrm{max}}$ from the numerical solution to Eq.~\eqref{eq:Moire_exciton_eigen_eqn}~\cite{Supplement}.
Therein, we assume the moir\'e potentials for both charges to be the same and have the following expression for simplicity:
\begin{equation}
\label{eq:moire_potential}
\Delta_c(\bm{r})
=
\Delta_v(\bm{r})
=
Re\left[
Z
\sum_{j=1}^3
e^{i\bm{r}\cdot\bm{G}_j}
\right]
,
\end{equation}
where $Z$ is a complex number characterizing these potentials.
$\bm{G}_{1,2,3}$ denote the reciprocal superlattice vectors, being rotations of $\frac{4\pi}{\sqrt{3}a_M}\bm{e}_y$ by multiples of $120^\circ$ ($\bm{e}_{x,y}$ being Cartesian unit vectors).
We focus on interlayer excitons, with parameters taken from the literature for MoSe\textsubscript{2}/WSe\textsubscript{2}, WSe\textsubscript{2}/WS\textsubscript{2}, and WSe\textsubscript{2}/MoS\textsubscript{2} (all materials R-stacked)~\cite{Supplement}.

Fig~\ref{Fig_exchange_integral_exp} shows $\Lambda$ and $\nu_{\mathrm{max}}$ for different bilayers at various superlattice spacings.
They generally give $\Lambda>1/3$ and thus $\nu_{\mathrm{max}}\leq 3$.

In addition, $\Lambda$ systematically
decreases with wider $a_M$.
Qualitatively, a larger superlattice corresponds to smaller charge moir\'e bandgaps, allowing Coulomb binding to mix more charge states to form an exciton.
Accordingly, the PSF effect is weaker for larger $a_M$, corresponding to a smaller $\Lambda$.

\begin{figure}[t]
\centering
\includegraphics[width=\columnwidth]{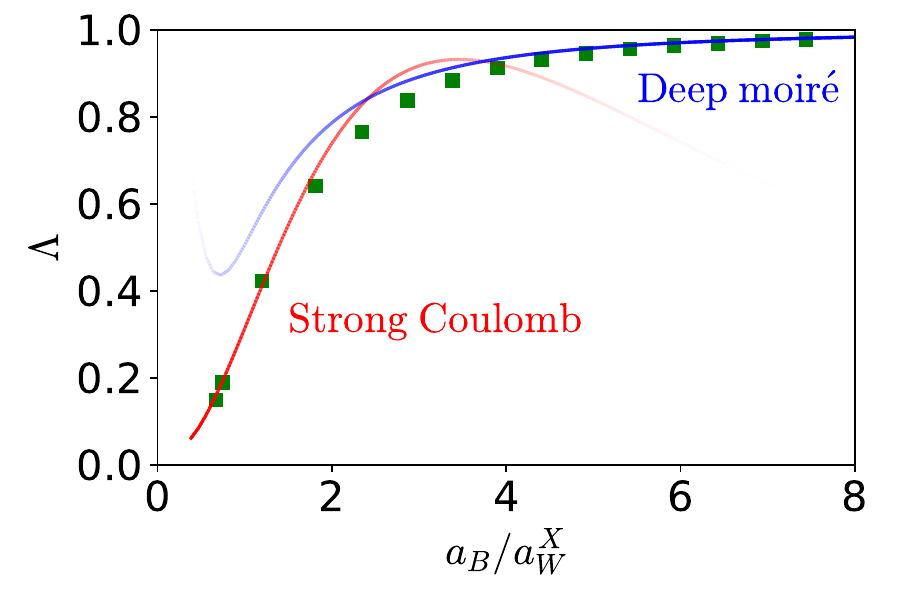}
\caption{
Exchange integral $\Lambda$ at various $a_B/a_W^x$ for interlayer distance $d_z=0$ (or equivalently intralayer excitons).
$a_W^x$ is tuned by $|Z|$, considered as a free parameter in this plot.
Data shown utilizes parameters (except $|Z|$) from WSe\textsubscript{2}/WS\textsubscript{2}~\cite{Supplement} at $a_M=8$nm.
Red and blue curves are from perturbative wavefunctions in the strong Coulomb and deep moir\'e regimes, respectively~\cite{Supplement}, with values in their regimes of validity (indicated by opacity) perfectly matching the numerics.
}
\label{Fig_exchange_integral}
\end{figure}

Comparison between length scales provides an alternative explanation. 
An exciton with its center-of-mass fluctuation $a_W^x$ more significant than $a_B$ possesses a strong electron-hole correlation.
As a consequence, charge exchange processes [see Fig.~\ref{Fig_illustration}(a)], the rate of which are captured by $\Lambda$, are weaker.
In contrast, $a_W^x\ll a_B$ implies negligible Coulomb binding, giving a nearly uncorrelated fermion pair.
In this situation, the amplitudes for processes Fig.~\ref{Fig_illustration}(c) and (d) are comparable such that $\Lambda\simeq 1$.
We confirm this understanding with Fig.~\ref{Fig_exchange_integral} (in which we set $d_z=0$ to focus on the length scales mentioned above), showing $\Lambda$ for a broader range of $a_W^x/a_B$ (achieved by manually tuning $|Z|$).
Thus, as larger $a_M$ provides wider $a_W^x$, $\Lambda$ drops, as confirmed in Fig.~\ref{Fig_exchange_integral_exp} for the more realistic $d_z \neq 0$ setting.

Finally, we benchmark our numerical results for $d_z = 0$ with perturbation theories in the strong Coulomb and deep moir\'e regimes~\cite{Supplement}.
In the strong Coulomb regime, electron-hole attraction dominates over moir\'e potential in the relative motion, whereas such binding interaction is perturbative in the deep moir\'e limit. 
As Fig.~\ref{Fig_exchange_integral} shows, our numerical results reproduce the analytical solutions in these regimes.

\textit{Conclusion and outlook.} --- We have demonstrated that moir\'e excitons in TMD bilayers can be very non-bosonic.
Due to their composite nature, they can experience a strong PSF effect from their constituent fermionic charges.
In particular, the occupancy of the lowest bound states cannot exceed $\nu_{\mathrm{max}}$, which depends on their commutation relation.
Nevertheless, we have shown they can be mapped to effective spin and bosonic operators, which leads the microscopic electron-hole Hamiltonian to an interacting spin model and an occupancy-constrained BH description, respectively, for moir\'e excitons.
Thus, we anticipate these composite particles being platforms for BH physics below the critical filling but not above it.

We expect this restriction to manifest in power-varying optical measurements, offering exciton filling tunability.
These experiments have demonstrated spectral jumps with increasing pumps, interpreted as Hubbard energy~\cite{Miao2021Strong,Gotting2022,Park2023Dipole,Gao2023,Xiong2023}.
Yet our results suggest this understanding is invalid at occupancy above $\nu_{\mathrm{max}}$.
Optical pumping above the corresponding critical power would introduce higher exciton states rather than adding the original ones, leading to distinct spectral jumps from those at lower power.
This prediction is consistent with experimental results for nearly untwisted WSe\textsubscript{2}/WS\textsubscript{2}~\cite{Park2023Dipole} --- our theory gives $\nu_{\mathrm{max}}=3$ so that such a shift occurs at the fourth jump.

We have also shown that $\nu_{\mathrm{max}}$ is smaller for a narrower exciton Wannier orbital or larger Bohr radius.
In particular, $a_W^x\ll a_B$ implies weak electron-hole correlation such that the fermionic nature of the constituent charges leads to a strong PSF effect.
Thus, the BH description is more constrained at deeper and narrower moir\'e potential and larger dielectric constant.
Notably, it also suggests a more restrictive bound for higher states in the relative degrees of freedom.

Such an occupation bound for excitons could be even more restrictive when the system involves doped charges~\cite{Huang2023,Gao2023,Miao2021Strong,Xiong2023,Park2023Dipole}.
This is qualitatively because these fermions already fill up a fraction of the phase space, which limits the available states for excitons.
This exciton-fermion PSF results microscopically from their charge exchange, leading to a non-trivial commutation relation between the two species.
We anticipate the presence of this in recent experiments aiming at optical signatures for the underlying electronic correlations~\cite{Gao2023,Miao2021Strong,Xiong2023,Park2023Dipole}, as excitons and doped charges coexist in these setups.

Finally, we discover that moir\'e excitons could serve as another platform for $(\nu_{\mathrm{max}}+1)$-order constrained particles.
For example, valley-polarized moir\'e excitons at $\Lambda\to1$ correspond to typical hardcore bosons, widely used to describe qubits.
Thus, we anticipate the corresponding TMDs being a platform for two-dimensional arrays of two-level emitters~\cite{Shahmoon2017,Rui2020,MorenoCardoner2021}.
At a lower $\Lambda$, moir\'e excitons become high-order constrained particles, providing many-body effects such as pair (dimer) superfluidity~\cite{Diehl2010,Bonnes2011,Singh2014}
and fractional quantum Hall physics~\cite{Paredes2007,Kapit2013,Hafezi2014}.
With these potential applications, we expect moir\'e excitons to broaden the scope of hardcore bosons.

\textit{Acknowledgements.} ---
We thank Daniel Suarez-Forero, Mahmoud Mehrabad, Michael Lindsey, Beini Gao, Supratik Sarkar, Yuxin Wang, Nigel Cooper, and Andrey Grankin for useful discussions. This work was supported by AFOSR MURI FA9550-19-1-0399, FA9550-22-1-0339, ARL W911NF1920181, DOE DE-AC02-05CH11231, Simons and Minta Martin Foundations.  P.L. acknowledges further support from the Harvard Quantum Initiative Postdoctoral Fellowship in Science and Engineering, and NSF DMR-2245246.
\bibliography{Biblio}

\begin{thebibliography}{55}%
\makeatletter
\providecommand \@ifxundefined [1]{%
 \@ifx{#1\undefined}
}%
\providecommand \@ifnum [1]{%
 \ifnum #1\expandafter \@firstoftwo
 \else \expandafter \@secondoftwo
 \fi
}%
\providecommand \@ifx [1]{%
 \ifx #1\expandafter \@firstoftwo
 \else \expandafter \@secondoftwo
 \fi
}%
\providecommand \natexlab [1]{#1}%
\providecommand \enquote  [1]{``#1''}%
\providecommand \bibnamefont  [1]{#1}%
\providecommand \bibfnamefont [1]{#1}%
\providecommand \citenamefont [1]{#1}%
\providecommand \href@noop [0]{\@secondoftwo}%
\providecommand \href [0]{\begingroup \@sanitize@url \@href}%
\providecommand \@href[1]{\@@startlink{#1}\@@href}%
\providecommand \@@href[1]{\endgroup#1\@@endlink}%
\providecommand \@sanitize@url [0]{\catcode `\\12\catcode `\$12\catcode
  `\&12\catcode `\#12\catcode `\^12\catcode `\_12\catcode `\%12\relax}%
\providecommand \@@startlink[1]{}%
\providecommand \@@endlink[0]{}%
\providecommand \url  [0]{\begingroup\@sanitize@url \@url }%
\providecommand \@url [1]{\endgroup\@href {#1}{\urlprefix }}%
\providecommand \urlprefix  [0]{URL }%
\providecommand \Eprint [0]{\href }%
\providecommand \doibase [0]{http://dx.doi.org/}%
\providecommand \selectlanguage [0]{\@gobble}%
\providecommand \bibinfo  [0]{\@secondoftwo}%
\providecommand \bibfield  [0]{\@secondoftwo}%
\providecommand \translation [1]{[#1]}%
\providecommand \BibitemOpen [0]{}%
\providecommand \bibitemStop [0]{}%
\providecommand \bibitemNoStop [0]{.\EOS\space}%
\providecommand \EOS [0]{\spacefactor3000\relax}%
\providecommand \BibitemShut  [1]{\csname bibitem#1\endcsname}%
\let\auto@bib@innerbib\@empty
\bibitem [{\citenamefont {Carusotto}\ and\ \citenamefont
  {Ciuti}(2013)}]{Carusotto2013}%
  \BibitemOpen
  \bibfield  {author} {\bibinfo {author} {\bibfnamefont {Iacopo}\ \bibnamefont
  {Carusotto}}\ and\ \bibinfo {author} {\bibfnamefont {Cristiano}\ \bibnamefont
  {Ciuti}},\ }\bibfield  {title} {\enquote {\bibinfo {title} {Quantum fluids of
  light},}\ }\href {\doibase 10.1103/RevModPhys.85.299} {\bibfield  {journal}
  {\bibinfo  {journal} {Reviews of Modern Physics}\ }\textbf {\bibinfo {volume}
  {85}},\ \bibinfo {pages} {299--366} (\bibinfo {year} {2013})}\BibitemShut
  {NoStop}%
\bibitem [{\citenamefont {Hartmann}(2016)}]{Hartmann2016}%
  \BibitemOpen
  \bibfield  {author} {\bibinfo {author} {\bibfnamefont {Michael~J}\
  \bibnamefont {Hartmann}},\ }\bibfield  {title} {\enquote {\bibinfo {title}
  {Quantum simulation with interacting photons},}\ }\href {\doibase
  10.1088/2040-8978/18/10/104005} {\bibfield  {journal} {\bibinfo  {journal}
  {Journal of Optics}\ }\textbf {\bibinfo {volume} {18}},\ \bibinfo {pages}
  {104005} (\bibinfo {year} {2016})}\BibitemShut {NoStop}%
\bibitem [{\citenamefont {Gross}\ and\ \citenamefont
  {Bloch}(2017)}]{Gross2017}%
  \BibitemOpen
  \bibfield  {author} {\bibinfo {author} {\bibfnamefont {Christian}\
  \bibnamefont {Gross}}\ and\ \bibinfo {author} {\bibfnamefont {Immanuel}\
  \bibnamefont {Bloch}},\ }\bibfield  {title} {\enquote {\bibinfo {title}
  {Quantum simulations with ultracold atoms in optical lattices},}\ }\href
  {\doibase 10.1126/science.aal3837} {\bibfield  {journal} {\bibinfo  {journal}
  {Science}\ }\textbf {\bibinfo {volume} {357}},\ \bibinfo {pages} {995--1001}
  (\bibinfo {year} {2017})}\BibitemShut {NoStop}%
\bibitem [{\citenamefont {Carusotto}\ \emph {et~al.}(2020)\citenamefont
  {Carusotto}, \citenamefont {Houck}, \citenamefont {Kollár}, \citenamefont
  {Roushan}, \citenamefont {Schuster},\ and\ \citenamefont
  {Simon}}]{Carusotto2020}%
  \BibitemOpen
  \bibfield  {author} {\bibinfo {author} {\bibfnamefont {Iacopo}\ \bibnamefont
  {Carusotto}}, \bibinfo {author} {\bibfnamefont {Andrew~A}\ \bibnamefont
  {Houck}}, \bibinfo {author} {\bibfnamefont {Alicia~J}\ \bibnamefont
  {Kollár}}, \bibinfo {author} {\bibfnamefont {Pedram}\ \bibnamefont
  {Roushan}}, \bibinfo {author} {\bibfnamefont {David~I}\ \bibnamefont
  {Schuster}}, \ and\ \bibinfo {author} {\bibfnamefont {Jonathan}\ \bibnamefont
  {Simon}},\ }\bibfield  {title} {\enquote {\bibinfo {title} {Photonic
  materials in circuit quantum electrodynamics},}\ }\href {\doibase
  10.1038/s41567-020-0815-y} {\bibfield  {journal} {\bibinfo  {journal} {Nature
  Physics}\ }\textbf {\bibinfo {volume} {16}},\ \bibinfo {pages} {268--279}
  (\bibinfo {year} {2020})}\BibitemShut {NoStop}%
\bibitem [{\citenamefont {Rivera}\ \emph {et~al.}(2018)\citenamefont {Rivera},
  \citenamefont {Yu}, \citenamefont {Seyler}, \citenamefont {Wilson},
  \citenamefont {Yao},\ and\ \citenamefont {Xu}}]{Rivera2018Interlayer}%
  \BibitemOpen
  \bibfield  {author} {\bibinfo {author} {\bibfnamefont {Pasqual}\ \bibnamefont
  {Rivera}}, \bibinfo {author} {\bibfnamefont {Hongyi}\ \bibnamefont {Yu}},
  \bibinfo {author} {\bibfnamefont {Kyle~L}\ \bibnamefont {Seyler}}, \bibinfo
  {author} {\bibfnamefont {Nathan~P}\ \bibnamefont {Wilson}}, \bibinfo {author}
  {\bibfnamefont {Wang}\ \bibnamefont {Yao}}, \ and\ \bibinfo {author}
  {\bibfnamefont {Xiaodong}\ \bibnamefont {Xu}},\ }\bibfield  {title} {\enquote
  {\bibinfo {title} {Interlayer valley excitons in heterobilayers of transition
  metal dichalcogenides},}\ }\href {\doibase 10.1038/s41565-018-0193-0}
  {\bibfield  {journal} {\bibinfo  {journal} {Nature Nanotechnology}\ }\textbf
  {\bibinfo {volume} {13}},\ \bibinfo {pages} {1004--1015} (\bibinfo {year}
  {2018})}\BibitemShut {NoStop}%
\bibitem [{\citenamefont {Wu}\ \emph {et~al.}(2018{\natexlab{a}})\citenamefont
  {Wu}, \citenamefont {Lovorn},\ and\ \citenamefont
  {MacDonald}}]{Wu2018Theory}%
  \BibitemOpen
  \bibfield  {author} {\bibinfo {author} {\bibfnamefont {Fengcheng}\
  \bibnamefont {Wu}}, \bibinfo {author} {\bibfnamefont {Timothy}\ \bibnamefont
  {Lovorn}}, \ and\ \bibinfo {author} {\bibfnamefont {A~H}\ \bibnamefont
  {MacDonald}},\ }\bibfield  {title} {\enquote {\bibinfo {title} {Theory of
  optical absorption by interlayer excitons in transition metal dichalcogenide
  heterobilayers},}\ }\href {\doibase 10.1103/PhysRevB.97.035306} {\bibfield
  {journal} {\bibinfo  {journal} {Physical Review B}\ }\textbf {\bibinfo
  {volume} {97}},\ \bibinfo {pages} {35306} (\bibinfo {year}
  {2018}{\natexlab{a}})}\BibitemShut {NoStop}%
\bibitem [{\citenamefont {Seyler}\ \emph {et~al.}(2019)\citenamefont {Seyler},
  \citenamefont {Rivera}, \citenamefont {Yu}, \citenamefont {Wilson},
  \citenamefont {Ray}, \citenamefont {Mandrus}, \citenamefont {Yan},
  \citenamefont {Yao},\ and\ \citenamefont {Xu}}]{Seyler2019Signatures}%
  \BibitemOpen
  \bibfield  {author} {\bibinfo {author} {\bibfnamefont {Kyle~L}\ \bibnamefont
  {Seyler}}, \bibinfo {author} {\bibfnamefont {Pasqual}\ \bibnamefont
  {Rivera}}, \bibinfo {author} {\bibfnamefont {Hongyi}\ \bibnamefont {Yu}},
  \bibinfo {author} {\bibfnamefont {Nathan~P}\ \bibnamefont {Wilson}}, \bibinfo
  {author} {\bibfnamefont {Essance~L}\ \bibnamefont {Ray}}, \bibinfo {author}
  {\bibfnamefont {David~G}\ \bibnamefont {Mandrus}}, \bibinfo {author}
  {\bibfnamefont {Jiaqiang}\ \bibnamefont {Yan}}, \bibinfo {author}
  {\bibfnamefont {Wang}\ \bibnamefont {Yao}}, \ and\ \bibinfo {author}
  {\bibfnamefont {Xiaodong}\ \bibnamefont {Xu}},\ }\bibfield  {title} {\enquote
  {\bibinfo {title} {Signatures of moiré-trapped valley excitons in mose2/wse2
  heterobilayers},}\ }\href {\doibase 10.1038/s41586-019-0957-1} {\bibfield
  {journal} {\bibinfo  {journal} {Nature}\ }\textbf {\bibinfo {volume} {567}},\
  \bibinfo {pages} {66--70} (\bibinfo {year} {2019})}\BibitemShut {NoStop}%
\bibitem [{\citenamefont {Tran}\ \emph {et~al.}(2019)\citenamefont {Tran},
  \citenamefont {Moody}, \citenamefont {Wu}, \citenamefont {Lu}, \citenamefont
  {Choi}, \citenamefont {Kim}, \citenamefont {Rai}, \citenamefont {Sanchez},
  \citenamefont {Quan}, \citenamefont {Singh}, \citenamefont {Embley},
  \citenamefont {Zepeda}, \citenamefont {Campbell}, \citenamefont {Autry},
  \citenamefont {Taniguchi}, \citenamefont {Watanabe}, \citenamefont {Lu},
  \citenamefont {Banerjee}, \citenamefont {Silverman}, \citenamefont {Kim},
  \citenamefont {Tutuc}, \citenamefont {Yang}, \citenamefont {MacDonald},\ and\
  \citenamefont {Li}}]{Tran2019Evidence}%
  \BibitemOpen
  \bibfield  {author} {\bibinfo {author} {\bibfnamefont {Kha}\ \bibnamefont
  {Tran}}, \bibinfo {author} {\bibfnamefont {Galan}\ \bibnamefont {Moody}},
  \bibinfo {author} {\bibfnamefont {Fengcheng}\ \bibnamefont {Wu}}, \bibinfo
  {author} {\bibfnamefont {Xiaobo}\ \bibnamefont {Lu}}, \bibinfo {author}
  {\bibfnamefont {Junho}\ \bibnamefont {Choi}}, \bibinfo {author}
  {\bibfnamefont {Kyounghwan}\ \bibnamefont {Kim}}, \bibinfo {author}
  {\bibfnamefont {Amritesh}\ \bibnamefont {Rai}}, \bibinfo {author}
  {\bibfnamefont {Daniel~A}\ \bibnamefont {Sanchez}}, \bibinfo {author}
  {\bibfnamefont {Jiamin}\ \bibnamefont {Quan}}, \bibinfo {author}
  {\bibfnamefont {Akshay}\ \bibnamefont {Singh}}, \bibinfo {author}
  {\bibfnamefont {Jacob}\ \bibnamefont {Embley}}, \bibinfo {author}
  {\bibfnamefont {André}\ \bibnamefont {Zepeda}}, \bibinfo {author}
  {\bibfnamefont {Marshall}\ \bibnamefont {Campbell}}, \bibinfo {author}
  {\bibfnamefont {Travis}\ \bibnamefont {Autry}}, \bibinfo {author}
  {\bibfnamefont {Takashi}\ \bibnamefont {Taniguchi}}, \bibinfo {author}
  {\bibfnamefont {Kenji}\ \bibnamefont {Watanabe}}, \bibinfo {author}
  {\bibfnamefont {Nanshu}\ \bibnamefont {Lu}}, \bibinfo {author} {\bibfnamefont
  {Sanjay~K}\ \bibnamefont {Banerjee}}, \bibinfo {author} {\bibfnamefont
  {Kevin~L}\ \bibnamefont {Silverman}}, \bibinfo {author} {\bibfnamefont
  {Suenne}\ \bibnamefont {Kim}}, \bibinfo {author} {\bibfnamefont {Emanuel}\
  \bibnamefont {Tutuc}}, \bibinfo {author} {\bibfnamefont {Li}~\bibnamefont
  {Yang}}, \bibinfo {author} {\bibfnamefont {Allan~H}\ \bibnamefont
  {MacDonald}}, \ and\ \bibinfo {author} {\bibfnamefont {Xiaoqin}\ \bibnamefont
  {Li}},\ }\bibfield  {title} {\enquote {\bibinfo {title} {Evidence for moiré
  excitons in van der waals heterostructures},}\ }\href {\doibase
  10.1038/s41586-019-0975-z} {\bibfield  {journal} {\bibinfo  {journal}
  {Nature}\ }\textbf {\bibinfo {volume} {567}},\ \bibinfo {pages} {71--75}
  (\bibinfo {year} {2019})}\BibitemShut {NoStop}%
\bibitem [{\citenamefont {Yuan}\ \emph {et~al.}(2020)\citenamefont {Yuan},
  \citenamefont {Zheng}, \citenamefont {Kunstmann}, \citenamefont {Brumme},
  \citenamefont {Kuc}, \citenamefont {Ma}, \citenamefont {Deng}, \citenamefont
  {Blach}, \citenamefont {Pan},\ and\ \citenamefont {Huang}}]{Yuan2020Twist}%
  \BibitemOpen
  \bibfield  {author} {\bibinfo {author} {\bibfnamefont {Long}\ \bibnamefont
  {Yuan}}, \bibinfo {author} {\bibfnamefont {Biyuan}\ \bibnamefont {Zheng}},
  \bibinfo {author} {\bibfnamefont {Jens}\ \bibnamefont {Kunstmann}}, \bibinfo
  {author} {\bibfnamefont {Thomas}\ \bibnamefont {Brumme}}, \bibinfo {author}
  {\bibfnamefont {Agnieszka~Beata}\ \bibnamefont {Kuc}}, \bibinfo {author}
  {\bibfnamefont {Chao}\ \bibnamefont {Ma}}, \bibinfo {author} {\bibfnamefont
  {Shibin}\ \bibnamefont {Deng}}, \bibinfo {author} {\bibfnamefont {Daria}\
  \bibnamefont {Blach}}, \bibinfo {author} {\bibfnamefont {Anlian}\
  \bibnamefont {Pan}}, \ and\ \bibinfo {author} {\bibfnamefont {Libai}\
  \bibnamefont {Huang}},\ }\bibfield  {title} {\enquote {\bibinfo {title}
  {Twist-angle-dependent interlayer exciton diffusion in ws2–wse2
  heterobilayers},}\ }\href {\doibase 10.1038/s41563-020-0670-3} {\bibfield
  {journal} {\bibinfo  {journal} {Nature Materials}\ }\textbf {\bibinfo
  {volume} {19}},\ \bibinfo {pages} {617--623} (\bibinfo {year}
  {2020})}\BibitemShut {NoStop}%
\bibitem [{\citenamefont {Karni}\ \emph {et~al.}(2022)\citenamefont {Karni},
  \citenamefont {Barré}, \citenamefont {Pareek}, \citenamefont {Georgaras},
  \citenamefont {Man}, \citenamefont {Sahoo}, \citenamefont {Bacon},
  \citenamefont {Zhu}, \citenamefont {Ribeiro}, \citenamefont {O’Beirne},
  \citenamefont {Hu}, \citenamefont {Al-Mahboob}, \citenamefont {Abdelrasoul},
  \citenamefont {Chan}, \citenamefont {Karmakar}, \citenamefont {Winchester},
  \citenamefont {Kim}, \citenamefont {Watanabe}, \citenamefont {Taniguchi},
  \citenamefont {Barmak}, \citenamefont {Madéo}, \citenamefont {da~Jornada},
  \citenamefont {Heinz},\ and\ \citenamefont {Dani}}]{Karni2022Structure}%
  \BibitemOpen
  \bibfield  {author} {\bibinfo {author} {\bibfnamefont {Ouri}\ \bibnamefont
  {Karni}}, \bibinfo {author} {\bibfnamefont {Elyse}\ \bibnamefont {Barré}},
  \bibinfo {author} {\bibfnamefont {Vivek}\ \bibnamefont {Pareek}}, \bibinfo
  {author} {\bibfnamefont {Johnathan~D}\ \bibnamefont {Georgaras}}, \bibinfo
  {author} {\bibfnamefont {Michael K~L}\ \bibnamefont {Man}}, \bibinfo {author}
  {\bibfnamefont {Chakradhar}\ \bibnamefont {Sahoo}}, \bibinfo {author}
  {\bibfnamefont {David~R}\ \bibnamefont {Bacon}}, \bibinfo {author}
  {\bibfnamefont {Xing}\ \bibnamefont {Zhu}}, \bibinfo {author} {\bibfnamefont
  {Henrique~B}\ \bibnamefont {Ribeiro}}, \bibinfo {author} {\bibfnamefont
  {Aidan~L}\ \bibnamefont {O’Beirne}}, \bibinfo {author} {\bibfnamefont
  {Jenny}\ \bibnamefont {Hu}}, \bibinfo {author} {\bibfnamefont {Abdullah}\
  \bibnamefont {Al-Mahboob}}, \bibinfo {author} {\bibfnamefont {Mohamed M~M}\
  \bibnamefont {Abdelrasoul}}, \bibinfo {author} {\bibfnamefont {Nicholas~S}\
  \bibnamefont {Chan}}, \bibinfo {author} {\bibfnamefont {Arka}\ \bibnamefont
  {Karmakar}}, \bibinfo {author} {\bibfnamefont {Andrew~J}\ \bibnamefont
  {Winchester}}, \bibinfo {author} {\bibfnamefont {Bumho}\ \bibnamefont {Kim}},
  \bibinfo {author} {\bibfnamefont {Kenji}\ \bibnamefont {Watanabe}}, \bibinfo
  {author} {\bibfnamefont {Takashi}\ \bibnamefont {Taniguchi}}, \bibinfo
  {author} {\bibfnamefont {Katayun}\ \bibnamefont {Barmak}}, \bibinfo {author}
  {\bibfnamefont {Julien}\ \bibnamefont {Madéo}}, \bibinfo {author}
  {\bibfnamefont {Felipe~H}\ \bibnamefont {da~Jornada}}, \bibinfo {author}
  {\bibfnamefont {Tony~F}\ \bibnamefont {Heinz}}, \ and\ \bibinfo {author}
  {\bibfnamefont {Keshav~M}\ \bibnamefont {Dani}},\ }\bibfield  {title}
  {\enquote {\bibinfo {title} {Structure of the moiré exciton captured by
  imaging its electron and hole},}\ }\href {\doibase
  10.1038/s41586-021-04360-y} {\bibfield  {journal} {\bibinfo  {journal}
  {Nature}\ }\textbf {\bibinfo {volume} {603}},\ \bibinfo {pages} {247--252}
  (\bibinfo {year} {2022})}\BibitemShut {NoStop}%
\bibitem [{\citenamefont {Naik}\ \emph {et~al.}(2022)\citenamefont {Naik},
  \citenamefont {Regan}, \citenamefont {Zhang}, \citenamefont {Chan},
  \citenamefont {Li}, \citenamefont {Wang}, \citenamefont {Yoon}, \citenamefont
  {Ong}, \citenamefont {Zhao}, \citenamefont {Zhao}, \citenamefont {Utama},
  \citenamefont {Gao}, \citenamefont {Wei}, \citenamefont {Sayyad},
  \citenamefont {Yumigeta}, \citenamefont {Watanabe}, \citenamefont
  {Taniguchi}, \citenamefont {Tongay}, \citenamefont {da~Jornada},
  \citenamefont {Wang},\ and\ \citenamefont {Louie}}]{Naik2022Intralayer}%
  \BibitemOpen
  \bibfield  {author} {\bibinfo {author} {\bibfnamefont {Mit~H}\ \bibnamefont
  {Naik}}, \bibinfo {author} {\bibfnamefont {Emma~C}\ \bibnamefont {Regan}},
  \bibinfo {author} {\bibfnamefont {Zuocheng}\ \bibnamefont {Zhang}}, \bibinfo
  {author} {\bibfnamefont {Yang-Hao}\ \bibnamefont {Chan}}, \bibinfo {author}
  {\bibfnamefont {Zhenglu}\ \bibnamefont {Li}}, \bibinfo {author}
  {\bibfnamefont {Danqing}\ \bibnamefont {Wang}}, \bibinfo {author}
  {\bibfnamefont {Yoseob}\ \bibnamefont {Yoon}}, \bibinfo {author}
  {\bibfnamefont {Chin~Shen}\ \bibnamefont {Ong}}, \bibinfo {author}
  {\bibfnamefont {Wenyu}\ \bibnamefont {Zhao}}, \bibinfo {author}
  {\bibfnamefont {Sihan}\ \bibnamefont {Zhao}}, \bibinfo {author}
  {\bibfnamefont {M~Iqbal~Bakti}\ \bibnamefont {Utama}}, \bibinfo {author}
  {\bibfnamefont {Beini}\ \bibnamefont {Gao}}, \bibinfo {author} {\bibfnamefont
  {Xin}\ \bibnamefont {Wei}}, \bibinfo {author} {\bibfnamefont {Mohammed}\
  \bibnamefont {Sayyad}}, \bibinfo {author} {\bibfnamefont {Kentaro}\
  \bibnamefont {Yumigeta}}, \bibinfo {author} {\bibfnamefont {Kenji}\
  \bibnamefont {Watanabe}}, \bibinfo {author} {\bibfnamefont {Takashi}\
  \bibnamefont {Taniguchi}}, \bibinfo {author} {\bibfnamefont {Sefaattin}\
  \bibnamefont {Tongay}}, \bibinfo {author} {\bibfnamefont {Felipe~H}\
  \bibnamefont {da~Jornada}}, \bibinfo {author} {\bibfnamefont {Feng}\
  \bibnamefont {Wang}}, \ and\ \bibinfo {author} {\bibfnamefont {Steven~G}\
  \bibnamefont {Louie}},\ }\bibfield  {title} {\enquote {\bibinfo {title}
  {Intralayer charge-transfer moiré excitons in van der waals
  superlattices},}\ }\href {\doibase 10.1038/s41586-022-04991-9} {\bibfield
  {journal} {\bibinfo  {journal} {Nature}\ }\textbf {\bibinfo {volume} {609}},\
  \bibinfo {pages} {52--57} (\bibinfo {year} {2022})}\BibitemShut {NoStop}%
\bibitem [{\citenamefont {Yu}\ \emph {et~al.}(2022)\citenamefont {Yu},
  \citenamefont {Liu}, \citenamefont {Tang}, \citenamefont {Xu},\ and\
  \citenamefont {Yao}}]{Yu2022Moire}%
  \BibitemOpen
  \bibfield  {author} {\bibinfo {author} {\bibfnamefont {Hongyi}\ \bibnamefont
  {Yu}}, \bibinfo {author} {\bibfnamefont {Gui-Bin}\ \bibnamefont {Liu}},
  \bibinfo {author} {\bibfnamefont {Jianju}\ \bibnamefont {Tang}}, \bibinfo
  {author} {\bibfnamefont {Xiaodong}\ \bibnamefont {Xu}}, \ and\ \bibinfo
  {author} {\bibfnamefont {Wang}\ \bibnamefont {Yao}},\ }\bibfield  {title}
  {\enquote {\bibinfo {title} {Moiré excitons: From programmable quantum
  emitter arrays to spin-orbit–coupled artificial lattices},}\ }\href
  {\doibase 10.1126/sciadv.1701696} {\bibfield  {journal} {\bibinfo  {journal}
  {Science Advances}\ }\textbf {\bibinfo {volume} {3}},\ \bibinfo {pages}
  {e1701696} (\bibinfo {year} {2022})}\BibitemShut {NoStop}%
\bibitem [{\citenamefont {Zeng}\ and\ \citenamefont
  {MacDonald}(2022)}]{Zeng2022Strong}%
  \BibitemOpen
  \bibfield  {author} {\bibinfo {author} {\bibfnamefont {Yongxin}\ \bibnamefont
  {Zeng}}\ and\ \bibinfo {author} {\bibfnamefont {Allan~H}\ \bibnamefont
  {MacDonald}},\ }\bibfield  {title} {\enquote {\bibinfo {title} {Strong
  modulation limit of excitons and trions in moir\'e materials},}\ }\href
  {\doibase 10.1103/PhysRevB.106.035115} {\bibfield  {journal} {\bibinfo
  {journal} {Physical Review B}\ }\textbf {\bibinfo {volume} {106}},\ \bibinfo
  {pages} {35115} (\bibinfo {year} {2022})}\BibitemShut {NoStop}%
\bibitem [{\citenamefont {Wu}\ \emph {et~al.}(2018{\natexlab{b}})\citenamefont
  {Wu}, \citenamefont {Lovorn}, \citenamefont {Tutuc},\ and\ \citenamefont
  {MacDonald}}]{Wu2018Hubbard}%
  \BibitemOpen
  \bibfield  {author} {\bibinfo {author} {\bibfnamefont {Fengcheng}\
  \bibnamefont {Wu}}, \bibinfo {author} {\bibfnamefont {Timothy}\ \bibnamefont
  {Lovorn}}, \bibinfo {author} {\bibfnamefont {Emanuel}\ \bibnamefont {Tutuc}},
  \ and\ \bibinfo {author} {\bibfnamefont {A.~H.}\ \bibnamefont {MacDonald}},\
  }\bibfield  {title} {\enquote {\bibinfo {title} {Hubbard model physics in
  transition metal dichalcogenide moir\'e bands},}\ }\href {\doibase
  10.1103/PhysRevLett.121.026402} {\bibfield  {journal} {\bibinfo  {journal}
  {Physical Review Letters}\ }\textbf {\bibinfo {volume} {121}},\ \bibinfo
  {pages} {26402} (\bibinfo {year} {2018}{\natexlab{b}})}\BibitemShut {NoStop}%
\bibitem [{\citenamefont {Götting}\ \emph {et~al.}(2022)\citenamefont
  {Götting}, \citenamefont {Lohof},\ and\ \citenamefont {Gies}}]{Gotting2022}%
  \BibitemOpen
  \bibfield  {author} {\bibinfo {author} {\bibfnamefont {Niclas}\ \bibnamefont
  {Götting}}, \bibinfo {author} {\bibfnamefont {Frederik}\ \bibnamefont
  {Lohof}}, \ and\ \bibinfo {author} {\bibfnamefont {Christopher}\ \bibnamefont
  {Gies}},\ }\bibfield  {title} {\enquote {\bibinfo {title}
  {Moir\'e-bose-hubbard model for interlayer excitons in twisted transition
  metal dichalcogenide heterostructures},}\ }\href {\doibase
  10.1103/PhysRevB.105.165419} {\bibfield  {journal} {\bibinfo  {journal}
  {Physical Review B}\ }\textbf {\bibinfo {volume} {105}},\ \bibinfo {pages}
  {165419} (\bibinfo {year} {2022})}\BibitemShut {NoStop}%
\bibitem [{\citenamefont {Miao}\ \emph {et~al.}(2021)\citenamefont {Miao},
  \citenamefont {Wang}, \citenamefont {Huang}, \citenamefont {Chen},
  \citenamefont {Lian}, \citenamefont {Wang}, \citenamefont {Blei},
  \citenamefont {Taniguchi}, \citenamefont {Watanabe}, \citenamefont {Tongay},
  \citenamefont {Wang}, \citenamefont {Xiao}, \citenamefont {Cui},\ and\
  \citenamefont {Shi}}]{Miao2021Strong}%
  \BibitemOpen
  \bibfield  {author} {\bibinfo {author} {\bibfnamefont {Shengnan}\
  \bibnamefont {Miao}}, \bibinfo {author} {\bibfnamefont {Tianmeng}\
  \bibnamefont {Wang}}, \bibinfo {author} {\bibfnamefont {Xiong}\ \bibnamefont
  {Huang}}, \bibinfo {author} {\bibfnamefont {Dongxue}\ \bibnamefont {Chen}},
  \bibinfo {author} {\bibfnamefont {Zhen}\ \bibnamefont {Lian}}, \bibinfo
  {author} {\bibfnamefont {Chong}\ \bibnamefont {Wang}}, \bibinfo {author}
  {\bibfnamefont {Mark}\ \bibnamefont {Blei}}, \bibinfo {author} {\bibfnamefont
  {Takashi}\ \bibnamefont {Taniguchi}}, \bibinfo {author} {\bibfnamefont
  {Kenji}\ \bibnamefont {Watanabe}}, \bibinfo {author} {\bibfnamefont
  {Sefaattin}\ \bibnamefont {Tongay}}, \bibinfo {author} {\bibfnamefont
  {Zenghui}\ \bibnamefont {Wang}}, \bibinfo {author} {\bibfnamefont
  {Di}~\bibnamefont {Xiao}}, \bibinfo {author} {\bibfnamefont {Yong-Tao}\
  \bibnamefont {Cui}}, \ and\ \bibinfo {author} {\bibfnamefont {Su-Fei}\
  \bibnamefont {Shi}},\ }\bibfield  {title} {\enquote {\bibinfo {title} {Strong
  interaction between interlayer excitons and correlated electrons in wse2/ws2
  moiré superlattice},}\ }\href {\doibase 10.1038/s41467-021-23732-6}
  {\bibfield  {journal} {\bibinfo  {journal} {Nature Communications}\ }\textbf
  {\bibinfo {volume} {12}},\ \bibinfo {pages} {3608} (\bibinfo {year}
  {2021})}\BibitemShut {NoStop}%
\bibitem [{\citenamefont {Park}\ \emph {et~al.}(2023)\citenamefont {Park},
  \citenamefont {Zhu}, \citenamefont {Wang}, \citenamefont {Wang},
  \citenamefont {Holtzmann}, \citenamefont {Taniguchi}, \citenamefont
  {Watanabe}, \citenamefont {Yan}, \citenamefont {Fu}, \citenamefont {Cao},
  \citenamefont {Xiao}, \citenamefont {Gamelin}, \citenamefont {Yu},
  \citenamefont {Yao},\ and\ \citenamefont {Xu}}]{Park2023Dipole}%
  \BibitemOpen
  \bibfield  {author} {\bibinfo {author} {\bibfnamefont {Heonjoon}\
  \bibnamefont {Park}}, \bibinfo {author} {\bibfnamefont {Jiayi}\ \bibnamefont
  {Zhu}}, \bibinfo {author} {\bibfnamefont {Xi}~\bibnamefont {Wang}}, \bibinfo
  {author} {\bibfnamefont {Yingqi}\ \bibnamefont {Wang}}, \bibinfo {author}
  {\bibfnamefont {William}\ \bibnamefont {Holtzmann}}, \bibinfo {author}
  {\bibfnamefont {Takashi}\ \bibnamefont {Taniguchi}}, \bibinfo {author}
  {\bibfnamefont {Kenji}\ \bibnamefont {Watanabe}}, \bibinfo {author}
  {\bibfnamefont {Jiaqiang}\ \bibnamefont {Yan}}, \bibinfo {author}
  {\bibfnamefont {Liang}\ \bibnamefont {Fu}}, \bibinfo {author} {\bibfnamefont
  {Ting}\ \bibnamefont {Cao}}, \bibinfo {author} {\bibfnamefont
  {Di}~\bibnamefont {Xiao}}, \bibinfo {author} {\bibfnamefont {Daniel~R}\
  \bibnamefont {Gamelin}}, \bibinfo {author} {\bibfnamefont {Hongyi}\
  \bibnamefont {Yu}}, \bibinfo {author} {\bibfnamefont {Wang}\ \bibnamefont
  {Yao}}, \ and\ \bibinfo {author} {\bibfnamefont {Xiaodong}\ \bibnamefont
  {Xu}},\ }\bibfield  {title} {\enquote {\bibinfo {title} {Dipole ladders with
  large hubbard interaction in a moiré exciton lattice},}\ }\href {\doibase
  10.1038/s41567-023-02077-5} {\bibfield  {journal} {\bibinfo  {journal}
  {Nature Physics}\ } (\bibinfo {year} {2023}),\
  10.1038/s41567-023-02077-5}\BibitemShut {NoStop}%
\bibitem [{\citenamefont {Gao}\ \emph {et~al.}(2023)\citenamefont {Gao},
  \citenamefont {Suárez-Forero}, \citenamefont {Sarkar}, \citenamefont
  {Huang}, \citenamefont {Session}, \citenamefont {Mehrabad}, \citenamefont
  {Ni}, \citenamefont {Xie}, \citenamefont {Vannucci}, \citenamefont {Mittal},
  \citenamefont {Watanabe}, \citenamefont {Taniguchi}, \citenamefont
  {Imamoglu}, \citenamefont {Zhou},\ and\ \citenamefont {Hafezi}}]{Gao2023}%
  \BibitemOpen
  \bibfield  {author} {\bibinfo {author} {\bibfnamefont {Beini}\ \bibnamefont
  {Gao}}, \bibinfo {author} {\bibfnamefont {Daniel~G.}\ \bibnamefont
  {Suárez-Forero}}, \bibinfo {author} {\bibfnamefont {Supratik}\ \bibnamefont
  {Sarkar}}, \bibinfo {author} {\bibfnamefont {Tsung-Sheng}\ \bibnamefont
  {Huang}}, \bibinfo {author} {\bibfnamefont {Deric}\ \bibnamefont {Session}},
  \bibinfo {author} {\bibfnamefont {Mahmoud~Jalali}\ \bibnamefont {Mehrabad}},
  \bibinfo {author} {\bibfnamefont {Ruihao}\ \bibnamefont {Ni}}, \bibinfo
  {author} {\bibfnamefont {Ming}\ \bibnamefont {Xie}}, \bibinfo {author}
  {\bibfnamefont {Jonathan}\ \bibnamefont {Vannucci}}, \bibinfo {author}
  {\bibfnamefont {Sunil}\ \bibnamefont {Mittal}}, \bibinfo {author}
  {\bibfnamefont {Kenji}\ \bibnamefont {Watanabe}}, \bibinfo {author}
  {\bibfnamefont {Takashi}\ \bibnamefont {Taniguchi}}, \bibinfo {author}
  {\bibfnamefont {Atac}\ \bibnamefont {Imamoglu}}, \bibinfo {author}
  {\bibfnamefont {You}\ \bibnamefont {Zhou}}, \ and\ \bibinfo {author}
  {\bibfnamefont {Mohammad}\ \bibnamefont {Hafezi}},\ }\href@noop {} {\enquote
  {\bibinfo {title} {Excitonic mott insulator in a bose-fermi-hubbard system of
  moir\'e ws2/wse2 heterobilayer},}\ } (\bibinfo {year} {2023}),\ \Eprint
  {http://arxiv.org/abs/2304.09731} {arXiv:2304.09731 [cond-mat.mes-hall]}
  \BibitemShut {NoStop}%
\bibitem [{\citenamefont {Xiong}\ \emph {et~al.}(2023)\citenamefont {Xiong},
  \citenamefont {Nie}, \citenamefont {Brantly}, \citenamefont {Hays},
  \citenamefont {Sailus}, \citenamefont {Watanabe}, \citenamefont {Taniguchi},
  \citenamefont {Tongay},\ and\ \citenamefont {Jin}}]{Xiong2023}%
  \BibitemOpen
  \bibfield  {author} {\bibinfo {author} {\bibfnamefont {Richen}\ \bibnamefont
  {Xiong}}, \bibinfo {author} {\bibfnamefont {Jacob~H}\ \bibnamefont {Nie}},
  \bibinfo {author} {\bibfnamefont {Samuel~L}\ \bibnamefont {Brantly}},
  \bibinfo {author} {\bibfnamefont {Patrick}\ \bibnamefont {Hays}}, \bibinfo
  {author} {\bibfnamefont {Renee}\ \bibnamefont {Sailus}}, \bibinfo {author}
  {\bibfnamefont {Kenji}\ \bibnamefont {Watanabe}}, \bibinfo {author}
  {\bibfnamefont {Takashi}\ \bibnamefont {Taniguchi}}, \bibinfo {author}
  {\bibfnamefont {Sefaattin}\ \bibnamefont {Tongay}}, \ and\ \bibinfo {author}
  {\bibfnamefont {Chenhao}\ \bibnamefont {Jin}},\ }\bibfield  {title} {\enquote
  {\bibinfo {title} {Correlated insulator of excitons in wse2/ws2 moiré
  superlattices},}\ }\href {\doibase 10.1126/science.add5574} {\bibfield
  {journal} {\bibinfo  {journal} {Science}\ }\textbf {\bibinfo {volume}
  {380}},\ \bibinfo {pages} {860--864} (\bibinfo {year} {2023})}\BibitemShut
  {NoStop}%
\bibitem [{\citenamefont {Camacho-Guardian}\ and\ \citenamefont
  {Cooper}(2022{\natexlab{a}})}]{CamachoGuardian2022Moire}%
  \BibitemOpen
  \bibfield  {author} {\bibinfo {author} {\bibfnamefont {A}~\bibnamefont
  {Camacho-Guardian}}\ and\ \bibinfo {author} {\bibfnamefont {N.~R.}\
  \bibnamefont {Cooper}},\ }\bibfield  {title} {\enquote {\bibinfo {title}
  {Moir\'e-induced optical nonlinearities: Single- and multiphoton
  resonances},}\ }\href {\doibase 10.1103/PhysRevLett.128.207401} {\bibfield
  {journal} {\bibinfo  {journal} {Physical Review Letters}\ }\textbf {\bibinfo
  {volume} {128}},\ \bibinfo {pages} {207401} (\bibinfo {year}
  {2022}{\natexlab{a}})}\BibitemShut {NoStop}%
\bibitem [{\citenamefont {Camacho-Guardian}\ and\ \citenamefont
  {Cooper}(2022{\natexlab{b}})}]{CamachoGuardian2022Optical}%
  \BibitemOpen
  \bibfield  {author} {\bibinfo {author} {\bibfnamefont {A}~\bibnamefont
  {Camacho-Guardian}}\ and\ \bibinfo {author} {\bibfnamefont {N~R}\
  \bibnamefont {Cooper}},\ }\bibfield  {title} {\enquote {\bibinfo {title}
  {Optical nonlinearities and spontaneous translational symmetry breaking in
  driven-dissipative moir\'e exciton-polaritons},}\ }\href {\doibase
  10.1103/PhysRevB.106.245402} {\bibfield  {journal} {\bibinfo  {journal}
  {Physical Review B}\ }\textbf {\bibinfo {volume} {106}},\ \bibinfo {pages}
  {245402} (\bibinfo {year} {2022}{\natexlab{b}})}\BibitemShut {NoStop}%
\bibitem [{\citenamefont {Remez}\ and\ \citenamefont
  {Cooper}(2022)}]{Remez2022Leaky}%
  \BibitemOpen
  \bibfield  {author} {\bibinfo {author} {\bibfnamefont {Benjamin}\
  \bibnamefont {Remez}}\ and\ \bibinfo {author} {\bibfnamefont {Nigel~R}\
  \bibnamefont {Cooper}},\ }\bibfield  {title} {\enquote {\bibinfo {title}
  {Leaky exciton condensates in transition metal dichalcogenide moir\'e
  bilayers},}\ }\href {\doibase 10.1103/PhysRevResearch.4.L022042} {\bibfield
  {journal} {\bibinfo  {journal} {Physical Review Research}\ }\textbf {\bibinfo
  {volume} {4}},\ \bibinfo {pages} {L022042--} (\bibinfo {year}
  {2022})}\BibitemShut {NoStop}%
\bibitem [{\citenamefont {Haug}\ and\ \citenamefont
  {Schmitt-Rink}(1984)}]{Haug1984Electron}%
  \BibitemOpen
  \bibfield  {author} {\bibinfo {author} {\bibfnamefont {H.}~\bibnamefont
  {Haug}}\ and\ \bibinfo {author} {\bibfnamefont {S.}~\bibnamefont
  {Schmitt-Rink}},\ }\bibfield  {title} {\enquote {\bibinfo {title} {Electron
  theory of the optical properties of laser-excited semiconductors},}\
  }\href@noop {} {\bibfield  {journal} {\bibinfo  {journal} {Progress in
  Quantum Electronics}\ }\textbf {\bibinfo {volume} {9}},\ \bibinfo {pages} {3
  -- 100} (\bibinfo {year} {1984})}\BibitemShut {NoStop}%
\bibitem [{\citenamefont {Combescot}\ \emph {et~al.}(2007)\citenamefont
  {Combescot}, \citenamefont {Betbeder-Matibet},\ and\ \citenamefont
  {Combescot}}]{Combescot2007}%
  \BibitemOpen
  \bibfield  {author} {\bibinfo {author} {\bibfnamefont {M}~\bibnamefont
  {Combescot}}, \bibinfo {author} {\bibfnamefont {O}~\bibnamefont
  {Betbeder-Matibet}}, \ and\ \bibinfo {author} {\bibfnamefont {R}~\bibnamefont
  {Combescot}},\ }\bibfield  {title} {\enquote {\bibinfo {title}
  {Exciton-exciton scattering: Composite boson versus elementary boson},}\
  }\href {\doibase 10.1103/PhysRevB.75.174305} {\bibfield  {journal} {\bibinfo
  {journal} {Physical Review B}\ }\textbf {\bibinfo {volume} {75}},\ \bibinfo
  {pages} {174305} (\bibinfo {year} {2007})}\BibitemShut {NoStop}%
\bibitem [{\citenamefont {Combescot}\ \emph {et~al.}(2008)\citenamefont
  {Combescot}, \citenamefont {Betbeder-Matibet},\ and\ \citenamefont
  {Dubin}}]{Combescot2008}%
  \BibitemOpen
  \bibfield  {author} {\bibinfo {author} {\bibfnamefont {Monique}\ \bibnamefont
  {Combescot}}, \bibinfo {author} {\bibfnamefont {Odile}\ \bibnamefont
  {Betbeder-Matibet}}, \ and\ \bibinfo {author} {\bibfnamefont {François}\
  \bibnamefont {Dubin}},\ }\bibfield  {title} {\enquote {\bibinfo {title} {The
  many-body physics of composite bosons},}\ }\href {\doibase
  https://doi.org/10.1016/j.physrep.2007.11.003} {\bibfield  {journal}
  {\bibinfo  {journal} {Physics Reports}\ }\textbf {\bibinfo {volume} {463}},\
  \bibinfo {pages} {215--320} (\bibinfo {year} {2008})}\BibitemShut {NoStop}%
\bibitem [{\citenamefont {Combescot}\ and\ \citenamefont
  {Pogosov}(2009)}]{Combescot2009}%
  \BibitemOpen
  \bibfield  {author} {\bibinfo {author} {\bibfnamefont {M}~\bibnamefont
  {Combescot}}\ and\ \bibinfo {author} {\bibfnamefont {W}~\bibnamefont
  {Pogosov}},\ }\bibfield  {title} {\enquote {\bibinfo {title} {Composite boson
  many-body theory for frenkel excitons},}\ }\href {\doibase
  10.1140/epjb/e2009-00086-6} {\bibfield  {journal} {\bibinfo  {journal} {The
  European Physical Journal B}\ }\textbf {\bibinfo {volume} {68}},\ \bibinfo
  {pages} {161--181} (\bibinfo {year} {2009})}\BibitemShut {NoStop}%
\bibitem [{Note1()}]{Note1}%
  \BibitemOpen
  \bibinfo {note} {In general, holes from two excitons could also exchange
  without swapping their electrons. Nevertheless, it is equivalent to the
  electron-exchange process if all incoming and outgoing excitons are the
  same.}\BibitemShut {Stop}%
\bibitem [{\citenamefont {Imamoğlu}(1998)}]{Imamoglu1998}%
  \BibitemOpen
  \bibfield  {author} {\bibinfo {author} {\bibfnamefont {A}~\bibnamefont
  {Imamoğlu}},\ }\bibfield  {title} {\enquote {\bibinfo {title} {Phase-space
  filling and stimulated scattering of composite bosons},}\ }\href {\doibase
  10.1103/PhysRevB.57.R4195} {\bibfield  {journal} {\bibinfo  {journal}
  {Physical Review B}\ }\textbf {\bibinfo {volume} {57}},\ \bibinfo {pages}
  {R4195--R4197} (\bibinfo {year} {1998})}\BibitemShut {NoStop}%
\bibitem [{\citenamefont {Thilagam}(2013)}]{Thilagam2013}%
  \BibitemOpen
  \bibfield  {author} {\bibinfo {author} {\bibfnamefont {A}~\bibnamefont
  {Thilagam}},\ }\bibfield  {title} {\enquote {\bibinfo {title} {Crossover from
  bosonic to fermionic features in composite boson systems},}\ }\href {\doibase
  10.1007/s10910-013-0190-3} {\bibfield  {journal} {\bibinfo  {journal}
  {Journal of Mathematical Chemistry}\ }\textbf {\bibinfo {volume} {51}},\
  \bibinfo {pages} {1897--1913} (\bibinfo {year} {2013})}\BibitemShut {NoStop}%
\bibitem [{\citenamefont {Thilagam}(2015)}]{Thilagam2015}%
  \BibitemOpen
  \bibfield  {author} {\bibinfo {author} {\bibfnamefont {A}~\bibnamefont
  {Thilagam}},\ }\bibfield  {title} {\enquote {\bibinfo {title} {Influence of
  the pauli exclusion principle on scattering properties of cobosons},}\ }\href
  {\doibase https://doi.org/10.1016/j.physb.2014.10.021} {\bibfield  {journal}
  {\bibinfo  {journal} {Physica B: Condensed Matter}\ }\textbf {\bibinfo
  {volume} {457}},\ \bibinfo {pages} {232--239} (\bibinfo {year}
  {2015})}\BibitemShut {NoStop}%
\bibitem [{\citenamefont {Haug}\ and\ \citenamefont {Koch}(2004)}]{Haug2004}%
  \BibitemOpen
  \bibfield  {author} {\bibinfo {author} {\bibfnamefont {H.}~\bibnamefont
  {Haug}}\ and\ \bibinfo {author} {\bibfnamefont {S.~W.}\ \bibnamefont
  {Koch}},\ }\href@noop {} {\emph {\bibinfo {title} {Quantum Theory of the
  Optical and Electronic Properties of Semiconductors}}}\ (\bibinfo
  {publisher} {World Scientific},\ \bibinfo {year} {2004})\BibitemShut
  {NoStop}%
\bibitem [{\citenamefont {Laussy}\ \emph {et~al.}(2006)\citenamefont {Laussy},
  \citenamefont {Glazov}, \citenamefont {Kavokin}, \citenamefont {Whittaker},\
  and\ \citenamefont {Malpuech}}]{Laussy2006}%
  \BibitemOpen
  \bibfield  {author} {\bibinfo {author} {\bibfnamefont {Fabrice~P}\
  \bibnamefont {Laussy}}, \bibinfo {author} {\bibfnamefont {Mikhail~M}\
  \bibnamefont {Glazov}}, \bibinfo {author} {\bibfnamefont {Alexey}\
  \bibnamefont {Kavokin}}, \bibinfo {author} {\bibfnamefont {David~M}\
  \bibnamefont {Whittaker}}, \ and\ \bibinfo {author} {\bibfnamefont
  {Guillaume}\ \bibnamefont {Malpuech}},\ }\bibfield  {title} {\enquote
  {\bibinfo {title} {Statistics of excitons in quantum dots and their effect on
  the optical emission spectra of microcavities},}\ }\href {\doibase
  10.1103/PhysRevB.73.115343} {\bibfield  {journal} {\bibinfo  {journal}
  {Physical Review B}\ }\textbf {\bibinfo {volume} {73}},\ \bibinfo {pages}
  {115343} (\bibinfo {year} {2006})}\BibitemShut {NoStop}%
\bibitem [{\citenamefont {Agranovich}\ and\ \citenamefont
  {Toshich}(1968)}]{Agranovich1968}%
  \BibitemOpen
  \bibfield  {author} {\bibinfo {author} {\bibfnamefont {VM}~\bibnamefont
  {Agranovich}}\ and\ \bibinfo {author} {\bibfnamefont {BS}~\bibnamefont
  {Toshich}},\ }\bibfield  {title} {\enquote {\bibinfo {title} {Collective
  properties of frenkel excitons},}\ }\href@noop {} {\bibfield  {journal}
  {\bibinfo  {journal} {Sov. Phys. JETP}\ }\textbf {\bibinfo {volume} {26}},\
  \bibinfo {pages} {104--112} (\bibinfo {year} {1968})}\BibitemShut {NoStop}%
\bibitem [{\citenamefont {Betzold}\ \emph {et~al.}(2020)\citenamefont
  {Betzold}, \citenamefont {Dusel}, \citenamefont {Kyriienko}, \citenamefont
  {Dietrich}, \citenamefont {Klembt}, \citenamefont {Ohmer}, \citenamefont
  {Fischer}, \citenamefont {Shelykh}, \citenamefont {Schneider},\ and\
  \citenamefont {Höfling}}]{Betzold2020}%
  \BibitemOpen
  \bibfield  {author} {\bibinfo {author} {\bibfnamefont {Simon}\ \bibnamefont
  {Betzold}}, \bibinfo {author} {\bibfnamefont {Marco}\ \bibnamefont {Dusel}},
  \bibinfo {author} {\bibfnamefont {Oleksandr}\ \bibnamefont {Kyriienko}},
  \bibinfo {author} {\bibfnamefont {Christof~P}\ \bibnamefont {Dietrich}},
  \bibinfo {author} {\bibfnamefont {Sebastian}\ \bibnamefont {Klembt}},
  \bibinfo {author} {\bibfnamefont {Jürgen}\ \bibnamefont {Ohmer}}, \bibinfo
  {author} {\bibfnamefont {Utz}\ \bibnamefont {Fischer}}, \bibinfo {author}
  {\bibfnamefont {Ivan~A}\ \bibnamefont {Shelykh}}, \bibinfo {author}
  {\bibfnamefont {Christian}\ \bibnamefont {Schneider}}, \ and\ \bibinfo
  {author} {\bibfnamefont {Sven}\ \bibnamefont {Höfling}},\ }\bibfield
  {title} {\enquote {\bibinfo {title} {Coherence and interaction in confined
  room-temperature polariton condensates with frenkel excitons},}\ }\href
  {\doibase 10.1021/acsphotonics.9b01300} {\bibfield  {journal} {\bibinfo
  {journal} {ACS Photonics}\ }\textbf {\bibinfo {volume} {7}},\ \bibinfo
  {pages} {384--392} (\bibinfo {year} {2020})}\BibitemShut {NoStop}%
\bibitem [{\citenamefont {Yagafarov}\ \emph {et~al.}(2020)\citenamefont
  {Yagafarov}, \citenamefont {Sannikov}, \citenamefont {Zasedatelev},
  \citenamefont {Georgiou}, \citenamefont {Baranikov}, \citenamefont
  {Kyriienko}, \citenamefont {Shelykh}, \citenamefont {Gai}, \citenamefont
  {Shen}, \citenamefont {Lidzey},\ and\ \citenamefont
  {Lagoudakis}}]{Yagafarov2020}%
  \BibitemOpen
  \bibfield  {author} {\bibinfo {author} {\bibfnamefont {Timur}\ \bibnamefont
  {Yagafarov}}, \bibinfo {author} {\bibfnamefont {Denis}\ \bibnamefont
  {Sannikov}}, \bibinfo {author} {\bibfnamefont {Anton}\ \bibnamefont
  {Zasedatelev}}, \bibinfo {author} {\bibfnamefont {Kyriacos}\ \bibnamefont
  {Georgiou}}, \bibinfo {author} {\bibfnamefont {Anton}\ \bibnamefont
  {Baranikov}}, \bibinfo {author} {\bibfnamefont {Oleksandr}\ \bibnamefont
  {Kyriienko}}, \bibinfo {author} {\bibfnamefont {Ivan}\ \bibnamefont
  {Shelykh}}, \bibinfo {author} {\bibfnamefont {Lizhi}\ \bibnamefont {Gai}},
  \bibinfo {author} {\bibfnamefont {Zhen}\ \bibnamefont {Shen}}, \bibinfo
  {author} {\bibfnamefont {David}\ \bibnamefont {Lidzey}}, \ and\ \bibinfo
  {author} {\bibfnamefont {Pavlos}\ \bibnamefont {Lagoudakis}},\ }\bibfield
  {title} {\enquote {\bibinfo {title} {Mechanisms of blueshifts in organic
  polariton condensates},}\ }\href {\doibase 10.1038/s42005-019-0278-6}
  {\bibfield  {journal} {\bibinfo  {journal} {Communications Physics}\ }\textbf
  {\bibinfo {volume} {3}},\ \bibinfo {pages} {18} (\bibinfo {year}
  {2020})}\BibitemShut {NoStop}%
\bibitem [{\citenamefont {Paredes}\ \emph {et~al.}(2007)\citenamefont
  {Paredes}, \citenamefont {Keilmann},\ and\ \citenamefont
  {Cirac}}]{Paredes2007}%
  \BibitemOpen
  \bibfield  {author} {\bibinfo {author} {\bibfnamefont {B}~\bibnamefont
  {Paredes}}, \bibinfo {author} {\bibfnamefont {T}~\bibnamefont {Keilmann}}, \
  and\ \bibinfo {author} {\bibfnamefont {J~I}\ \bibnamefont {Cirac}},\
  }\bibfield  {title} {\enquote {\bibinfo {title} {Pfaffian-like ground state
  for three-body hard-core bosons in one-dimensional lattices},}\ }\href
  {\doibase 10.1103/PhysRevA.75.053611} {\bibfield  {journal} {\bibinfo
  {journal} {Physical Review A}\ }\textbf {\bibinfo {volume} {75}},\ \bibinfo
  {pages} {53611} (\bibinfo {year} {2007})}\BibitemShut {NoStop}%
\bibitem [{\citenamefont {Daley}\ \emph {et~al.}(2009)\citenamefont {Daley},
  \citenamefont {Taylor}, \citenamefont {Diehl}, \citenamefont {Baranov},\ and\
  \citenamefont {Zoller}}]{Daley2009}%
  \BibitemOpen
  \bibfield  {author} {\bibinfo {author} {\bibfnamefont {A~J}\ \bibnamefont
  {Daley}}, \bibinfo {author} {\bibfnamefont {J~M}\ \bibnamefont {Taylor}},
  \bibinfo {author} {\bibfnamefont {S}~\bibnamefont {Diehl}}, \bibinfo {author}
  {\bibfnamefont {M}~\bibnamefont {Baranov}}, \ and\ \bibinfo {author}
  {\bibfnamefont {P}~\bibnamefont {Zoller}},\ }\bibfield  {title} {\enquote
  {\bibinfo {title} {Atomic three-body loss as a dynamical three-body
  interaction},}\ }\href {\doibase 10.1103/PhysRevLett.102.040402} {\bibfield
  {journal} {\bibinfo  {journal} {Physical Review Letters}\ }\textbf {\bibinfo
  {volume} {102}},\ \bibinfo {pages} {40402} (\bibinfo {year}
  {2009})}\BibitemShut {NoStop}%
\bibitem [{\citenamefont {Diehl}\ \emph {et~al.}(2010)\citenamefont {Diehl},
  \citenamefont {Baranov}, \citenamefont {Daley},\ and\ \citenamefont
  {Zoller}}]{Diehl2010}%
  \BibitemOpen
  \bibfield  {author} {\bibinfo {author} {\bibfnamefont {S}~\bibnamefont
  {Diehl}}, \bibinfo {author} {\bibfnamefont {M}~\bibnamefont {Baranov}},
  \bibinfo {author} {\bibfnamefont {A~J}\ \bibnamefont {Daley}}, \ and\
  \bibinfo {author} {\bibfnamefont {P}~\bibnamefont {Zoller}},\ }\bibfield
  {title} {\enquote {\bibinfo {title} {Quantum field theory for the three-body
  constrained lattice bose gas. ii. application to the many-body problem},}\
  }\href {\doibase 10.1103/PhysRevB.82.064510} {\bibfield  {journal} {\bibinfo
  {journal} {Physical Review B}\ }\textbf {\bibinfo {volume} {82}},\ \bibinfo
  {pages} {64510} (\bibinfo {year} {2010})}\BibitemShut {NoStop}%
\bibitem [{\citenamefont {Mazza}\ \emph {et~al.}(2010)\citenamefont {Mazza},
  \citenamefont {Rizzi}, \citenamefont {Lewenstein},\ and\ \citenamefont
  {Cirac}}]{Mazza2010}%
  \BibitemOpen
  \bibfield  {author} {\bibinfo {author} {\bibfnamefont {L}~\bibnamefont
  {Mazza}}, \bibinfo {author} {\bibfnamefont {M}~\bibnamefont {Rizzi}},
  \bibinfo {author} {\bibfnamefont {M}~\bibnamefont {Lewenstein}}, \ and\
  \bibinfo {author} {\bibfnamefont {J~I}\ \bibnamefont {Cirac}},\ }\bibfield
  {title} {\enquote {\bibinfo {title} {Emerging bosons with three-body
  interactions from spin-1 atoms in optical lattices},}\ }\href {\doibase
  10.1103/PhysRevA.82.043629} {\bibfield  {journal} {\bibinfo  {journal}
  {Physical Review A}\ }\textbf {\bibinfo {volume} {82}},\ \bibinfo {pages}
  {43629} (\bibinfo {year} {2010})}\BibitemShut {NoStop}%
\bibitem [{\citenamefont {Bonnes}\ and\ \citenamefont
  {Wessel}(2011)}]{Bonnes2011}%
  \BibitemOpen
  \bibfield  {author} {\bibinfo {author} {\bibfnamefont {Lars}\ \bibnamefont
  {Bonnes}}\ and\ \bibinfo {author} {\bibfnamefont {Stefan}\ \bibnamefont
  {Wessel}},\ }\bibfield  {title} {\enquote {\bibinfo {title} {Pair
  superfluidity of three-body constrained bosons in two dimensions},}\ }\href
  {\doibase 10.1103/PhysRevLett.106.185302} {\bibfield  {journal} {\bibinfo
  {journal} {Physical Review Letters}\ }\textbf {\bibinfo {volume} {106}},\
  \bibinfo {pages} {185302} (\bibinfo {year} {2011})}\BibitemShut {NoStop}%
\bibitem [{\citenamefont {Kapit}\ and\ \citenamefont
  {Simon}(2013)}]{Kapit2013}%
  \BibitemOpen
  \bibfield  {author} {\bibinfo {author} {\bibfnamefont {Eliot}\ \bibnamefont
  {Kapit}}\ and\ \bibinfo {author} {\bibfnamefont {Steven~H}\ \bibnamefont
  {Simon}},\ }\bibfield  {title} {\enquote {\bibinfo {title} {Three- and
  four-body interactions from two-body interactions in spin models: A route to
  abelian and non-abelian fractional chern insulators},}\ }\href {\doibase
  10.1103/PhysRevB.88.184409} {\bibfield  {journal} {\bibinfo  {journal}
  {Physical Review B}\ }\textbf {\bibinfo {volume} {88}},\ \bibinfo {pages}
  {184409} (\bibinfo {year} {2013})}\BibitemShut {NoStop}%
\bibitem [{\citenamefont {Hafezi}\ \emph {et~al.}(2014)\citenamefont {Hafezi},
  \citenamefont {Adhikari},\ and\ \citenamefont {Taylor}}]{Hafezi2014}%
  \BibitemOpen
  \bibfield  {author} {\bibinfo {author} {\bibfnamefont {M}~\bibnamefont
  {Hafezi}}, \bibinfo {author} {\bibfnamefont {P}~\bibnamefont {Adhikari}}, \
  and\ \bibinfo {author} {\bibfnamefont {J~M}\ \bibnamefont {Taylor}},\
  }\bibfield  {title} {\enquote {\bibinfo {title} {Engineering three-body
  interaction and pfaffian states in circuit qed systems},}\ }\href {\doibase
  10.1103/PhysRevB.90.060503} {\bibfield  {journal} {\bibinfo  {journal}
  {Physical Review B}\ }\textbf {\bibinfo {volume} {90}},\ \bibinfo {pages}
  {60503} (\bibinfo {year} {2014})}\BibitemShut {NoStop}%
\bibitem [{\citenamefont {Singh}\ \emph {et~al.}(2014)\citenamefont {Singh},
  \citenamefont {Mishra}, \citenamefont {Pai},\ and\ \citenamefont
  {Das}}]{Singh2014}%
  \BibitemOpen
  \bibfield  {author} {\bibinfo {author} {\bibfnamefont {Manpreet}\
  \bibnamefont {Singh}}, \bibinfo {author} {\bibfnamefont {Tapan}\ \bibnamefont
  {Mishra}}, \bibinfo {author} {\bibfnamefont {Ramesh~V}\ \bibnamefont {Pai}},
  \ and\ \bibinfo {author} {\bibfnamefont {B~P}\ \bibnamefont {Das}},\
  }\bibfield  {title} {\enquote {\bibinfo {title} {Quantum phases of attractive
  bosons on a bose-hubbard ladder with three-body constraint},}\ }\href
  {\doibase 10.1103/PhysRevA.90.013625} {\bibfield  {journal} {\bibinfo
  {journal} {Physical Review A}\ }\textbf {\bibinfo {volume} {90}},\ \bibinfo
  {pages} {13625} (\bibinfo {year} {2014})}\BibitemShut {NoStop}%
\bibitem [{\citenamefont {Petrov}(2014)}]{Petrov2014}%
  \BibitemOpen
  \bibfield  {author} {\bibinfo {author} {\bibfnamefont {D.~S.}\ \bibnamefont
  {Petrov}},\ }\bibfield  {title} {\enquote {\bibinfo {title} {Three-body
  interacting bosons in free space},}\ }\href {\doibase
  10.1103/PhysRevLett.112.103201} {\bibfield  {journal} {\bibinfo  {journal}
  {Physical Review Letters}\ }\textbf {\bibinfo {volume} {112}},\ \bibinfo
  {pages} {103201} (\bibinfo {year} {2014})}\BibitemShut {NoStop}%
\bibitem [{\citenamefont {Bulgac}(2002)}]{Bulgac2022}%
  \BibitemOpen
  \bibfield  {author} {\bibinfo {author} {\bibfnamefont {Aurel}\ \bibnamefont
  {Bulgac}},\ }\bibfield  {title} {\enquote {\bibinfo {title} {Dilute quantum
  droplets},}\ }\href {\doibase 10.1103/PhysRevLett.89.050402} {\bibfield
  {journal} {\bibinfo  {journal} {Physical Review Letters}\ }\textbf {\bibinfo
  {volume} {89}},\ \bibinfo {pages} {50402} (\bibinfo {year}
  {2002})}\BibitemShut {NoStop}%
\bibitem [{\citenamefont {Huang}\ \emph {et~al.}(2023)\citenamefont {Huang},
  \citenamefont {Chou}, \citenamefont {Baldwin}, \citenamefont {Wu},\ and\
  \citenamefont {Hafezi}}]{Huang2023}%
  \BibitemOpen
  \bibfield  {author} {\bibinfo {author} {\bibfnamefont {Tsung-Sheng}\
  \bibnamefont {Huang}}, \bibinfo {author} {\bibfnamefont {Yang-Zhi}\
  \bibnamefont {Chou}}, \bibinfo {author} {\bibfnamefont {Christopher~L}\
  \bibnamefont {Baldwin}}, \bibinfo {author} {\bibfnamefont {Fengcheng}\
  \bibnamefont {Wu}}, \ and\ \bibinfo {author} {\bibfnamefont {Mohammad}\
  \bibnamefont {Hafezi}},\ }\bibfield  {title} {\enquote {\bibinfo {title}
  {Mott-moir\'e excitons},}\ }\href {\doibase 10.1103/PhysRevB.107.195151}
  {\bibfield  {journal} {\bibinfo  {journal} {Physical Review B}\ }\textbf
  {\bibinfo {volume} {107}},\ \bibinfo {pages} {195151} (\bibinfo {year}
  {2023})}\BibitemShut {NoStop}%
\bibitem [{Sup()}]{Supplement}%
  \BibitemOpen
  \href@noop {} {}\bibinfo {note} {See supplementary material of this
  paper.}\BibitemShut {Stop}%
\bibitem [{\citenamefont {Auerbach}(1994)}]{Auerbach1994}%
  \BibitemOpen
  \bibfield  {author} {\bibinfo {author} {\bibfnamefont {A.}~\bibnamefont
  {Auerbach}},\ }\href@noop {} {\emph {\bibinfo {title} {Interacting Electrons
  and Quantum Magnetism}}}\ (\bibinfo  {publisher} {Springer},\ \bibinfo {year}
  {1994})\BibitemShut {NoStop}%
\bibitem [{Note2()}]{Note2}%
  \BibitemOpen
  \bibinfo {note} {In the derivation of $\protect \hat {\protect \mathcal
  {H}}_{\protect \mathrm {eff}}$ in the emergent boson representation, we do
  not implement the large-spin (small $\Lambda $) approximation that is
  typically applied after the Holstein-Primakoff map}\BibitemShut {NoStop}%
\bibitem [{\citenamefont {Santiago-García}\ and\ \citenamefont
  {Camacho-Guardian}(2023)}]{Santiago-Garcia2023}%
  \BibitemOpen
  \bibfield  {author} {\bibinfo {author} {\bibfnamefont {Moroni}\ \bibnamefont
  {Santiago-García}}\ and\ \bibinfo {author} {\bibfnamefont {Arturo}\
  \bibnamefont {Camacho-Guardian}},\ }\bibfield  {title} {\enquote {\bibinfo
  {title} {Collective excitations of a bose–einstein condensate of hard-core
  bosons and their mediated interactions: from two-body bound states to
  mediated superfluidity},}\ }\href {\doibase 10.1088/1367-2630/acf72d}
  {\bibfield  {journal} {\bibinfo  {journal} {New Journal of Physics}\ }\textbf
  {\bibinfo {volume} {25}},\ \bibinfo {pages} {093032} (\bibinfo {year}
  {2023})}\BibitemShut {NoStop}%
\bibitem [{\citenamefont {Conti}\ \emph {et~al.}(2020)\citenamefont {Conti},
  \citenamefont {Neilson}, \citenamefont {Peeters},\ and\ \citenamefont
  {Perali}}]{Conti2020}%
  \BibitemOpen
  \bibfield  {author} {\bibinfo {author} {\bibfnamefont {Sara}\ \bibnamefont
  {Conti}}, \bibinfo {author} {\bibfnamefont {David}\ \bibnamefont {Neilson}},
  \bibinfo {author} {\bibfnamefont {François~M}\ \bibnamefont {Peeters}}, \
  and\ \bibinfo {author} {\bibfnamefont {Andrea}\ \bibnamefont {Perali}},\
  }\bibfield  {title} {\enquote {\bibinfo {title} {Transition metal
  dichalcogenides as strategy for high temperature electron-hole
  superfluidity},}\ }\href@noop {} {\bibfield  {journal} {\bibinfo  {journal}
  {Condensed Matter}\ }\textbf {\bibinfo {volume} {5}},\ \bibinfo {pages} {22}
  (\bibinfo {year} {2020})}\BibitemShut {NoStop}%
\bibitem [{\citenamefont {Zhang}\ \emph {et~al.}(2020)\citenamefont {Zhang},
  \citenamefont {Yuan},\ and\ \citenamefont {Fu}}]{Zhang2020}%
  \BibitemOpen
  \bibfield  {author} {\bibinfo {author} {\bibfnamefont {Yang}\ \bibnamefont
  {Zhang}}, \bibinfo {author} {\bibfnamefont {Noah F~Q}\ \bibnamefont {Yuan}},
  \ and\ \bibinfo {author} {\bibfnamefont {Liang}\ \bibnamefont {Fu}},\
  }\bibfield  {title} {\enquote {\bibinfo {title} {Moir\'e quantum chemistry:
  Charge transfer in transition metal dichalcogenide superlattices},}\ }\href
  {\doibase 10.1103/PhysRevB.102.201115} {\bibfield  {journal} {\bibinfo
  {journal} {Physical Review B}\ }\textbf {\bibinfo {volume} {102}},\ \bibinfo
  {pages} {201115} (\bibinfo {year} {2020})}\BibitemShut {NoStop}%
\bibitem [{\citenamefont {Shahmoon}\ \emph {et~al.}(2017)\citenamefont
  {Shahmoon}, \citenamefont {Wild}, \citenamefont {Lukin},\ and\ \citenamefont
  {Yelin}}]{Shahmoon2017}%
  \BibitemOpen
  \bibfield  {author} {\bibinfo {author} {\bibfnamefont {Ephraim}\ \bibnamefont
  {Shahmoon}}, \bibinfo {author} {\bibfnamefont {Dominik~S}\ \bibnamefont
  {Wild}}, \bibinfo {author} {\bibfnamefont {Mikhail~D}\ \bibnamefont {Lukin}},
  \ and\ \bibinfo {author} {\bibfnamefont {Susanne~F}\ \bibnamefont {Yelin}},\
  }\bibfield  {title} {\enquote {\bibinfo {title} {Cooperative resonances in
  light scattering from two-dimensional atomic arrays},}\ }\href {\doibase
  10.1103/PhysRevLett.118.113601} {\bibfield  {journal} {\bibinfo  {journal}
  {Physical Review Letters}\ }\textbf {\bibinfo {volume} {118}},\ \bibinfo
  {pages} {113601} (\bibinfo {year} {2017})}\BibitemShut {NoStop}%
\bibitem [{\citenamefont {Rui}\ \emph {et~al.}(2020)\citenamefont {Rui},
  \citenamefont {Wei}, \citenamefont {Rubio-Abadal}, \citenamefont {Hollerith},
  \citenamefont {Zeiher}, \citenamefont {Stamper-Kurn}, \citenamefont {Gross},\
  and\ \citenamefont {Bloch}}]{Rui2020}%
  \BibitemOpen
  \bibfield  {author} {\bibinfo {author} {\bibfnamefont {Jun}\ \bibnamefont
  {Rui}}, \bibinfo {author} {\bibfnamefont {David}\ \bibnamefont {Wei}},
  \bibinfo {author} {\bibfnamefont {Antonio}\ \bibnamefont {Rubio-Abadal}},
  \bibinfo {author} {\bibfnamefont {Simon}\ \bibnamefont {Hollerith}}, \bibinfo
  {author} {\bibfnamefont {Johannes}\ \bibnamefont {Zeiher}}, \bibinfo {author}
  {\bibfnamefont {Dan~M}\ \bibnamefont {Stamper-Kurn}}, \bibinfo {author}
  {\bibfnamefont {Christian}\ \bibnamefont {Gross}}, \ and\ \bibinfo {author}
  {\bibfnamefont {Immanuel}\ \bibnamefont {Bloch}},\ }\bibfield  {title}
  {\enquote {\bibinfo {title} {A subradiant optical mirror formed by a single
  structured atomic layer},}\ }\href {\doibase 10.1038/s41586-020-2463-x}
  {\bibfield  {journal} {\bibinfo  {journal} {Nature}\ }\textbf {\bibinfo
  {volume} {583}},\ \bibinfo {pages} {369--374} (\bibinfo {year}
  {2020})}\BibitemShut {NoStop}%
\bibitem [{\citenamefont {Moreno-Cardoner}\ \emph {et~al.}(2021)\citenamefont
  {Moreno-Cardoner}, \citenamefont {Goncalves},\ and\ \citenamefont
  {Chang}}]{MorenoCardoner2021}%
  \BibitemOpen
  \bibfield  {author} {\bibinfo {author} {\bibfnamefont {M}~\bibnamefont
  {Moreno-Cardoner}}, \bibinfo {author} {\bibfnamefont {D}~\bibnamefont
  {Goncalves}}, \ and\ \bibinfo {author} {\bibfnamefont {D.~E.}\ \bibnamefont
  {Chang}},\ }\bibfield  {title} {\enquote {\bibinfo {title} {Quantum nonlinear
  optics based on two-dimensional rydberg atom arrays},}\ }\href {\doibase
  10.1103/PhysRevLett.127.263602} {\bibfield  {journal} {\bibinfo  {journal}
  {Physical Review Letters}\ }\textbf {\bibinfo {volume} {127}},\ \bibinfo
  {pages} {263602} (\bibinfo {year} {2021})}\BibitemShut {NoStop}%
\end{thebibliography}%


\begin{thebibliography}{11}%
\makeatletter
\providecommand \@ifxundefined [1]{%
 \@ifx{#1\undefined}
}%
\providecommand \@ifnum [1]{%
 \ifnum #1\expandafter \@firstoftwo
 \else \expandafter \@secondoftwo
 \fi
}%
\providecommand \@ifx [1]{%
 \ifx #1\expandafter \@firstoftwo
 \else \expandafter \@secondoftwo
 \fi
}%
\providecommand \natexlab [1]{#1}%
\providecommand \enquote  [1]{``#1''}%
\providecommand \bibnamefont  [1]{#1}%
\providecommand \bibfnamefont [1]{#1}%
\providecommand \citenamefont [1]{#1}%
\providecommand \href@noop [0]{\@secondoftwo}%
\providecommand \href [0]{\begingroup \@sanitize@url \@href}%
\providecommand \@href[1]{\@@startlink{#1}\@@href}%
\providecommand \@@href[1]{\endgroup#1\@@endlink}%
\providecommand \@sanitize@url [0]{\catcode `\\12\catcode `\$12\catcode `\&12\catcode `\#12\catcode `\^12\catcode `\_12\catcode `\%12\relax}%
\providecommand \@@startlink[1]{}%
\providecommand \@@endlink[0]{}%
\providecommand \url  [0]{\begingroup\@sanitize@url \@url }%
\providecommand \@url [1]{\endgroup\@href {#1}{\urlprefix }}%
\providecommand \urlprefix  [0]{URL }%
\providecommand \Eprint [0]{\href }%
\providecommand \doibase [0]{http://dx.doi.org/}%
\providecommand \selectlanguage [0]{\@gobble}%
\providecommand \bibinfo  [0]{\@secondoftwo}%
\providecommand \bibfield  [0]{\@secondoftwo}%
\providecommand \translation [1]{[#1]}%
\providecommand \BibitemOpen [0]{}%
\providecommand \bibitemStop [0]{}%
\providecommand \bibitemNoStop [0]{.\EOS\space}%
\providecommand \EOS [0]{\spacefactor3000\relax}%
\providecommand \BibitemShut  [1]{\csname bibitem#1\endcsname}%
\let\auto@bib@innerbib\@empty
\bibitem [{\citenamefont {Haug}\ and\ \citenamefont {Schmitt-Rink}(1984)}]{Haug1984Electron}%
  \BibitemOpen
  \bibfield  {author} {\bibinfo {author} {\bibfnamefont {H.}~\bibnamefont {Haug}}\ and\ \bibinfo {author} {\bibfnamefont {S.}~\bibnamefont {Schmitt-Rink}},\ }\bibfield  {title} {\enquote {\bibinfo {title} {Electron theory of the optical properties of laser-excited semiconductors},}\ }\href@noop {} {\bibfield  {journal} {\bibinfo  {journal} {Progress in Quantum Electronics}\ }\textbf {\bibinfo {volume} {9}},\ \bibinfo {pages} {3 -- 100} (\bibinfo {year} {1984})}\BibitemShut {NoStop}%
\bibitem [{\citenamefont {Rivera}\ \emph {et~al.}(2018)\citenamefont {Rivera}, \citenamefont {Yu}, \citenamefont {Seyler}, \citenamefont {Wilson}, \citenamefont {Yao},\ and\ \citenamefont {Xu}}]{Rivera2018Interlayer}%
  \BibitemOpen
  \bibfield  {author} {\bibinfo {author} {\bibfnamefont {Pasqual}\ \bibnamefont {Rivera}}, \bibinfo {author} {\bibfnamefont {Hongyi}\ \bibnamefont {Yu}}, \bibinfo {author} {\bibfnamefont {Kyle~L}\ \bibnamefont {Seyler}}, \bibinfo {author} {\bibfnamefont {Nathan~P}\ \bibnamefont {Wilson}}, \bibinfo {author} {\bibfnamefont {Wang}\ \bibnamefont {Yao}}, \ and\ \bibinfo {author} {\bibfnamefont {Xiaodong}\ \bibnamefont {Xu}},\ }\bibfield  {title} {\enquote {\bibinfo {title} {Interlayer valley excitons in heterobilayers of transition metal dichalcogenides},}\ }\href {\doibase 10.1038/s41565-018-0193-0} {\bibfield  {journal} {\bibinfo  {journal} {Nature Nanotechnology}\ }\textbf {\bibinfo {volume} {13}},\ \bibinfo {pages} {1004--1015} (\bibinfo {year} {2018})}\BibitemShut {NoStop}%
\bibitem [{Note1()}]{Note1}%
  \BibitemOpen
  \bibinfo {note} {We show later that the physical subspace excludes $\nu _{\tau ;\protect \bm {R}}$ above a certain value.}\BibitemShut {Stop}%
\bibitem [{\citenamefont {Auerbach}(1994)}]{Auerbach1994}%
  \BibitemOpen
  \bibfield  {author} {\bibinfo {author} {\bibfnamefont {A.}~\bibnamefont {Auerbach}},\ }\href@noop {} {\emph {\bibinfo {title} {Interacting Electrons and Quantum Magnetism}}}\ (\bibinfo  {publisher} {Springer},\ \bibinfo {year} {1994})\BibitemShut {NoStop}%
\bibitem [{\citenamefont {Wu}\ \emph {et~al.}(2018)\citenamefont {Wu}, \citenamefont {Lovorn},\ and\ \citenamefont {MacDonald}}]{Wu2018Theory}%
  \BibitemOpen
  \bibfield  {author} {\bibinfo {author} {\bibfnamefont {Fengcheng}\ \bibnamefont {Wu}}, \bibinfo {author} {\bibfnamefont {Timothy}\ \bibnamefont {Lovorn}}, \ and\ \bibinfo {author} {\bibfnamefont {A~H}\ \bibnamefont {MacDonald}},\ }\bibfield  {title} {\enquote {\bibinfo {title} {Theory of optical absorption by interlayer excitons in transition metal dichalcogenide heterobilayers},}\ }\href {\doibase 10.1103/PhysRevB.97.035306} {\bibfield  {journal} {\bibinfo  {journal} {Physical Review B}\ }\textbf {\bibinfo {volume} {97}},\ \bibinfo {pages} {35306} (\bibinfo {year} {2018})}\BibitemShut {NoStop}%
\bibitem [{\citenamefont {Tran}\ \emph {et~al.}(2019)\citenamefont {Tran}, \citenamefont {Moody}, \citenamefont {Wu}, \citenamefont {Lu}, \citenamefont {Choi}, \citenamefont {Kim}, \citenamefont {Rai}, \citenamefont {Sanchez}, \citenamefont {Quan}, \citenamefont {Singh}, \citenamefont {Embley}, \citenamefont {Zepeda}, \citenamefont {Campbell}, \citenamefont {Autry}, \citenamefont {Taniguchi}, \citenamefont {Watanabe}, \citenamefont {Lu}, \citenamefont {Banerjee}, \citenamefont {Silverman}, \citenamefont {Kim}, \citenamefont {Tutuc}, \citenamefont {Yang}, \citenamefont {MacDonald},\ and\ \citenamefont {Li}}]{Tran2019Evidence}%
  \BibitemOpen
  \bibfield  {author} {\bibinfo {author} {\bibfnamefont {Kha}\ \bibnamefont {Tran}}, \bibinfo {author} {\bibfnamefont {Galan}\ \bibnamefont {Moody}}, \bibinfo {author} {\bibfnamefont {Fengcheng}\ \bibnamefont {Wu}}, \bibinfo {author} {\bibfnamefont {Xiaobo}\ \bibnamefont {Lu}}, \bibinfo {author} {\bibfnamefont {Junho}\ \bibnamefont {Choi}}, \bibinfo {author} {\bibfnamefont {Kyounghwan}\ \bibnamefont {Kim}}, \bibinfo {author} {\bibfnamefont {Amritesh}\ \bibnamefont {Rai}}, \bibinfo {author} {\bibfnamefont {Daniel~A}\ \bibnamefont {Sanchez}}, \bibinfo {author} {\bibfnamefont {Jiamin}\ \bibnamefont {Quan}}, \bibinfo {author} {\bibfnamefont {Akshay}\ \bibnamefont {Singh}}, \bibinfo {author} {\bibfnamefont {Jacob}\ \bibnamefont {Embley}}, \bibinfo {author} {\bibfnamefont {André}\ \bibnamefont {Zepeda}}, \bibinfo {author} {\bibfnamefont {Marshall}\ \bibnamefont {Campbell}}, \bibinfo {author} {\bibfnamefont {Travis}\ \bibnamefont {Autry}}, \bibinfo {author} {\bibfnamefont {Takashi}\ \bibnamefont {Taniguchi}},
  \bibinfo {author} {\bibfnamefont {Kenji}\ \bibnamefont {Watanabe}}, \bibinfo {author} {\bibfnamefont {Nanshu}\ \bibnamefont {Lu}}, \bibinfo {author} {\bibfnamefont {Sanjay~K}\ \bibnamefont {Banerjee}}, \bibinfo {author} {\bibfnamefont {Kevin~L}\ \bibnamefont {Silverman}}, \bibinfo {author} {\bibfnamefont {Suenne}\ \bibnamefont {Kim}}, \bibinfo {author} {\bibfnamefont {Emanuel}\ \bibnamefont {Tutuc}}, \bibinfo {author} {\bibfnamefont {Li}~\bibnamefont {Yang}}, \bibinfo {author} {\bibfnamefont {Allan~H}\ \bibnamefont {MacDonald}}, \ and\ \bibinfo {author} {\bibfnamefont {Xiaoqin}\ \bibnamefont {Li}},\ }\bibfield  {title} {\enquote {\bibinfo {title} {Evidence for moiré excitons in van der waals heterostructures},}\ }\href {\doibase 10.1038/s41586-019-0975-z} {\bibfield  {journal} {\bibinfo  {journal} {Nature}\ }\textbf {\bibinfo {volume} {567}},\ \bibinfo {pages} {71--75} (\bibinfo {year} {2019})}\BibitemShut {NoStop}%
\bibitem [{\citenamefont {Conti}\ \emph {et~al.}(2020)\citenamefont {Conti}, \citenamefont {Neilson}, \citenamefont {Peeters},\ and\ \citenamefont {Perali}}]{Conti2020}%
  \BibitemOpen
  \bibfield  {author} {\bibinfo {author} {\bibfnamefont {Sara}\ \bibnamefont {Conti}}, \bibinfo {author} {\bibfnamefont {David}\ \bibnamefont {Neilson}}, \bibinfo {author} {\bibfnamefont {François~M}\ \bibnamefont {Peeters}}, \ and\ \bibinfo {author} {\bibfnamefont {Andrea}\ \bibnamefont {Perali}},\ }\bibfield  {title} {\enquote {\bibinfo {title} {Transition metal dichalcogenides as strategy for high temperature electron-hole superfluidity},}\ }\href@noop {} {\bibfield  {journal} {\bibinfo  {journal} {Condensed Matter}\ }\textbf {\bibinfo {volume} {5}},\ \bibinfo {pages} {22} (\bibinfo {year} {2020})}\BibitemShut {NoStop}%
\bibitem [{\citenamefont {Park}\ \emph {et~al.}(2023)\citenamefont {Park}, \citenamefont {Zhu}, \citenamefont {Wang}, \citenamefont {Wang}, \citenamefont {Holtzmann}, \citenamefont {Taniguchi}, \citenamefont {Watanabe}, \citenamefont {Yan}, \citenamefont {Fu}, \citenamefont {Cao}, \citenamefont {Xiao}, \citenamefont {Gamelin}, \citenamefont {Yu}, \citenamefont {Yao},\ and\ \citenamefont {Xu}}]{Park2023Dipole}%
  \BibitemOpen
  \bibfield  {author} {\bibinfo {author} {\bibfnamefont {Heonjoon}\ \bibnamefont {Park}}, \bibinfo {author} {\bibfnamefont {Jiayi}\ \bibnamefont {Zhu}}, \bibinfo {author} {\bibfnamefont {Xi}~\bibnamefont {Wang}}, \bibinfo {author} {\bibfnamefont {Yingqi}\ \bibnamefont {Wang}}, \bibinfo {author} {\bibfnamefont {William}\ \bibnamefont {Holtzmann}}, \bibinfo {author} {\bibfnamefont {Takashi}\ \bibnamefont {Taniguchi}}, \bibinfo {author} {\bibfnamefont {Kenji}\ \bibnamefont {Watanabe}}, \bibinfo {author} {\bibfnamefont {Jiaqiang}\ \bibnamefont {Yan}}, \bibinfo {author} {\bibfnamefont {Liang}\ \bibnamefont {Fu}}, \bibinfo {author} {\bibfnamefont {Ting}\ \bibnamefont {Cao}}, \bibinfo {author} {\bibfnamefont {Di}~\bibnamefont {Xiao}}, \bibinfo {author} {\bibfnamefont {Daniel~R}\ \bibnamefont {Gamelin}}, \bibinfo {author} {\bibfnamefont {Hongyi}\ \bibnamefont {Yu}}, \bibinfo {author} {\bibfnamefont {Wang}\ \bibnamefont {Yao}}, \ and\ \bibinfo {author} {\bibfnamefont {Xiaodong}\ \bibnamefont {Xu}},\ }\bibfield
  {title} {\enquote {\bibinfo {title} {Dipole ladders with large hubbard interaction in a moiré exciton lattice},}\ }\href {\doibase 10.1038/s41567-023-02077-5} {\bibfield  {journal} {\bibinfo  {journal} {Nature Physics}\ } (\bibinfo {year} {2023}),\ 10.1038/s41567-023-02077-5}\BibitemShut {NoStop}%
\bibitem [{\citenamefont {Zhang}\ \emph {et~al.}(2020)\citenamefont {Zhang}, \citenamefont {Yuan},\ and\ \citenamefont {Fu}}]{Zhang2020}%
  \BibitemOpen
  \bibfield  {author} {\bibinfo {author} {\bibfnamefont {Yang}\ \bibnamefont {Zhang}}, \bibinfo {author} {\bibfnamefont {Noah F~Q}\ \bibnamefont {Yuan}}, \ and\ \bibinfo {author} {\bibfnamefont {Liang}\ \bibnamefont {Fu}},\ }\bibfield  {title} {\enquote {\bibinfo {title} {Moir\'e quantum chemistry: Charge transfer in transition metal dichalcogenide superlattices},}\ }\href {\doibase 10.1103/PhysRevB.102.201115} {\bibfield  {journal} {\bibinfo  {journal} {Physical Review B}\ }\textbf {\bibinfo {volume} {102}},\ \bibinfo {pages} {201115} (\bibinfo {year} {2020})}\BibitemShut {NoStop}%
\bibitem [{\citenamefont {Karni}\ \emph {et~al.}(2022)\citenamefont {Karni}, \citenamefont {Barré}, \citenamefont {Pareek}, \citenamefont {Georgaras}, \citenamefont {Man}, \citenamefont {Sahoo}, \citenamefont {Bacon}, \citenamefont {Zhu}, \citenamefont {Ribeiro}, \citenamefont {O’Beirne}, \citenamefont {Hu}, \citenamefont {Al-Mahboob}, \citenamefont {Abdelrasoul}, \citenamefont {Chan}, \citenamefont {Karmakar}, \citenamefont {Winchester}, \citenamefont {Kim}, \citenamefont {Watanabe}, \citenamefont {Taniguchi}, \citenamefont {Barmak}, \citenamefont {Madéo}, \citenamefont {da~Jornada}, \citenamefont {Heinz},\ and\ \citenamefont {Dani}}]{Karni2022Structure}%
  \BibitemOpen
  \bibfield  {author} {\bibinfo {author} {\bibfnamefont {Ouri}\ \bibnamefont {Karni}}, \bibinfo {author} {\bibfnamefont {Elyse}\ \bibnamefont {Barré}}, \bibinfo {author} {\bibfnamefont {Vivek}\ \bibnamefont {Pareek}}, \bibinfo {author} {\bibfnamefont {Johnathan~D}\ \bibnamefont {Georgaras}}, \bibinfo {author} {\bibfnamefont {Michael K~L}\ \bibnamefont {Man}}, \bibinfo {author} {\bibfnamefont {Chakradhar}\ \bibnamefont {Sahoo}}, \bibinfo {author} {\bibfnamefont {David~R}\ \bibnamefont {Bacon}}, \bibinfo {author} {\bibfnamefont {Xing}\ \bibnamefont {Zhu}}, \bibinfo {author} {\bibfnamefont {Henrique~B}\ \bibnamefont {Ribeiro}}, \bibinfo {author} {\bibfnamefont {Aidan~L}\ \bibnamefont {O’Beirne}}, \bibinfo {author} {\bibfnamefont {Jenny}\ \bibnamefont {Hu}}, \bibinfo {author} {\bibfnamefont {Abdullah}\ \bibnamefont {Al-Mahboob}}, \bibinfo {author} {\bibfnamefont {Mohamed M~M}\ \bibnamefont {Abdelrasoul}}, \bibinfo {author} {\bibfnamefont {Nicholas~S}\ \bibnamefont {Chan}}, \bibinfo {author} {\bibfnamefont
  {Arka}\ \bibnamefont {Karmakar}}, \bibinfo {author} {\bibfnamefont {Andrew~J}\ \bibnamefont {Winchester}}, \bibinfo {author} {\bibfnamefont {Bumho}\ \bibnamefont {Kim}}, \bibinfo {author} {\bibfnamefont {Kenji}\ \bibnamefont {Watanabe}}, \bibinfo {author} {\bibfnamefont {Takashi}\ \bibnamefont {Taniguchi}}, \bibinfo {author} {\bibfnamefont {Katayun}\ \bibnamefont {Barmak}}, \bibinfo {author} {\bibfnamefont {Julien}\ \bibnamefont {Madéo}}, \bibinfo {author} {\bibfnamefont {Felipe~H}\ \bibnamefont {da~Jornada}}, \bibinfo {author} {\bibfnamefont {Tony~F}\ \bibnamefont {Heinz}}, \ and\ \bibinfo {author} {\bibfnamefont {Keshav~M}\ \bibnamefont {Dani}},\ }\bibfield  {title} {\enquote {\bibinfo {title} {Structure of the moiré exciton captured by imaging its electron and hole},}\ }\href {\doibase 10.1038/s41586-021-04360-y} {\bibfield  {journal} {\bibinfo  {journal} {Nature}\ }\textbf {\bibinfo {volume} {603}},\ \bibinfo {pages} {247--252} (\bibinfo {year} {2022})}\BibitemShut {NoStop}%
\bibitem [{\citenamefont {Zeng}\ and\ \citenamefont {MacDonald}(2022)}]{Zeng2022Strong}%
  \BibitemOpen
  \bibfield  {author} {\bibinfo {author} {\bibfnamefont {Yongxin}\ \bibnamefont {Zeng}}\ and\ \bibinfo {author} {\bibfnamefont {Allan~H}\ \bibnamefont {MacDonald}},\ }\bibfield  {title} {\enquote {\bibinfo {title} {Strong modulation limit of excitons and trions in moir\'e materials},}\ }\href {\doibase 10.1103/PhysRevB.106.035115} {\bibfield  {journal} {\bibinfo  {journal} {Physical Review B}\ }\textbf {\bibinfo {volume} {106}},\ \bibinfo {pages} {35115} (\bibinfo {year} {2022})}\BibitemShut {NoStop}%
\end{thebibliography}%
\end{document}


\title{Supplementary material: Non-bosonic moir\'e excitons}

\author{Tsung-Sheng Huang}
\affiliation{Joint Quantum Institute, University of Maryland, College Park, MD 20742, USA}

\author{Peter Lunts}
\affiliation{Department of Physics, Harvard University, Cambridge MA 02138, USA}
\affiliation{Joint Quantum Institute, University of Maryland, College Park, MD 20742, USA}

\author{Mohammad Hafezi}
\affiliation{Joint Quantum Institute, University of Maryland, College Park, MD 20742, USA}

\date{\today}

\maketitle

\section{Two-body Schroedinger equation for a single exciton} 
\label{Appendix:BSE}

This section reviews the derivation for the two-body Schroedinger equation (SE)~\cite{Haug1984Electron}, giving the eigenvalue problem for a single exciton.
We start from the microscopic electron-hole Hamiltonian $\hat{H}_{\mathrm{eh}}=\hat{H}_{\mathrm{eh}}^0+\hat{V}$ used in the main text. $\hat{H}_{\mathrm{eh}}^0$ is the sector with non-interacting charges, which we reproduce below:
\begin{equation}
\hat{H}_{\mathrm{eh}}^0
=
\sum_{\tau}
\int d\bm{r}
\hat{c}_{\tau}^\dagger(\bm{r})
\hat{h}_c(\bm{r})
\hat{c}_{\tau}(\bm{r})
+
\hat{v}_{\tau}^\dagger(\bm{r})
\hat{h}_v(\bm{r})
\hat{v}_{\tau}(\bm{r})
,\quad
\hat{h}_c(\bm{r})
=
-
\frac{\hbar^2\nabla_{\bm{r}}^2}{2m_c}
+
\Delta_c(\bm{r})
,\quad
\hat{h}_v(\bm{r})
=
-
\frac{\hbar^2\nabla_{\bm{r}}^2}{2m_v}
+
\Delta_v(\bm{r})
,
\end{equation}
with valley pseudospin $\tau$ and position $\bm{r}$.
$\hat{c}_{\tau}(\bm{r})$ and $\hat{v}_{\tau}(\bm{r})$ are annihilation operators of conduction band electron and valence band hole (shorthand notation of $\hat{\psi}_{c,\tau}(\bm{r})$ and $\hat{\psi}_{v,\tau}(\bm{r})$ used in the main text).
$m_c$ and $m_v$ are effective masses of the corresponding charges, and $\Delta_c(\bm{r})$ and $\Delta_v(\bm{r})$ are their moir\'e potentials.
We also reproduce the interaction between fermions as follows:
\begin{equation}
\hat{V}
=
\frac{e^2}{2\epsilon_r}
\int d\bm{r} d\bm{r}'
\left[
\frac{
\hat{\rho}_{c}(\bm{r})
\hat{\rho}_{c}(\bm{r}')
+
\hat{\rho}_{v}(\bm{r})
\hat{\rho}_{v}(\bm{r}')
}{|\bm{r}-\bm{r}'|}
-
\frac{
2
\hat{\rho}_c(\bm{r})
\hat{\rho}_v(\bm{r}')
}{|\bm{r}-\bm{r}'+d_z\bm{e}_z|}
\right]
,\;
\hat{\rho}_c(\bm{r})
=
\sum_{\tau}
\hat{c}_{\tau}^\dagger(\bm{r})
\hat{c}_{\tau}(\bm{r})
,\;
\hat{\rho}_v(\bm{r})
=
\sum_{\tau}
\hat{v}_{\tau}^\dagger(\bm{r})
\hat{v}_{\tau}(\bm{r})
,
\end{equation}
with $e$, $\epsilon_r$, and $d_z\bm{e}_z$ being the electron charge, the dielectric constant, and the displacement between the two layers ($\bm{e}_z$ denotes the unit vector perpendicular to the layers), respectively.
To describe the two-body bound state, we consider the two-particle Green's function following $\hat{H}_{\mathrm{eh}}$:
\begin{equation}
\begin{gathered}
{\cal{G}}_{\tau}(\bm{r},\bm{r}',E)
=
-
\int dt
e^{iEt}
\langle \mathrm{vac}|
e^{i\hat{H}_{\mathrm{eh}}t}
\hat{v}_{\tau}(\bm{r}')
\hat{c}_{\tau}(\bm{r})
e^{-i\hat{H}_{\mathrm{eh}}t}
|cv\rangle
,
\end{gathered}
\end{equation}
where $E$ represents the total energy of the two particles, and $t$ denotes time. 
$|\mathrm{vac}\rangle$ and $|cv\rangle$ label the vacuum and a state with one electron-hole pair, respectively. 
We consider the two charges to be at the same pseudospin $\tau$ as it is more relevant to moir\'e excitons~\cite{Rivera2018Interlayer}.
Within the ladder approximation~\cite{Haug1984Electron} and assuming $E$ is away from the particle-hole continuum, we find:
\begin{equation}
\label{eq:Bethe-Salpeter}
{\cal{G}}_{\tau}(\bm{r},\bm{r}',E)
\simeq
\sum_{\tilde{\tau}}
\int
d\bm{\tilde{r}}
d\bm{\tilde{r}}'
\langle\mathrm{vac}|
\hat{v}_{\tau}(\bm{r}')
\hat{c}_{\tau}(\bm{r})
\frac{\hat{V}}{E-\hat{H}_{\mathrm{eh}}^0}
\hat{c}_{\tilde{\tau}}^\dagger(\bm{\tilde{r}})
\hat{v}_{\tilde{\tau}}^\dagger(\bm{\tilde{r}}')
|\mathrm{vac}\rangle
{\cal{G}}_{\tilde{\tau}}(\bm{\tilde{r}},\bm{\tilde{r}}',E)
,
\end{equation}
which holds when the Kernel equals the product of delta functions in positions.
This condition could occur if $E$ is an eigenvalue of the matrix $\langle\mathrm{vac}|
\hat{v}_{\tau}(\bm{r}_v)
\hat{c}_{\tau}(\bm{r}_c)
\hat{H}_{\mathrm{eh}}
\hat{c}_{\tilde{\tau}}^\dagger(\bm{\tilde{r}}_c)
\hat{v}_{\tilde{\tau}}^\dagger(\bm{\tilde{r}}_v)
|\mathrm{vac}\rangle$, or equivalently:
\begin{equation}
\label{eq:2_body_SE}
\left[
\hat{h}_c(\bm{r}_c)
+
\hat{h}_v(\bm{r}_v)
-
\frac{e^2}{\epsilon_r \sqrt{(\bm{r}_c-\bm{r}_v)^2 + d_z^2}}
-E_{n,\bm{Q}}
\right]
\phi_{n,\bm{Q}}(\bm{r}_c,\bm{r}_v)
=0
,
\end{equation}
where $\hat{H}_{\mathrm{eh}}^0$ and $\hat{V}$ yields the first two and the third terms, respectively.
$\phi_{n,\bm{Q}}(\bm{r}_c,\bm{r}_v)=\langle\mathrm{vac}|\hat{v}_{\tau}(\bm{r}_v)
\hat{c}_{\tau}(\bm{r}_c)|\phi_{n,\bm{Q}}\rangle$ is the eigenfunction corresponding to energy $E_{n,\bm{Q}}$ and eigenstate $|\phi_{n,\bm{Q}}\rangle$, which are labeled by total momenta $\bm{Q}$ (since $\Delta_c(\bm{r})$ and $\Delta_v(\bm{r})$ are invariant under translation over moir\'e periods) and internal state index $n$.

\section{Moir\'e potential} 
\label{Appendix:moire_potential}

In this section, we discuss details of the moir\'e potential used in the main text, which we reproduce below:
\begin{equation}
\label{eq:moire_potential}
\Delta_c(\bm{r})
=
\Delta_v(\bm{r})
\equiv
\Delta(\bm{r})
=
Re\left[
Z
\sum_{j=1}^3
e^{i\bm{r}\cdot\bm{G}_j}
\right]
,\;
\bm{G}_1=\frac{2\pi}{a_M}\left(\bm{e}_x-\frac{\bm{e}_y}{\sqrt 3}\right),\;
\bm{G}_2=\frac{4\pi\bm{e}_y}{\sqrt{3}a_M},\;
\bm{G}_3=-\frac{2\pi}{a_M}\left(\bm{e}_x+\frac{\bm{e}_y}{\sqrt 3}\right)
,
\end{equation}
where $\bm{e}_x$ and $\bm{e}_y$ are in-plane Cartesian unit vectors and $a_M$ is the moir\'e period.
The complex number $Z$ characterizes this potential, with its argument determining the profile, e.g., extremum, of $\Delta(\bm{r})$ (while $|Z|$ only controls the depth).
In particular, local minima of $\Delta(\bm{r})$, $\bm{R}$, appear as superlattice translations of:
\begin{equation}
\label{eq:moire_potential_minimum}
\bm{R}
\supset
\begin{cases}
-\frac{a_M}{\sqrt{3}}\bm{e}_y, &\text{if $0^\circ<\arg (Z)<120^\circ$}
\\
0, &\text{if $120^\circ<\arg (Z)<240^\circ$}
\\
\frac{a_M}{\sqrt{3}}\bm{e}_y, &\text{if $240^\circ<\arg (Z)<360^\circ$}
.
\end{cases}
\end{equation}
Note that when $\arg(Z)$ is an integer multiple of $120^\circ$ the corresponding superlattice is hexagonal rather than triangular.
We do not focus on those values in this work.

Next, we elaborate on expansion of $\Delta(\bm{r})$ near $\bm{R}$ to quadratic order of $a_M^{-1}(\bm{r}-\bm{R})$.
This treatment is legitimate within the large-$a_M$ approximation, which assumes $\min_{\bm{R}}|\bm{r}-\bm{R}|\ll a_M$ for all positions $\bm{r}$ of interest and yields:
\begin{equation}
\label{eq:large_aM_expansion}
\Delta(\bm{r})|_{\bm{r}\simeq\bm{R}}
\simeq
\Delta(\bm{R})
+
\frac{\Delta''}{2}
(\bm{r}-\bm{R})^2
,\quad
\Delta''
=
-
\frac{16\pi^2}{a_M^2}
Re\left(Ze^{i\vartheta}\right)
,\quad
\vartheta
=
\begin{cases}
\frac{2\pi}{3}, &\text{if $0^\circ<\arg(Z)<120^\circ$}
\\
0, &\text{if $120^\circ<\arg(Z)<240^\circ$}
\\
-\frac{2\pi}{3}, &\text{if $240^\circ<\arg(Z)<360^\circ$}
\end{cases}
.
\end{equation}

In addition, the quadratic expansion on moir\'e potential simplifies the two-body SE Eq.~\eqref{eq:2_body_SE}.
Performing the following Fourier transformation into the supersite representation (with $N$ denoting the total number of supersites):
\begin{equation}
\label{eq:Exciton_Wannier_state}
\phi_{n,\bm{Q}}(\bm{r}_c,\bm{r}_v)
=
\frac{1}{\sqrt{N}}
\sum_{\bm{R}}
e^{i\bm{Q}\cdot\bm{R}}
W_{n,\bm{R}}(\bm{r}_c,\bm{r}_v)
,
\end{equation}
and assuming the orbitals $W_{n,\bm{R}}(\bm{r}_c,\bm{r}_v)$ are local (i.e., $|W_{n,\bm{R}}(\bm{r}_c,\bm{r}_v)|$ is negligible at $|\bm{r}_c-\bm{R}|$ and $|\bm{r}_v-\bm{R}|$ greater than $a_M$), we find:
\begin{equation}
\label{eq:2_body_SE_large_aM}
\left[
-\frac{\hbar^2\nabla_{\bm{r}_c}^2}{2m_c}
-\frac{\hbar^2\nabla_{\bm{r}_v}^2}{2m_v}
+
\frac{\Delta''}{2}
\big[
(\bm{r}_c-\bm{R})^2
+
(\bm{r}_v-\bm{R})^2
\big]
-
\frac{e^2}{\epsilon_r\sqrt{(\bm{r}_c-\bm{r}_v)^2 + d_z^2}}
-E_n
\right]
W_{n,\bm{R}}(\bm{r}_c,\bm{r}_v)
\simeq
0
,
\end{equation}
where the large-$a_M$ approximation suppresses the total momentum dependence of $E_{n,\bm{Q}}\to E_n$ (shifted such that $2\Delta(\bm{R})$ is aborbed).
We will perform perturbation analysis on this equation in Section~\ref{Appendix:Perturbation}.

\section{Details on many-body formulation of excitons}
\label{Appendix_Exciton_Formalism}

In this section, we elaborate on the derivation for exciton commutation relation, its emergent boson, and the effective Hamiltonian from the microscopic electron-hole model.
We focus on the lowest exciton $\hat{x}_{\tau,\bm{R}}$ with the definition reproduced below:
\begin{equation}
\label{eq:exciton_operator}
\hat{x}_{\tau;\bm{R}}^\dagger
=
\int
d\bm{r}_c
d\bm{r}_v
w_{\bm{R}}(\bm{r}_c,\bm{r}_v)
\hat{c}_{\tau}^\dagger(\bm{r}_c)
\hat{v}_{\tau}^\dagger(\bm{r}_v)
,
\end{equation}
where $w_{\bm{R}}(\bm{r}_c,\bm{r}_v)=W_{0,\bm{R}}(\bm{r}_c,\bm{r}_v)$.
More specifically, we aim at a many-body formulation for excitons within the following subspace~\footnote{We show later that the physical subspace excludes $\nu_{\tau;\bm{R}}$ above a certain value. 
}:
\begin{equation}
\label{eq:exciton_Fock}
{\mathcal{V}}
=
\mathrm{Span}
\left\{
\prod_{\tau;\bm{R}}
\frac{
(
\hat{x}_{\tau;\bm{R}}^{\dagger}
)^{\nu_{\tau;\bm{R}}}
}{\sqrt{C^{(\nu_{\tau;\bm{R}})}}}
|\mathrm{vac}\rangle
,\quad
\forall
\nu_{\tau;\bm{R}}\in\{0,1,2,...\}
\right\}
,\quad
C^{(\nu)}
=
\langle\mathrm{vac}|
(
\hat{x}_{\tau;\bm{R}}
)^{\nu}
(
\hat{x}_{\tau;\bm{R}}^{\dagger}
)^{\nu}
|\mathrm{vac}\rangle
.
\end{equation}
For this purpose, a systematic approach to project a generic operator in terms of fermions, denoted as $\hat{O}(\hat{c},\hat{v})$, into $\mathcal{V}$ is required.
Direct implementation of such projection might be complicated as it involves high-order expectation values such as $\langle\mathrm{vac}|
\hat{x}_{\tau;\bm{R}}^{\nu'}
\hat{O}(\hat{c},\hat{v})
\hat{x}_{\tau;\bm{R}}^{\dagger\nu}
|\mathrm{vac}\rangle$.

We can proceed with an alternative approach, which utilizes the fact that $\hat{\mathcal{O}}(\hat{x})$, an operator in terms of excitons, faithfully represents $\hat{O}(\hat{c},\hat{v})$ in $\mathcal{V}$ if the following relation holds:
\begin{equation}
[
\hat{\mathcal{O}}(\hat{x})
,
\hat{x}_{\tau;\bm{R}}^\dagger
]
=
[
\hat{O}(\hat{c},\hat{v})
,
\hat{x}_{\tau;\bm{R}}^\dagger
]_{\mathcal{V}}
,\quad
\hat{\mathcal{O}}(\hat{x})
|\mathrm{vac}\rangle
=
\hat{O}(\hat{c},\hat{v})
|\mathrm{vac}\rangle
=
O|\mathrm{vac}\rangle
,
\end{equation}
where projection of the commutator into ${\mathcal{V}}$ is indicated by the subscript and $O$ is the eigenvalue of both $\hat{\mathcal{O}}(\hat{x})$ and $\hat{O}(\hat{c},\hat{v})$ for vacuum state.
We illustrate the above statement with the following example:
\begin{equation}
\langle\mathrm{vac}|
\hat{x}_{\tau;\bm{R}}^{\nu'}
\hat{O}(\hat{c},\hat{v})
\hat{x}_{\tau;\bm{R}}^{\dagger\nu}
|\mathrm{vac}\rangle
=
\sum_{k=0}^{\nu-1}
\langle\mathrm{vac}|
\hat{x}_{\tau;\bm{R}}^{\nu'}
\hat{x}_{\tau;\bm{R}}^{\dagger k}
[
\hat{O}(\hat{c},\hat{v})
,
\hat{x}_{\tau;\bm{R}}^{\dagger}
]
\hat{x}_{\tau;\bm{R}}^{\dagger(\nu-1-k)}
|\mathrm{vac}\rangle
+
O
\langle\mathrm{vac}|
\hat{x}_{\tau;\bm{R}}^{\nu'}
\hat{x}_{\tau;\bm{R}}^{\dagger\nu}
|\mathrm{vac}\rangle
,
\end{equation}
suggesting the expectation value on the left-hand side be uniquely determined by $[\hat{O}(\hat{c},\hat{v}),\hat{x}_{\tau;\bm{R}}^{\dagger}]_{\mathcal{V}}$ and $O$.
Therefore, $\hat{\mathcal{O}}(\hat{x})$ is an equivalent representation of $\hat{O}(\hat{c},\hat{v})$ if it reproduces these quantities.
In principle, one can construct $\hat{\mathcal{O}}(\hat{x})$ from $[
\hat{O}(\hat{c},\hat{v}),\hat{x}_{\tau;\bm{R}}^\dagger]$ (together with $O$) if this commutator has a simple expression in terms of $\hat{x}_{\tau;\bm{R}}^\dagger$ and $\hat{x}_{\tau;\bm{R}}$ (while $\hat{O}(\hat{c},\hat{v})$ itself may not be as simple).
In the following sections, we will utilize this method to derive the exciton commutation relations and Hamiltonian.

\subsection{The large moir\'e period limit}

We begin with a superlattice in the large-$a_M$ limit for its simplicity in analytics.
More specifically, negligible overlapping between off-site Wannier orbitals would lead to commuting excitons at different sites and suppressed tunneling.
Thus, the many-body problem of excitons reduces to evaluating on-site statistics and Hamiltonian, which we discuss in the following.

\subsubsection{On-site exciton commutator and the effective boson representation} 
\label{Appendix_Exciton_comm}

\begin{figure}[t]
\centering
\includegraphics[width=\columnwidth]{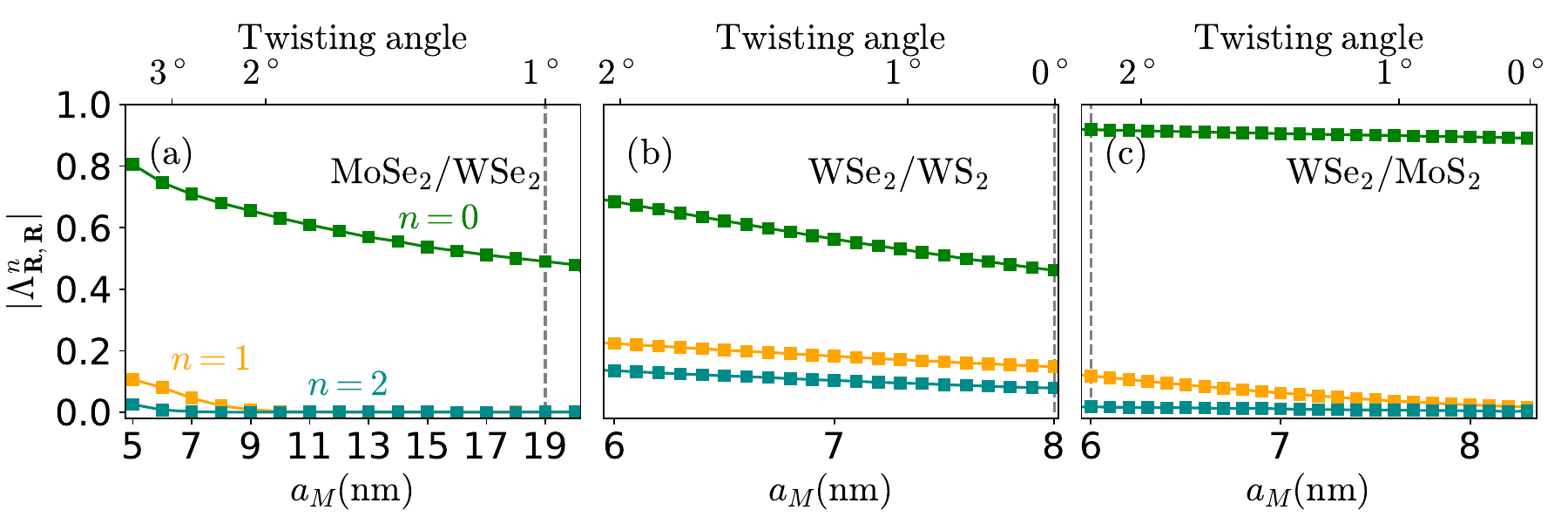}
\caption{
Mutual exchange integral $\Lambda_{\bm{R},\bm{R}}^n$ between the lowest and higher orbitals on the same site. 
The parameters used for the three materials are elaborated in Section~\ref{Appendix_Parameters}.
}
\label{Fig_mutual_exchange_exp}
\end{figure}

In this section, we aim at an equivalent representation of:
\begin{equation}
\begin{aligned}
\label{eq:X_commutator}
\hat{y}_{\tau;\bm{R}}
\equiv
\left[
\hat{x}_{\tau;\bm{R}}
,
\hat{x}_{\tau;\bm{R}}^\dagger
\right]
,
\end{aligned}
\end{equation}
in $\mathcal{V}$ utilizing the approach discussed above, which requires $(\hat{y}_{\tau;\bm{R}}-1)|\mathrm{vac}\rangle=0$ and the knowledge of $[\hat{y}_{\tau;\bm{R}},\hat{x}_{\tau;\bm{R}}^\dagger]$.
The former follows from direct computation in the fermion basis and always holds regardless of exciton statistics, and the latter is evaluated (again in the fermion basis) as:
\begin{equation}
[\hat{y}_{\tau;\bm{R}},\hat{x}_{\tau;\bm{R}}^\dagger]
=
-
2
\sum_{n,\bm{R}'}
\Lambda_{\bm{R},\bm{R}'}^n
\hat{X}_{\tau;n,\bm{R}'}^\dagger
,
\end{equation}
where $\Lambda_{\bm{R},\bm{R}'}^n$ is the exchange integral generalized to involve higher orbitals:
\begin{equation}
\Lambda_{\bm{R},\bm{R}'}^n
=
\int
d\bm{r}_c
d\bm{r}_v
d\bm{r}'_c
d\bm{r}'_v
W_{n,\bm{R}'}^\ast(\bm{r}_c,\bm{r}'_v)
w_{\bm{R}}(\bm{r}_c,\bm{r}_v)
w_{\bm{R}}(\bm{r}'_c,\bm{r}'_v)
w_{\bm{R}}^\ast(\bm{r}'_c,\bm{r}_v)
,
\end{equation}
and excited states at internal state $n$ are described by the following exciton operator:
\begin{equation}
\hat{X}_{\tau;n,\bm{R}}^\dagger
=
\int
d\bm{r}_c
d\bm{r}_v
W_{n,\bm{R}}(\bm{r}_c,\bm{r}_v)
\hat{c}_{\tau}^\dagger(\bm{r}_c)
\hat{v}_{\tau}^\dagger(\bm{r}_v)
.
\end{equation}
Note that $\hat{x}_{\tau;\bm{R}}^\dagger=\hat{X}_{\tau;0,\bm{R}}^\dagger$.
For simplicity, we assume $\Lambda\equiv\Lambda_{\bm{R},\bm{R}}^0$ to dominate over all other $\Lambda_{\bm{R},\bm{R}'}^n$ with $\bm{R}\neq\bm{R}'$ or $n\neq0$.
The former is consistent with the large-$a_M$ approximation, and the latter is justified in Fig.~\ref{Fig_mutual_exchange_exp}.
Dropping minor corrections, we find:
\begin{equation}[\hat{y}_{\tau;\bm{R}},\hat{x}_{\tau;\bm{R}}^\dagger]_{\mathcal{V}}
\simeq
-2\Lambda
\hat{x}_{\tau;\bm{R}}^\dagger
,
\end{equation}
which, together with Eq.~\eqref{eq:X_commutator}, become commutation relations for angular momentum operators $(\hat{\mathcal{S}}_{\tau;\bm{R}}^+,\hat{\mathcal{S}}_{\tau;\bm{R}}^-,\hat{\mathcal{S}}_{\tau;\bm{R}}^z)$:
\begin{equation}
\left[
\hat{\mathcal{S}}_{\tau;\bm{R}}^+
,
\hat{\mathcal{S}}_{\tau;\bm{R}}^-
\right]
=
2\hat{\mathcal{S}}_{\tau;\bm{R}}^z
,\quad
\left[
\hat{\mathcal{S}}_{\tau;\bm{R}}^z
,
\hat{\mathcal{S}}_{\tau;\bm{R}}^-
\right]
\simeq
-
\hat{\mathcal{S}}_{\tau;\bm{R}}^-
,
\end{equation}
provided the following scaling:
\begin{equation}
\hat{x}_{\tau;\bm{R}}
=
\sqrt{\Lambda}
\hat{\mathcal{S}}_{\tau;\bm{R}}^+
,\quad
\hat{x}_{\tau;\bm{R}}^\dagger
=
\sqrt{\Lambda}
\hat{\mathcal{S}}_{\tau;\bm{R}}^-
,\quad
\hat{y}_{\tau;\bm{R}}
=
2\Lambda
\hat{\mathcal{S}}_{\tau;\bm{R}}^z
.
\end{equation}
Combining the above relations, $\hat{\mathcal{S}}_{\tau;\bm{R}}^+|\mathrm{vac}\rangle=0$, and $\hat{\mathcal{S}}_{\tau;\bm{R}}^z|\mathrm{vac}\rangle=(2\Lambda)^{-1}|\mathrm{vac}\rangle$, we find the excitons behaving as spin-$(2\Lambda)^{-1}$ operators.
Note that $(2\Lambda)^{-1}$ does not have to be multiple of $\frac{1}{2}$ because these fictitious spins are not generators of rotations.

Finally, the similarity between exciton and angular momentum operators suggests the following Holstein-Primakoff (HP) transformation for these composite particles:
\begin{equation}
\label{eq:effective_boson_rep}
\hat{x}_{\tau;\bm{R}}
\simeq
\theta(1-\Lambda\hat{a}_{\tau;\bm{R}}^\dagger\hat{a}_{\tau;\bm{R}}) \,
\sqrt{1-\Lambda\hat{a}_{\tau;\bm{R}}^\dagger\hat{a}_{\tau;\bm{R}}}
\; \hat{a}_{\tau;\bm{R}}
,\quad
\hat{\mathcal{S}}_{\tau;\bm{R}}^z
=
\theta(1-\Lambda\hat{a}_{\tau;\bm{R}}^\dagger\hat{a}_{\tau;\bm{R}})
\frac{1-2\Lambda
\hat{a}_{\tau;\bm{R}}^\dagger\hat{a}_{\tau;\bm{R}}}{2\Lambda}
,
\end{equation}
with the emergent bosonic operators $\hat{a}_{\tau;\bm{R}}$ and $\hat{a}_{\tau;\bm{R}}^\dagger$.
Note that the step function $\theta(1-\Lambda\hat{a}_{\tau;\bm{R}}^\dagger\hat{a}_{\tau;\bm{R}})$ is implemented because $\langle
\hat{x}_{\tau;\bm{R}}\hat{x}_{\tau;\bm{R}}^\dagger
\rangle\geq0$ for any state, giving an occupancy bound.

\subsubsection{On-site exciton Hamiltonian below occupancy bound} 
\label{Appendix:exciton Hamiltonian}

\begin{figure}[t]
\centering
\includegraphics[width=\columnwidth]{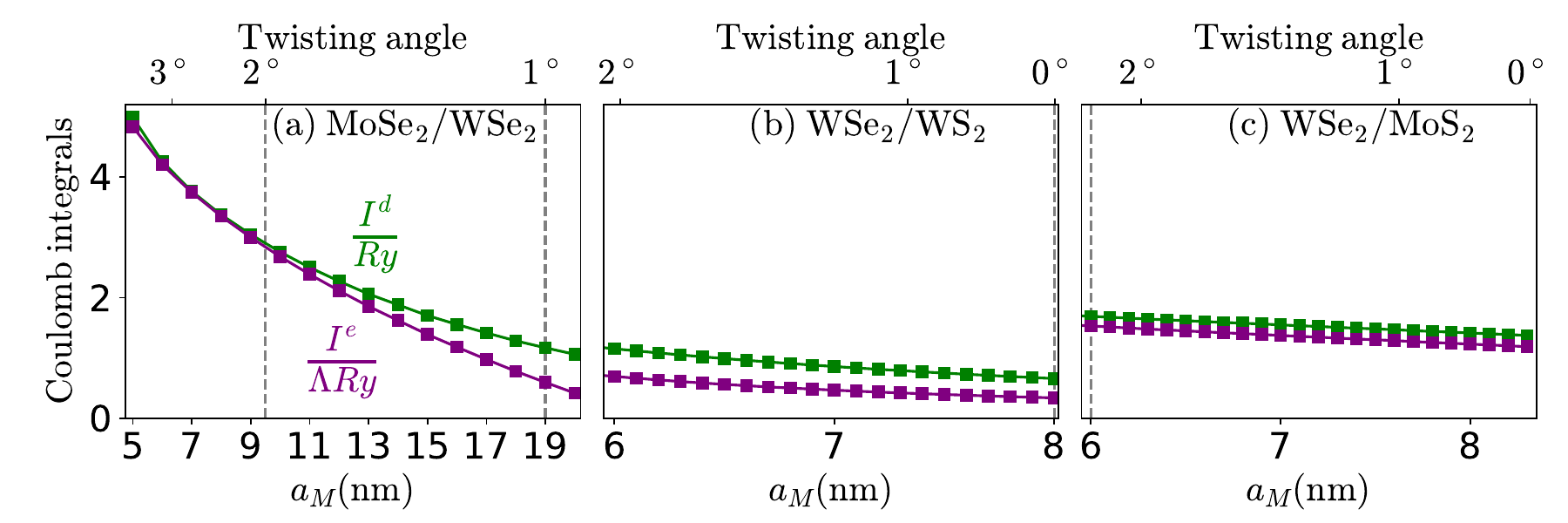}
\caption{
Coulomb integrals in units of Rydberg constant $Ry=\frac{\hbar^2}{2\mu a_B^2}$ ($\mu=\frac{m_cm_v}{m_c+m_v}$ is the reduced mass and $a_B=\frac{\epsilon_r\hbar^2}{\mu e^2}$ is the Bohr radius) for three bilayers.
The parameters used are elaborated in Section~\ref{Appendix_Parameters}.
}
\label{Fig_Coulomb_integral_exp}
\end{figure}

We proceed to derive the equivalent description of $\hat{H}_{\mathrm{eh}}$ in $\mathcal{V}$, $\hat{H}_{\mathrm{eff}}$, in the large-$a_M$ limit utilizing the aforementioned approach.
Such method requires $\hat{H}_{\mathrm{eh}}|\mathrm{vac}\rangle=0$ and:
\begin{equation}
\label{eq:H_eh_x_comm}
\begin{aligned}\relax
[\hat{H}_{\mathrm{eh}},\hat{x}_{\tau;\bm{R}}^\dagger]
&\simeq
E_0
\hat{x}_{\tau;\bm{R}}^\dagger
-
\sum_{\bm{R}'\neq\bm{R}}
t_{\bm{R}',\bm{R}}
\hat{x}_{\tau;\bm{R}'}^\dagger
+
\hat{F}_{\tau;\bm{R}}
,
\end{aligned}
\end{equation}
where we dropped higher orbitals as in section~\ref{Appendix_Exciton_comm}.
The first term captures the single exciton occupation energy:
\begin{equation}
E_0
=
\int 
d\bm{r}_c
d\bm{r}_v
w_{\bm{R}}^\ast(\bm{r}_c,\bm{r}_v)
\left[
\hat{h}_c(\bm{r}_c)
+
\hat{h}_v(\bm{r}_v)
-
\frac{e^2}{\epsilon_r\sqrt{(\bm{r}_c-\bm{r}_v)^2+d_z^2}}
\right]
w_{\bm{R}}(\bm{r}_c,\bm{r}_v)
.
\end{equation}
The second term contains the hopping integral for the lowest exciton:
\begin{equation}
t_{\bm{R}',\bm{R}}
=
-
\int 
d\bm{r}_c
d\bm{r}_v
w_{\bm{R}'}^\ast(\bm{r}_c,\bm{r}_v)
\left[
\hat{h}_c(\bm{r}_c)
+
\hat{h}_v(\bm{r}_v)
-
\frac{e^2}{\epsilon_r\sqrt{(\bm{r}_c-\bm{r}_v)^2+d_z^2}}
\right]
w_{\bm{R}}(\bm{r}_c,\bm{r}_v)
.
\end{equation}
We drop $t_{\bm{R}',\bm{R}}$ in this section (and discuss this term in section~\ref{Appendix_tunneling}) because it involves overlap between off-site orbitals and thus are negligible in the large-$a_M$ limit.
The last term results from the Coulomb interaction:
\begin{equation}
\hat{F}_{\tau;\bm{R}}
=
\int
d\bm{r}_c
d\bm{r}_v
w_{\bm{R}}(\bm{r}_c,\bm{r}_v)
\hat{c}_{\tau}^\dagger(\bm{r}_c)
\hat{v}_{\tau}^\dagger(\bm{r}_v)
\int
d\bm{r}
\left[
\mathcal{U}(\bm{r}_c,\bm{r}_v;\bm{r},\infty)
\hat{\rho}_c(\bm{r})
+
\mathcal{U}(\bm{r}_c,\bm{r}_v;\infty,\bm{r})
\hat{\rho}_v(\bm{r})
\right]
,
\end{equation}
\begin{equation}
\label{eq:V_dipole_dipole}
\begin{aligned}
\mathcal{U}(\bm{r}_c,\bm{r}_v;\bm{r}'_c,\bm{r}'_v)
=
\frac{e^2}{\epsilon_r}
\bigg[
\frac{1}{|\bm{r}_c-\bm{r}'_c|}
+
\frac{1}{|\bm{r}_v-\bm{r}'_v|}
-
\frac{1}{|\bm{r}_c-\bm{r}'_v+d_z\bm{e}_z|}
-
\frac{1}{|\bm{r}_v-\bm{r}'_c+d_z\bm{e}_z|}
\bigg]
,
\end{aligned}
\end{equation}
whose equivalent representation in $\mathcal{V}$ is not straightforward.
Thus, we obtain such description by repeating the aforementioned procedure, which requires $[\hat{F}_{\tau;\bm{R}},\hat{x}_{\tau';\bm{R}}^\dagger]$ (and $\hat{F}_{\tau;\bm{R}}|\mathrm{vac}\rangle=0$).
In the large-$a_M$ limit, this commutator becomes:
\begin{equation}
[\hat{F}_{\tau;\bm{R}},\hat{x}_{\tau';\bm{R}}^\dagger]_{\mathcal{V}}
\simeq
\delta_{\tau,\tau'}
\frac{I^d-I^e}{1-\Lambda}
\hat{x}_{\tau;\bm{R}}^{\dagger2}
+
(1-\delta_{\tau,\tau'})
I^d
\hat{x}_{\tau;\bm{R}}^\dagger
\hat{x}_{-\tau;\bm{R}}^\dagger
,
\end{equation}
where $I^d$ and $I^e$ are the direct and exchange Coulomb integrals:
\begin{equation}
\label{eq:Id}
\begin{aligned}
I^d
=
\int
d\bm{r}_c
d\bm{r}_v
d\bm{r}'_c
d\bm{r}'_v
\mathcal{U}(\bm{r}_c,\bm{r}_v;\bm{r}'_c,\bm{r}'_v)
\left|
w_{\bm{R}}(\bm{r}_c,\bm{r}_v)
w_{\bm{R}}(\bm{r}'_c,\bm{r}'_v)
\right|^2
\end{aligned}
\end{equation}
\begin{equation}
\label{eq:Ie}
\begin{aligned}
I^e
=
\int
d\bm{r}_c
d\bm{r}_v
d\bm{r}'_c
d\bm{r}'_v
w_{\bm{R}}^\ast(\bm{r}_c,\bm{r}'_v)
w_{\bm{R}}(\bm{r}_c,\bm{r}_v)
\mathcal{U}(\bm{r}_c,\bm{r}_v;\bm{r}'_c,\bm{r}'_v)
w_{\bm{R}}(\bm{r}'_c,\bm{r}'_v)
w_{\bm{R}}^\ast(\bm{r}'_c,\bm{r}_v)
.
\end{aligned}
\end{equation}
With these expressions, we find the equivalent representation of $\hat{F}_{\tau;\bm{R}}$ in $\mathcal{V}$ as:
\begin{equation}
\hat{F}_{\tau;\bm{R}}\big|_{\mathcal{V}}
=
\hat{x}_{\tau;\bm{R}}^\dagger
\left[
\frac{I^d-I^e}{1-\Lambda}
\left(
\frac{1}{2\Lambda}
-
\hat{\mathcal{S}}_{\tau;\bm{R}}^z
\right)
+
I^d
\left(
\frac{1}{2\Lambda}
-
\hat{\mathcal{S}}_{-\tau;\bm{R}}^z
\right)
\right]
,
\end{equation}
which directly yields the effective Hamiltonian in the large-$a_M$ limit:
\begin{equation}
\hat{H}_{\mathrm{eh}}\big|_{\mathcal{V}}
=
\hat{H}_{\mathrm{eff}}
\simeq
-
E_0
\sum_{\bm{R},\tau}
\hat{\mathcal{S}}_{\tau;\bm{R}}^z
+
\sum_{\bm{R},\tau,\tau'}
\frac{U_{\tau,\tau'}}{2}
\left(
\frac{1}{2\Lambda}
-
\hat{\mathcal{S}}_{\tau;\bm{R}}^z
\right)
\left(
\frac{1}{2\Lambda}
-
\hat{\mathcal{S}}_{\tau';\bm{R}}^z
\right)
,\quad
U_{\tau,\tau}
=
\frac{I^d-I^e}{1-\Lambda}
,\quad
U_{\tau,-\tau}
=
I^d
.
\end{equation}
This Hamiltonian can be further simplified by considering approximations on $I^e$, which becomes the following expression utilizing the completeness relation of the Wannier orbitals:
\begin{equation}
\begin{aligned}
I^e
=
\sum_{n,n'}
(\Lambda_{n}^{n'})^\ast
\int
d\bm{r}_cd\bm{r}_vd\bm{r}'_cd\bm{r}'_v
W_{n,\bm{R}}^\ast(\bm{r}_c,\bm{r}_v)
w_{\bm{R}}(\bm{r}_c,\bm{r}_v)
\mathcal{U}(\bm{r}_c,\bm{r}_v;\bm{r}'_c,\bm{r}'_v)
w_{\bm{R}}(\bm{r}'_c,\bm{r}'_v)
W_{n',\bm{R}}^\ast(\bm{r}'_c,\bm{r}'_v)
,
\end{aligned}
\end{equation}
where $\Lambda_{n}^{n'}$ is the exchange integral between different on-site orbitals:
\begin{equation}
\begin{aligned}
\Lambda_{n}^{n'}
=
\int
d\bm{r}_cd\bm{r}_vd\bm{r}'_cd\bm{r}'_v
W_{n,\bm{R}}^\ast(\bm{r}_c,\bm{r}'_v)
w_{\bm{R}}(\bm{r}_c,\bm{r}_v)
w_{\bm{R}}(\bm{r}'_c,\bm{r}'_v)
W_{n',\bm{R}}^\ast(\bm{r}'_c,\bm{r}_v)
.
\end{aligned}
\end{equation}
Similar to the derivation in section~\ref{Appendix_Exciton_comm}, we assume $\Lambda=\Lambda_0^0$ dominating over all other $\Lambda_{n}^{n'}$, yielding:
\begin{equation}
\label{eq:Ie_approx}
I^e
\simeq
\Lambda
I^d
,
\end{equation}
which is benchmarked for different materials and twisting angles in Fig.~\ref{Fig_Coulomb_integral_exp}.
We find Eq.~\eqref{eq:Ie_approx} a good approximation for MoSe\textsubscript{2}/WSe\textsubscript{2} and WSe\textsubscript{2}/MoS\textsubscript{2} but rather fragile for WSe\textsubscript{2}/WS\textsubscript{2}.
This observation is consistent with Fig.~\ref{Fig_mutual_exchange_exp}, suggesting $|\Lambda_n^0|$ with $n\neq0$ is generally more significant for such a material.
With Eq.~\eqref{eq:Ie_approx}, we obtain $U_{\tau,\tau}\simeq I^d\equiv U$ and:
\begin{equation}
\hat{H}_{\mathrm{eff}}
\simeq
-\mathcal{E}_0
\sum_{\bm{R},\tau}
\hat{\mathcal{S}}_{\tau;\bm{R}}^z
+
\frac{U}{2}
\sum_{\bm{R},\tau,\tau'}
\hat{\mathcal{S}}_{\tau;\bm{R}}^z
\hat{\mathcal{S}}_{\tau';\bm{R}}^z
,\quad
\mathcal{E}_0=E_0+\frac{U}{\Lambda}
,
\end{equation}
up to a constant shift.

Finally, utilizing the HP representation Eq.~\eqref{eq:effective_boson_rep}, we find the following expression for the large-$a_M$ effective Hamiltonian:
\begin{equation}
\label{eq:H_onsite_a_rep}
\hat{H}_{\mathrm{eff}}
\simeq
E_0
\sum_{\bm{R},\tau}\hat{a}_{\tau;\bm{R}}^\dagger\hat{a}_{\tau;\bm{R}}
+
\sum_{\bm{R},\tau,\tau'}
\frac{U}{2}
\hat{a}_{\tau;\bm{R}}^\dagger
\hat{a}_{\tau';\bm{R}}^\dagger
\hat{a}_{\tau';\bm{R}}
\hat{a}_{\tau;\bm{R}}
,
\end{equation}
provided that exciton occupancy at each $(\bm{R},\tau)$ does not exceed $\Lambda^{-1}$.

\subsection{Beyond the large moir\'e period limit: Exciton tunneling}
\label{Appendix_tunneling}

In this section, we incorporate tunneling term $\hat{H}_t$ as a correction to the effective exciton Hamiltonian $\hat{H}_{\mathrm{eff}}$, which requires a finite $a_M$. 
This situation is nevertheless complicated because (strictly speaking) excitons at different sites would not commute.
To make analytical progress, we will still assume commuting off-site excitons and derive $\hat{H}_t$, which satisfies:
\begin{equation}\relax
\label{eq:Ht_xdag_comm}
[\hat{H}_t,\hat{x}_{\tau;\bm{R}}^\dagger]
=
-
\sum_{\bm{R}'\neq\bm{R}}
t_{\bm{R}',\bm{R}}
\hat{x}_{\tau;\bm{R}'}^\dagger
,
\end{equation}
which directly implies $[\hat{H}_t,\sum_{\bm{R}}\hat{x}_{\tau;\bm{R}}^\dagger\hat{x}_{\tau;\bm{R}}]=0$. 
This condition divides all tunneling processes into two categories based on whether they conserve $\sum_{\bm{R}}\hat{x}_{\tau;\bm{R}}^\dagger\hat{x}_{\tau;\bm{R}}$, or equivalently $\sum_{\bm{R}}\hat{\mathcal{S}}_{\tau;\bm{R}}^z(\hat{\mathcal{S}}_{\tau;\bm{R}}^z+1)$.
To be more specific, we consider hopping of a $\tau$-valley exciton from site $\bm{R}$ to $\bm{R}'$, which couples $|\nu,\nu'\rangle_{\bm{R},\bm{R}';\tau}$ and $|\nu-1,\nu'+1\rangle_{\bm{R},\bm{R}';\tau}$.
These states and the corresponding tunneling matrix element are defined as:
\begin{equation}
|\nu,\nu'\rangle_{\bm{R},\bm{R}';\tau}
=
\frac{
(\hat{x}_{\tau;\bm{R}}^\dagger)^\nu
(\hat{x}_{\tau;\bm{R}'}^\dagger)^{\nu'}
}{\sqrt{C^{(\nu)}C^{(\nu')}}}
|\Phi\rangle
,\quad
\mathcal{T}_{\bm{R};\nu\to\nu-1}^{\bm{R}';\nu'\to\nu'+1}
=
{}_{\bm{R},\bm{R}';\tau}
\langle\nu-1,\nu'+1|
\hat{H}_t
|\nu,\nu'\rangle_{\bm{R},\bm{R}';\tau}
,
\end{equation}
with $|\Phi\rangle$ being a state containing generic excitons at sites and valleys other than $(\bm{R},\tau)$ and $(\bm{R}',\tau)$.
From these expressions, we find that the processes with $\nu'=\nu-1$ conserves $\sum_{\bm{R}}\hat{\mathcal{S}}_{\tau;\bm{R}}^z(\hat{\mathcal{S}}_{\tau;\bm{R}}^z+1)$, while the others do not.
Tunneling matrix elements of the former category follow directly from Eq.~\eqref{eq:Ht_xdag_comm} as:
\begin{equation}
\mathcal{T}_{\bm{R};\nu\to\nu-1}^{\bm{R}';\nu-1\to\nu}
=
-
\nu
t_{\bm{R'},\bm{R}}
,
\end{equation}
which can be captured by the following tunneling Hamiltonian in the emergent boson representation:
\begin{equation}
\label{eq:Ht_a_rep}
\hat{H}_t
=
-
\sum_{\tau,\bm{R}\neq\bm{R}'}
t_{\bm{R'},\bm{R}}
\hat{a}_{\tau;\bm{R}'}^\dagger
\hat{a}_{\tau;\bm{R}}
.
\end{equation}

We nevertheless note that Eq.~\eqref{eq:Ht_a_rep} is insufficient to capture $\mathcal{T}_{\bm{R};\nu\to\nu-1}^{\bm{R}';\nu'\to\nu'+1}$ with $\nu'\neq\nu-1$.
More specifically, the description of these processes is beyond the assumption of commuting off-site excitons (and hence the emergent boson representation), which leads to a contradiction: 
$[\hat{H}_t,\sum_{\bm{R}}\hat{x}_{\tau;\bm{R}}^\dagger\hat{x}_{\tau;\bm{R}}]=0$ implies $\mathcal{T}_{\bm{R};\nu\to\nu-1}^{\bm{R}';\nu'\to\nu'+1}\sim\delta_{\nu',\nu-1}$, but the ones with $\nu'\neq\nu-1$ are generally non-zero according to direct computation within the charge basis.
This indicates the necessity of off-site exciton commutators (which are beyond the scope of this work) to capture these tunneling processes.

Such complexity can nonetheless be neglected in the presence of a large on-site repulsion $U\gg|\mathcal{T}_{\bm{R};\nu\to\nu-1}^{\bm{R}';\nu'\to\nu'+1}|$, which provides a significant energy separation $\simeq (\nu-\nu'-1)U$ between $|\nu,\nu'\rangle_{\bm{R},\bm{R}';\tau}$ and $|\nu-1,\nu'+1\rangle_{\bm{R},\bm{R}';\tau}$ with $\nu'\neq\nu-1$.
Thus, although the tunneling process between them is captured by $\mathcal{T}_{\bm{R};\nu\to\nu-1}^{\bm{R}';\nu'\to\nu'+1}\sim t_{\bm{R}',\bm{R}}$, they only contribute as a second order perturbation $\sim t_{\bm{R}',\bm{R}}^2/U$ at low energy (close to that of initial state)~\cite{Auerbach1994}.
With this consideration, we can neglect the tunnelings with $\nu'\neq\nu-1$ and focus on the ones with $\nu'=\nu-1$ provided $U\gg|t_{\bm{R},\bm{R}'}|$, which suggests the validity of the following model in this regime:
\begin{equation}
\hat{H}_{\mathrm{eff}}
=
E_0
\sum_{\bm{R},\tau}\hat{a}_{\tau;\bm{R}}^\dagger\hat{a}_{\tau;\bm{R}}
-
t
\sum_{\langle\bm{R}',\bm{R}\rangle}
(
\hat{a}_{\tau;\bm{R}'}^\dagger
\hat{a}_{\tau;\bm{R}}
+
\hat{a}_{\tau;\bm{R}}^\dagger
\hat{a}_{\tau;\bm{R}'}
)
+
\sum_{\bm{R},\tau,\tau'}
\frac{U}{2}
\hat{a}_{\tau;\bm{R}}^\dagger
\hat{a}_{\tau';\bm{R}}^\dagger
\hat{a}_{\tau';\bm{R}}
\hat{a}_{\tau;\bm{R}}
,
\end{equation}
where we include only the nearest-neighbor hopping $t$ among all $t_{\bm{R},\bm{R}'}$.

Finally, we note that $\hat{H}_{\mathrm{eff}}$ generally possesses a rather complicated expression in terms of emergent spins when the hopping term is incorporated, which can nevertheless be simplified under various conditions.
For instance, in the dilute limit where occupancy of emergent bosons per $(\bm{R},\tau)$ is at most one (or equivalently, the expectation value of $\hat{\mathcal{S}}_{\tau;\bm{R}}^z$ is close to $\frac{1}{2\Lambda}$), the HP transformation becomes $\hat{a}_{\tau;\bm{R}}\simeq\sqrt{\Lambda}\hat{\mathcal{S}}_{\tau;\bm{R}}^+$ such that:
\begin{equation}
\label{eq:H_eff_dilute}
\begin{aligned}
\hat{\mathcal{H}}_{\mathrm{eff}}
\simeq
-\mathcal{E}_0
\sum_{\bm{R},\tau}
\hat{\mathcal{S}}_{\tau;\bm{R}}^z
+
\frac{U}{2}
\sum_{\bm{R},\tau,\tau'}
\hat{\mathcal{S}}_{\tau;\bm{R}}^z
\hat{\mathcal{S}}_{\tau';\bm{R}}^z
-
t\Lambda
\sum_{\langle\bm{R}',\bm{R}\rangle}
(
\hat{\mathcal{S}}_{\tau;\bm{R}'}^-
\hat{\mathcal{S}}_{\tau;\bm{R}}^+
+
\mathrm{H.c.}
)
.
\end{aligned}
\end{equation}
Another simplification can be achieved near $\Lambda\simeq0$, where the large-spin expansion for the HP representation gives $\hat{a}_{\tau;\bm{R}}\simeq(\frac{5}{4}-\frac{\Lambda}{2}\hat{\mathcal{S}}_{\tau;\bm{R}}^z)\sqrt{\Lambda}\hat{\mathcal{S}}_{\tau;\bm{R}}^+$ such that:
\begin{equation}
\begin{aligned}
\hat{\mathcal{H}}_{\mathrm{eff}}
\simeq
-\mathcal{E}_0
\sum_{\bm{R},\tau}
\hat{\mathcal{S}}_{\tau;\bm{R}}^z
+
\frac{U}{2}
\sum_{\bm{R},\tau,\tau'}
\hat{\mathcal{S}}_{\tau;\bm{R}}^z
\hat{\mathcal{S}}_{\tau';\bm{R}}^z
-
t\Lambda
\sum_{\langle\bm{R}',\bm{R}\rangle}
\left[
\hat{\mathcal{S}}_{\tau;\bm{R}'}^-
\left(\frac{5}{4}-\frac{\Lambda}{2}\hat{\mathcal{S}}_{\tau;\bm{R}'}^z\right)
\left(\frac{5}{4}-\frac{\Lambda}{2}\hat{\mathcal{S}}_{\tau;\bm{R}}^z\right)
\hat{\mathcal{S}}_{\tau;\bm{R}}^+
+
\mathrm{H.c.}
\right]
.
\end{aligned}
\end{equation}
Note that Eq.~\eqref{eq:H_eff_dilute} is recovered by replacing $\hat{\mathcal{S}}_{\tau;\bm{R}}^z$ and $\hat{\mathcal{S}}_{\tau;\bm{R}'}^z$ with $\frac{1}{2\Lambda}$ in the above expression.

\section{Numerical details}

\subsection{Eigenfunctions of the two-body Schroedinger equation}

We start by rewriting the two-body Schroedinger equation in center-of-mass (COM) and relative coordinates (which we denote as $\bm{r}_x=\frac{1}{M}(m_c\bm{r}_c+m_v\bm{r_v})$ and $\bm{r}_l=\bm{r}_c-\bm{r}_v$, respectively, with total mass $M=m_c+m_v$):
\begin{equation}
\label{eq:eigenvalue_equation_exciton}
\left[
-\frac{\hbar^2\nabla_{\bm{r}_x}^2}{2M}
- \frac{\hbar^2\nabla_{\bm{r}_l}^2}{2\mu}
-
\frac{\hbar^2}{\mu a_B \sqrt{\bm{r}_l^2 + d_z^2}}
+
\hat{\Delta}(\bm{r}_x,\bm{r}_l)
-
E_{n,\bm{Q}}
\right]
\tilde{\phi}_{n,\bm{Q}}(\bm{r}_x,\bm{r}_l)
=
0
,
\end{equation}
where $\mu=\frac{m_cm_v}{M}$ is the reduced mass, $a_B=\frac{\epsilon_r\hbar^2}{\mu e^2}$ is the Bohr radius, $\tilde{\phi}_{n,\bm{Q}}(\bm{r}_x,\bm{r}_l)=\phi_{n,\bm{Q}}(\bm{r}_c,\bm{r}_v)$, and the total moir\'e potential is just the sum of the two:
\begin{equation}
\label{eq:moire_potential_exciton}
\hat{\Delta}(\bm{r}_x,\bm{r}_l)
=
\Delta_c(\bm{r}_c)
+
\Delta_v(\bm{r}_v).
\end{equation}
We now show how we solve this equation numerically with the spectral method.

The exciton wavefunction is periodic in $\bm{r}_x$, so we use Bloch's theorem, $\tilde{\phi}_{n,\bm{Q}}(\bm{r}_x,\bm{r}_l)=e^{i\bm{Q}\cdot \bm{r}_x}u_{n,\bm{Q}}(\bm{r}_x,\bm{r}_l)$, where $u_{n,\bm{Q}}(\bm{r}_x,\bm{r}_l)$ has the periodicity of the triangular lattice in $\bm{r}_x$. 
Therefore, we consider only one primitive unit cell as the domain of $\bm{r}_x$, which is a parallelogram.
For $\bm{r}_l$, we choose another domain, which we take to be a square with size $a_l$. 
Furthermore, since for any finite $a_B$ the exciton wavefunction decays exponentially in $|\bm{r}_l|/a_B$, if we take a large enough $a_l$, the wavefunction at the boundary of the domain will vanish, and we can take the wavefunction to be periodic with the periodicity of the domain ($a_l\approx5a_B$ is large enough for the parameters we run, while arbitrarily accurate results can always be obtained by approaching the limit of $a_l/a_B \rightarrow \infty$). 
This allows us to use a Fourier decomposition in $\bm{r}_l$ as well.

The new SE for a given $\bm{Q}$ is now given by:
\begin{equation}
\label{eq:eigenvalue_equation_exciton_Bloch}
\left[
\frac{-\hbar^2(\nabla_{\bm{r}_x}+i\bm{Q})^2}{2M}
- \frac{\hbar^2\nabla_l^2}{2\mu}
-
\frac{\hbar^2}{\mu a_B \sqrt{r_l^2 + d_z^2}}
+
\hat{\Delta}(\bm{r}_x,\bm{r}_l)
\right]
u_{n,\bm{Q}}(\bm{r}_x,\bm{r}_l)
=
E_{n,\bm{Q}} \,
u_{n,\bm{Q}}(\bm{r}_x,\bm{r}_l)
.
\end{equation}
The wavefunction's periodicity is given by:
\begin{equation}
\begin{gathered}
u_{n,\bm{Q}}(\bm{r}_x+a_M\bm{e}_1,\bm{r}_l)
= 
u_{n,\bm{Q}}(\bm{r}_x+a_M\bm{e}_2,\bm{r}_l)
= 
u_{n,\bm{Q}}(\bm{r}_x,\bm{r}_l),
\\
u_{n,\bm{Q}}(\bm{r}_x,\bm{r}_l+a_l\bm{e}_x)
= 
u_{n,\bm{Q}}(\bm{r}_x,\bm{r}_l+a_l\bm{e}_y)
= 
u_{n,\bm{Q}}(\bm{r}_x,\bm{r}_l)
,
\end{gathered}
\end{equation}
where $\bm{e}_1 =\bm{e}_x, \bm{e}_2 = \frac{1}{2} \bm{e}_x + \frac{\sqrt{3}}{2} \bm{e}_y$ are the two primitive translation vectors for the COM coordinate on the triangular lattice. 
These periodicities allow us to define the wavefunctions in the Fourier basis $a^{(n,\bm{Q})}_{j,k,l,m}$ (in terms of four integer variables $j,k,l,m$) from Bloch functions $u_{n,\bm{Q}}(\bm{r}_x,\bm{r}_l)$ via:
\begin{equation}
\label{eq:Fourier decomp}
u_{n,\bm{Q}}(\bm{r}_x,\bm{r}_l)
= 
\sum_{j,k,l,m = - \infty}^{\infty} 
a^{(n,\bm{Q})}_{j,k,l,m} \;
e^{i (j \bm{G}_1 + k \bm{G}_2) \cdot \bm{r}_x}
e^{\frac{2\pi i}{a_l} (l r_{l,x} + m r_{l,y})}
,
\end{equation}
where we express $\bm{r}_l=r_{l,x}\bm{e}_x+r_{l,y}\bm{e}_y$.

Now we turn to evaluate the matrix elements of the Hamiltonian operator in Eq.~\eqref{eq:eigenvalue_equation_exciton_Bloch} in the Fourier basis, starting with the moir\'e potential:
\begin{equation}
\hat{\Delta}_{(j',k',l',m'),(j,k,l,m)} 
= 
\int_{\Omega_M} 
\frac{d\bm{r}_x}{\Omega_M} 
\int_{-\frac{a_l}{2}}^{\frac{a_l}{2}} 
\int_{-\frac{a_l}{2}}^{\frac{a_l}{2}}
\frac{dr_{l,x}dr_{l,y}}{\Omega_l}
\hat{\Delta}(\bm{r}_x,\bm{r}_l)
e^{i [(j-j') \bm{G}_1 + (k-k') \bm{G}_2] \cdot \bm{r}_x}
e^{\frac{2\pi i}{a_l}[(l-l') r_{l,x} + (m-m') r_{l,y}]}
,
\end{equation}
where $\Omega_M = \frac{\sqrt 3}{2} a_M^2, \Omega_l = a_l^2$ are the unit cell volumes. 
Evaluating the integrals, we find:
\begin{equation}
\begin{gathered}
\hat{\Delta}_{(j',k',l',m'),(j,k,l,m)}
= 
|Z|
\sum_{\beta=1}^3 
\sum_{q=0}^1
\sum_{\lambda=c,v}
\Delta_{j,k}^{(\beta,q)}
e^{(-1)^q i\mathrm{arg}(Z)}
\left[ (1 - \delta_{\beta,2} \delta_{l,l'})
\eta_{q,l-l'}^\lambda(\tilde G_{\beta,x})
+ \delta_{\beta,2} \delta_{l,l'}
\right]
\eta_{q,m-m'}^\lambda(\tilde G_{\beta,y})
,
\end{gathered}
\end{equation}
where:
\begin{equation}
\Delta_{j,k}^{(\beta,q)}
=
\delta_{q,0}\left[\delta_{\beta,1} \delta_{j,-1} \delta_{k,0}
+ \delta_{\beta,2} \delta_{j,0} \delta_{k,-1}
+ \delta_{\beta,3} \delta_{j,1} \delta_{k,1}\right]
+ \delta_{q,1}\left[\delta_{\beta,1} \delta_{j,1} \delta_{k,0}
+ \delta_{\beta,2} \delta_{j,0} \delta_{k,1}
+ \delta_{\beta,3} \delta_{j,-1} \delta_{k,-1}\right]
,
\end{equation}
\begin{equation}
\eta_{q,l}^c(x)
=
\frac{(-1)^l}{\pi}
\frac{\sin(\pi \frac{m_v}{M}\frac{a_l}{a_M}x)}{(-1)^{q}l+\frac{m_v}{M}\frac{a_l}{a_M}x}
,\quad
\eta_{q,l}^v(x)
=
\frac{(-1)^l}{\pi}
\frac{\sin(\pi\frac{m_c}{M}\frac{a_l}{a_M}x)}{(-1)^{q+1}l+\frac{m_c}{M}\frac{a_l}{a_M}x}
,\quad
\bm{\tilde G}_\beta
\equiv
\bm{G}_\beta \frac{a_M}{2\pi}
=
\tilde{G}_{\beta,x}
\bm{e}_x
+
\tilde{G}_{\beta,y}
\bm{e}_y
.
\end{equation}
Importantly, the matrix $\hat{\Delta}_{(j',k',l',m'),(j,k,l,m)}$ is 4d Toeplitz and can be fast multiplied via a 4d FFT.
Next, we compute the matrix elements for the Coulomb potential term in Eq.~\eqref{eq:eigenvalue_equation_exciton_Bloch}, which becomes:
\begin{equation}
\begin{split}
\hat{U}_{(j',k',l',m'),(j,k,l,m)}
& =
\frac{2\hbar^2\delta_{j,j'}\delta_{k,k'}}{\mu a_B a_l}
\int_{0}^{\frac{1}{2}} d r_{l,x}
\int_{-\frac{1}{2}}^{\frac{1}{2}} d r_{l,y}
\; 
\frac{\cos(2\pi [ (l-l')r_{l,x} + (m-m')r_{l,y}])}{\sqrt{r^2_{l,x} + r^2_{l,y} + (\frac{d_z}{a_l})^2}}.
\end{split}
\end{equation}
These integrals can now be computed numerically. To simplify matters, we note that they are symmetric under both $l-l' \rightarrow -l+l'$ and $m-m' \rightarrow -m+m'$.
Finally, the matrix elements of the kinetic terms are:
\begin{align}
- \frac{\hbar^2[(\nabla_{\bm{r}_x} + i \bm{Q})^2]_{(j',k',l',m'),(j,k,l,m)}}{2 M}
& = \frac{\hbar^2\delta_{j',j} \delta_{k',k} \delta_{l',l} \delta_{m',m}}{2 M}
\left[\left(\frac{2\pi}{a_M} j + Q_x\right)^2 + \left(\frac{2\pi}{a_M}\left(-\frac{j}{\sqrt{3}} + \frac{2k}{\sqrt{3}}\right) + Q_y\right)^2\right],
\\ 
-\frac{\hbar^2[\nabla_{\bm{r}_l}^2]_{(j',k',l',m'),(j,k,l,m)}}{2\mu}
& = 
\frac{\hbar^2\delta_{j',j} \delta_{k',k} \delta_{l',l} \delta_{m',m}}{2\mu} \left(\frac{2\pi}{a_l}\right)^2 \; 
[l^2 + m^2].
\end{align}
We note that $\bm{Q}$ lives in the BZ of the triangular lattice, which is a hexagon, and therefore $Q_x \in (-\frac{2}{3} \frac{2\pi}{a_M},\frac{2}{3} \frac{2\pi}{a_M})$ and $Q_y \in (-\frac{1}{\sqrt{3}} \frac{2\pi}{a_M},\frac{1}{\sqrt{3}} \frac{2\pi}{a_M})$.

Putting everything together, the SE now becomes a matrix equation.
We now truncate the sums over $j,k,l,m$ to $N_X$ for $j,k$ and $N_l$ for $l,m$:
\begin{equation}
\label{eq:matrix_equation_exciton_Bloch_truncated}
\sum_{j,k = - N_X}^{N_X}
\sum_{l,m = - N_l}^{N_l}
\left[
- \frac{\hbar^2(\hat{\nabla}_{\bm{r}_x} + i \bm{Q})^2}{2M}
- \frac{\hbar^2\hat{\nabla}_{\bm{r}_l}^2}{2\mu}
-
\hat U
+
\hat{\Delta}
\right]_{(j',k',l',m'),(j,k,l,m)}
a_{(j,k,l,m)}^{(n,\bm{Q})}
=
E_{n,\bm{Q}} \,
a_{(j',k',l',m')}^{(n,\bm{Q})}
\end{equation}
The problem is now reduced to an eigenvalue problem with a finite 4d Toeplitz matrix. This is solved using the LOBPCG algorithm, running on a GPU, and matrix-vector multiplication being done with FFTs. We increase $N_X,N_l$ until we see convergence and find that $N_X = N_l = 20$ is large enough in all the parameter regimes we study.

\subsection{Wannier orbitals and the resulting integrals}

The Wannier orbitals are FT of the Bloch wavefunctions:
\begin{equation}
\label{eq:Bloch_to_Wannier}
W_{n,\bm{R}}(\bm{r}_c,\bm{r}_v)
=
\frac{1}{\sqrt{N}}
\sum_{\bm{Q}}
e^{i\bm{Q}\cdot (\bm{r}_x-\bm{R})}
u_{n,\bm{Q}}(\bm{r}_x,\bm{r}_l)
,
\end{equation}
with $u_{n,\bm{Q}}(\bm{r}_x,\bm{r}_l)$ given by Eq.~\eqref{eq:Fourier decomp} together with the solution of Eq.~\eqref{eq:matrix_equation_exciton_Bloch_truncated}.
$N$ denotes the number of moir\'e sites (note the difference from $N_X$ and $N_l$).

First, we note that the phase of $u_{n,\bm{Q}}(\bm{r}_x,\bm{r}_l)$ is arbitrary and cannot be determined by the two-body Schroedinger equation.
Moreover, a generic choice of the phase would not give localized orbitals near $\bm{R}$ as it affects the $\bm{Q}$ summation.
To address this issue, we assume that the Wannier orbitals are concentrated within a moir\'e unit cell (or alternatively, corrections to the orbitals from inter-site tunneling are negligible due to the flat exciton moir\'e bands), which is consistent with the large-$a_M$ approximation utilized in our analytics.
This approximation leads the inverse of Eq.~\eqref{eq:Bloch_to_Wannier} to:
\begin{equation}
e^{i\bm{Q}\cdot\bm{r}_x}
u_{n,\bm{Q}}(\bm{r}_x,\bm{r}_l)
\simeq
\frac{1}{\sqrt{N}}
e^{i\bm{Q}\cdot\bm{R}}
W_{n,\bm{R}}(\bm{r}_c,\bm{r}_v)
,\quad
\forall
|\bm{r}_c-\bm{R}|,|\bm{r}_v-\bm{R}|\ll a_M
,
\end{equation}
which gives Wannier orbitals that are independent of phase of $u_{n,\bm{Q}}(\bm{r}_x,\bm{r}_l)$.
If the approximation is valid, orbitals obtained from distinct $\bm{Q}$ should only differ by a phase.
We confirm this by computing the orbitals from a set of $\bm{Q}$ and checking their overlap within a moir\'e unit cell.

With such a way to obtain the Wannier orbitals, we sample the following set of discretized $\bm{r}_c$ and $\bm{r}_v$ to compute integrals such as $\Lambda$:
\begin{equation}
\bm{r}_c,\bm{r}_v
\in
\left\{
\frac{a_M}{2N_X+1}
\left[
\left(
\chi_1+\frac{\chi_2}{2}
\right)
\bm{e}_x
+
\frac{\sqrt{3}\chi_2}{2}\bm{e}_y
\right]
\;\bigg|\;
\forall
\chi_{1,2}\in\{-N_X,-N_X+1,...,N_X-1,N_X\}
\right\}
.
\end{equation}
It is rather complicated to calculate $\Lambda$ with a collection of $\bm{r}_x$ and $\bm{r}_l$, because the notion of CM and relative coordinates are messed up after charge exchange.
Nevertheless, it is $\{\bm{r}_x,\bm{r}_l\}$ that yields directly as the reciprocal lattice of the momentum set used in the previous section, and that normalization of the orbitals is guaranteed only with such a set of $\bm{r}_x$ and $\bm{r}_l$.
Owing to this issue, we normalize the wavefunctions afterward such that summation of $|W_{n,\bm{R}}(\bm{r}_c,\bm{r}_v)|^2$ over $\{\bm{r}_c,\bm{r}_v\}$ equals one before computing the integrals.

\subsection{Parameters}
\label{Appendix_Parameters}

For MoSe\textsubscript{2}/WSe\textsubscript{2}, we set the charge masses as $(m_c,m_v)=(0.49,0.35)m_0$~\cite{Wu2018Theory} with $m_0$ being free electron mass, the Bohr radius $a_B=1$nm, and moir\'e potential parameters $\arg(Z)=\pi$ and $2|Z|=18$meV~\cite{Tran2019Evidence}.
For WSe\textsubscript{2}/WS\textsubscript{2}, we use $(m_c,m_v)=(0.33,0.3)m_0$~\cite{Conti2020}, $a_B=2$nm~\cite{Park2023Dipole}, and $\arg(Z)=\frac{5}{4}\pi$~\cite{Zhang2020}.
For WSe\textsubscript{2}/MoS\textsubscript{2}, $(m_c,m_v)=(0.7,0.42)m_0$, $a_B=2$nm~\cite{Karni2022Structure}, and $\arg(Z)=220^\circ$~\cite{Zhang2020}. 

For the last two materials, we estimate $|Z|$ from the exciton Wannier orbital size $a_W^x$, as they are provided in the literature --- $a_W^x=2.8$nm for WSe\textsubscript{2}/WS\textsubscript{2} with $a_M=8$nm~\cite{Park2023Dipole} and $a_W^x=0.9$nm for WSe\textsubscript{2}/MoS\textsubscript{2} with $a_M=6$nm~\cite{Karni2022Structure}.
$a_W^x$ and $|Z|\sim|\Delta''|$ are (approximately) related as:
\begin{equation}
a_W^x
\equiv
\sqrt{
\int d\bm{r}_cd\bm{r}_v
|w_{\bm{R}}(\bm{r}_c,\bm{r}_v)|^2
(\bm{r}_x-\bm{R})^2
}
\simeq
\left(\frac{\hbar^2}{2M\Delta''}\right)^{1/4}
,
\end{equation}
which originates from approximating $\Delta_c(\bm{r}_c)+\Delta_v(\bm{r}_v)$ following Eq.~\eqref{eq:large_aM_expansion} and dropping all $\bm{r}_l$-dependent terms therein for simplicity.

\section{Perturbation approaches for eigenvalue equation of moir\'e exciton} 
\label{Appendix:Perturbation}

In this section, we will discuss perturbation schemes within the large-$a_M$ limit, which simplifies the two-body SE to Eq.~\eqref{eq:2_body_SE_large_aM}. We will explore different approximations to this equation under the strong Coulomb and deep moir\'e regimes.
In the strong Coulomb limit, the relative motion of an exciton depends mainly on Coulomb attraction rather than moir\'e potential, whereas the opposite scenario occurs in the deep moir\'e regime.
In the following sections, we will elaborate on these two situations but restrict our analysis to $d_z=0$ for simplicity.

\subsection{Strong Coulomb regime}

We start by expressing the approximated two-body SE as:
\begin{equation}
\left[
-\frac{\hbar^2\nabla_{\bm{r}_x}^2}{2M}
+
\Delta''
(\bm{r}_x-\bm{R})^2
-\frac{\hbar^2\nabla_{\bm{r}_l}^2}{2\mu}
-
\frac{\hbar^2}{\mu a_B r_l}
+\delta\hat{h}_{\bm{R}}^{SC}(\bm{r}_x,\bm{r}_l)
-E_n
\right]
W_{n,\bm{R}}(\bm{r}_c,\bm{r}_v)
\simeq
0
,
\end{equation}
with the coupling between COM and relative degrees of freedom treated as perturbation:
\begin{equation}
\begin{aligned}
\delta\hat{h}_{\bm{R}}^{SC}
=
\frac{m_v-m_c}{M}
\Delta''
(\bm{r}_x-\bm{R})\cdot\bm{r}_l
+
\frac{m_v^2+m_c^2}{2M^2}
\Delta''
r_l^2
.
\end{aligned}
\end{equation}
Exciton wavefunctions to generic order correction follow directly from standard perturbation theory.
Here we consider only the unperturbed states (as they already contain electron-hole correlation).
The lowest orbital is then:
\begin{equation}
w_{\bm{R}}(\bm{r}_c,\bm{r}_v)
\simeq
\frac{2\sqrt{2}}{\pi a_Ba_W^x}
\exp\left[-\frac{(\bm{r}_x-\bm{R})^2}{2(a_W^x)^2}-\frac{2r_l}{a_B}\right]
,
\end{equation}
which gives the length scales:
\begin{equation}
\label{eq:length_scales_SC}
\int d\bm{r}_cd\bm{r}_v
|w_{\bm{R}}(\bm{r}_c,\bm{r}_v)|^2
r_l^2
=
\frac{3a_B^2}{8}
,\quad
\int d\bm{r}_cd\bm{r}_v
|w_{\bm{R}}(\bm{r}_c,\bm{r}_v)|^2
(\bm{r}_x-\bm{R})^2
=
(a_W^x)^2
=
\frac{\hbar}{\sqrt{2M\Delta''}}
.
\end{equation}
Accordingly, the two terms in $\delta\hat{h}_{\bm{R}}^{SC}$ scale as $\sim\frac{\hbar^2a_B}{M(a_W^x)^3}$ and $\sim\frac{\hbar^2a_B^2}{M(a_W^x)^4}$, respectively, which are required to be smaller than the Coulomb binding $\sim\frac{\hbar^2}{\mu a_B^2}$.
Thus, together with the large-$a_M$ assumption, we find that this perturbation scheme is valid when $a_B\ll a_W^x\ll a_M$.

\subsection{Deep moir\'e regime}

In the deep moir\'e regime, we treat the Coulomb interaction in Eq.~\eqref{eq:2_body_SE_large_aM} as a perturbation.
Accordingly, the zeroth order wavefunction is the product state of 2-dimensional harmonic oscillators from the two charges.
The lowest unperturbed orbital is then:
\begin{equation}
w_{\bm{R}}(\bm{r}_c,\bm{r}_v)
\simeq
\frac{
\exp\left[-\frac{(\bm{r}_c-\bm{R})^2}{2(a_W^c)^2}-\frac{(\bm{r}_v-\bm{R})^2}{2(a_W^v)^2}\right]
}{\pi a_W^ca_W^v}
,\quad
a_W^c
=
\left(\frac{\hbar^2}{m_c\Delta''}\right)^{1/4}
,\quad
a_W^v
=
\left(\frac{\hbar^2}{m_v\Delta''}\right)^{1/4}
,
\end{equation}
which gives the following length scales:
\begin{equation}
\int d\bm{r}_cd\bm{r}_v
|w_{\bm{R}}(\bm{r}_c,\bm{r}_v)|^2
r_l^2
=
(a_W^c)^2
+
(a_W^v)^2
,\;
(a_W^x)^2
\equiv
\int d\bm{r}_cd\bm{r}_v
|w_{\bm{R}}(\bm{r}_c,\bm{r}_v)|^2
(\bm{r}_x-\bm{R})^2
=
\frac{m_c^2(a_W^c)^2+m_v^2(a_W^v)^2}{M^2}
,
\end{equation}
telling that $a_W^x$ is comparable to $a_W^c$ and $a_W^v$ (for comparable charge masses) and that the relative separation does not scale with $a_B$ (unlike in the Strong Coulomb case).

Such a zeroth order wavefunction is insufficient to describe an exciton because it lacks electron-hole correlation.
Thus, we seek for the wavefunction to first order correction:
\begin{equation}
\begin{aligned}
w_{\bm{R}}(\bm{r}_c,\bm{r}_v)
\simeq
\frac{1}{\sqrt{\cal{N}}}
\sum_{\bm{n}_c,\bm{n}_v}
\gamma_{\bm{n}_c}^{\bm{n}_v}
\varphi_{n_c^x}(x_c)
\varphi_{n_c^y}(y_c)
\varphi_{n_v^x}(x_v)
\varphi_{n_v^y}(y_v)
,\quad
\bm{r}_c
=
x_c\bm{e}_x+y_c\bm{e}_y
,\quad
\bm{r}_v
=
x_v\bm{e}_x+y_v\bm{e}_y
,
\end{aligned}
\end{equation}
where $\bm{n}_c=(n_c^x,n_c^y)$ and $\bm{n}_v=(n_v^x,n_v^y)$ are non-negative integers labeling the harmonic ladders.
$\varphi_{n}(x)$ are eigenfunctions of one-dimensional harmonic oscillator with energy ladders labeled by $n$.
${\cal{N}}=\sum_{\bm{n}_c,\bm{n}_v}\left|\gamma_{\bm{n}_c}^{\bm{n}_v}\right|^2$ is the normalization constant with:
\begin{equation}
\gamma_{\bm{n}_c}^{\bm{n}_v}
=
\begin{cases}
1,\quad\mathrm{if}\;\bm{n}_c,\bm{n}_v=0
,
\\
-\frac{V_{\bm{n}_c,\bm{n}_v}^{0,0}
}{
\hbar[\omega_c(n_c^x+n_c^y)+\omega_v(n_v^x+n_v^y)]
}
,\quad
\mathrm{else}
\end{cases}
,\quad
\omega_c
=
\frac{\hbar}{m_c(a_W^c)^2}
,\quad
\omega_v
=
\frac{\hbar}{m_v(a_W^v)^2}
,
\end{equation}
where $V_{\bm{n}_c,\bm{n}_v}^{0,0}$ are matrix elements of the Coulomb attraction in the 2-dimensional harmonic ladder basis.
Following standard procedure~\cite{Zeng2022Strong}, we obtain the expression below when $n_c^x+n_v^x$ and $n_c^y+n_v^y$ are both even numbers:
\begin{equation}
\begin{aligned}
V_{\bm{n}_c,\bm{n}_v}^{0,0}
&=
-
\frac{\hbar^2\sqrt{2^{n_t}}}{\pi\mu a_B}
\frac{
\left(-1\right)^{\frac{n_c-n_v}{2}}
}{
\sqrt{n_c^x!n_c^y!n_v^x!n_v^y!}
}
\frac{
\left(a_W^c\right)^{n_c}
\left(a_W^v\right)^{n_v}
}{[(a_W^c)^2+(a_W^v)^2]^{\frac{1+n_t}{2}}}
\frac{
\Gamma(\frac{n_c^x+n_v^x+1}{2})
\Gamma(\frac{n_c^y+n_v^y+1}{2})
\Gamma\left(\frac{n_t+1}{2}\right)
}{\Gamma(\frac{n_t}{2}+1)}
,
\end{aligned}
\end{equation}
and $V_{\bm{n}_c,\bm{n}_v}^{0,0}=0$ otherwise.
We denote the total level for each respective charge as $n_c=n_c^x+n_c^y$ and $n_v=n_v^x+n_v^y$, and define $n_t=n_c+n_v$.
$\Gamma(x)$ is the gamma function.
Assuming $a_W^c\simeq a_W^v$ and $m_c\simeq m_v$, these expressions imply that the first order corrections scale with $\sim a_W^x/a_B$.
Together with the large-$a_M$ assumption, we find that this perturbation theory is valid when $a_W^x\ll a_B\ll a_M$.

\bibliography{Biblio}